\documentclass[pdftex, 11pt]{article}
\usepackage[english]{babel}
\usepackage{algorithm}
\usepackage{algorithmic}
\usepackage{amsfonts}
\usepackage{amsmath}
\usepackage{mathtools}
\usepackage{amsthm}
\usepackage{amssymb}
\usepackage{bbm}
\usepackage{blindtext}
\usepackage{booktabs}
\usepackage{xcolor}
\usepackage{geometry}
\usepackage{tabularx}
\usepackage{setspace}

\usepackage{times}
\usepackage{url}
\usepackage{graphicx}
\usepackage{bmpsize}
\usepackage{natbib}
\usepackage[normalem]{ulem}
\usepackage{caption}

\usepackage{titlesec}
\titleformat*{\section}{\large\bfseries}
\titleformat*{\subsection}{\it}

\def\zero{{\text{\boldmath$0$}}}

\def\d{{\text{\boldmath $d$}}}
\def\m{{\text{\boldmath $m$}}}
\def\p{{\text{\boldmath $p$}}}
\def\s{{\text{\boldmath $s$}}}
\def\x{{\text{\boldmath $x$}}}
\def\y{{\text{\boldmath $y$}}}

\def\C{{\text{\boldmath $C$}}}
\def\D{{\text{\boldmath $D$}}}
\def\I{{\text{\boldmath $I$}}}

\def\Q{{\text{\boldmath $Q$}}}

\def\X{{\text{\boldmath $X$}}}
\def\V{{\text{\boldmath $V$}}}

\def\S{{\text{\boldmath $S$}}}
\def\u{{\text{\boldmath $u$}}}
\def\W{{\text{\boldmath $W$}}}
\def\Y{{\text{\boldmath $Y$}}}

\def\bbe{{\text{\boldmath $\beta$}}}
\def\bet{{\text{\boldmath $\eta$}}}
\def\biota{{\text{\boldmath $\iota$}}}
\def\bOmega{{\text{\boldmath $\Omega$}}}
\def\bSigma{{\text{\boldmath $\Sigma$}}}

\def\diag{\text{diag}}

\title{\Large{\bf Spatio-temporal Smoothing, Interpolation and Prediction of Income Distributions based on Grouped Data}}

\date{}
\author{
}

\begin{document}

\maketitle
\doublespacing

\vspace{-1.5cm}
\begin{center}
Genya Kobayashi$^{1}$\footnote{Author of correspondance: \texttt{gkobayashi@meiji.ac.jp}}, Shonosuke Sugasawa$^{2}$ and Yuki Kawakubo$^{3}$

\end{center}

\noindent
$^1$School of Commerce, Meiji University\\
$^2$Center for Spatial Information Science, The University of Tokyo\\
$^3$Graduate School of Social Sciences, Chiba University\\

\abstract{
\vspace{-0cm}
The Housing and Land Survey (HLS) of Japan provides municipality-level grouped data on household incomes. 
Although such data can be invaluable for effective local policymaking, their analysis is often hindered by several challenges, including limited information inherent in the grouped format, the presence of missing areas, and the low frequency of survey implementation.  
To address these issues, we propose a novel grouped-data-based spatio-temporal finite mixture model for estimating income distributions across multiple spatial units and time points. 
A unique feature of the proposed method is that all areas share common latent distributions, while the mixing proportions, incorporating spatial and temporal effects, capture the potential area-wise heterogeneity. 
Consequently, the inclusion of these effects enables smoothing  quantities of interest over space and time, imputing missing values, and predicting future trends. 
By applying the proposed method to the HLS data, we generate complete maps of income and inequality measures at any given time, thereby facilitating rapid and efficient policymaking with fine granularity. 
}

\noindent\textbf{keywords:} {Log-normal distribution, Markov chain Monte Carlo, Mixture-of-experts, P\'olya-gamma augmentation,  State space model}


\maketitle

\section{Introduction}
Income data are typically provided as grouped data, where individual incomes are grouped into a set of predefined intervals (e.g., annual income between $3$ and $4$ million Japanese Yen; JPY) and the number of individuals or households within each class is recorded. 
As the amount of information in the grouped data is considerably limited by the grouping mechanism, the underlying distribution structure cannot be expressly observed. 
Nevertheless, numerous studies in the literature on income distribution estimation have explored various statistical methods and income models to reconstruct these income distributions \citep[see][for comprehensive reviews]{kleiber03book, choti08book, jorda}.

In Japan, the Housing and Land Survey (HLS) includes municipality-level grouped income data, where the number of income classes varies over time. 
These data offer detailed information for effective policymaking at fine granularity. 
However, as the published HLS data do not cover every municipality, spatial interpolation is required to produce complete maps of income or inequality measures. 
Furthermore, the HLS is conducted only every five years and the results become accessible only a year after the survey. 
Therefore, temporal prediction of the latest income status is necessary for any given time point to facilitate policymaking based on the most current information on income status.  

Existing studies on income distribution have predominantly focused on estimating the income distribution of a single area at a single time point, for instance, the national-level income distribution estimated from a single round of an income survey\footnote{Due to space limitations, the specific existing methods and countries to which they have been applied are reviewed in the Supplementary Material.}. 

National or single-area income distributions and their associated inequality measures can be estimated with reasonable precision using grouped data from surveys covering a large number of households or individuals. 
However, these estimates do not provide sufficient information for local-level policymaking, such as at the state or city level, which is crucial for implementing policies with fine granularity. 
Furthermore, even when local-level data are available, as in the HLS, estimating the income distribution for each area separately is challenging due to the inherent loss of information from grouping. 
Therefore, a joint statistical model that borrows strength across datasets is desirable to enhance statistical efficiency.

The joint estimation of income distributions across multiple areas and/or periods represents a burgeoning stream in the literature on the statistical analysis of income distributions. 
This approach is increasingly viable due to the recent availability of abundant data.
However, it remains under-explored despite its statistical and economic significance. 
For example, \cite{nishino12} and \cite{nishino15} considered the log-normal income model with time-varying scale parameters. 
\cite{kobayashi2021bayesian} considered a state space model based on grouped Lorenz curve data to accurately estimate Lorenz curves and Gini indices.  
These models are also capable of temporal prediction. 
In a spatial context, \cite{sugasawa2020estimation} considered spatially varying income distributions based on a parametric family, producing spatially smoothed estimates and enabling the interpolation of missing areas. 
Although these studies demonstrate how joint estimation enhances efficiency, they typically consider only either spatial or temporal dependence, partly due to data constraints. 
Furthermore, reliance on a specific parametric family can be restrictive, as the most suitable distribution may vary across areas or periods.

Based on the aforementioned considerations, we present a novel grouped-data-based framework for spatio-temporal smoothing, interpolation, and prediction of areal income distributions. 
To overcome the limitations of the parametric approach, we employ a flexible finite mixture model. 
A unique feature of our framework is that all the areas share common component distributions that are invariant across space and time, except for the influence of covariate values.
The area-wise heterogeneity of income distributions is characterised by mixing proportions, which are hierarchically modelled to incorporate spatial and temporal effects. 
This approach is advantageous as it reduces the number of parameters and latent variables through the shared component distributions, thereby ensuring stable estimation even from grouped data. 
Furthermore, it offers an intuitive interpretation. 
The income distribution of each area and period is represented as a weighted combination of template income distributions, with the weights evolving over space and time. 

While \cite{sugasawa2019latent} and \cite{sugasawa2020grouped} adopted a similar mixture-modelling strategy, though neither study considered the spatio-temporal setting. 
Through these spatio-temporal mixing proportions, our proposed model can facilitate spatial interpolation and temporal prediction for any spatial unit at any time point, which are the essential requirements for analysing HLS data. 
Although \cite{flach} and \cite{lubrano16} also considered  finite mixture approaches for modelling income distributions, their models were designed for national-level individual income data rather than grouped or municipality-level data.  
Moreover, while they analysed the income data of the United Kingdom over multiple periods, they estimated the mixture model independently for each period, as their models lacked a formal temporal structure.

Some relevant studies have explored mixture models within spatio-temporal settings. 
\cite{fernandez02} considered a spatial mixture model for areal data featuring a variable number of components and spatially varying mixing proportions. 
\cite{neelon14} considered a multivariate normal mixture model for areal data, incorporating spatial random effects into both the component distributions and mixing proportions. 
Regarding the Poisson mixture, \cite{hossai14} proposed algorithms for variable and model selection, and relabelling posterior simulation. 
\cite{viroli11} proposed a general finite mixture model for three-way arrays based on matrix normal distributions. 
A model for analysing spatio-temporal multivariate data presents a special case in which the spatial effects are incorporated in the mixing proportions. 
In the context of point-referenced data, \cite{paci18} and \cite{vanhatalo20} considered spatio-temporal mixture models. 
Additionally, several studies employed Bayesian nonparametric mixture approaches to analyse spatio-temporal data, typically built upon or connected to Dirichlet process mixture models \citep{kottas08,zhang16,lee20,wang22}.

The remainder of this paper is organised as follows. 
Section~\ref{sec:data} describes the challenges inherent in analysing the HLS data of Japan.
Section~\ref{sec:method} introduces the proposed spatio-temporal mixture modelling framework for the grouped income data.
In Section~\ref{sec:sim}, we present a simulation studies to demonstrate the performance of the proposed method, followed by 
an application to the HLS data in Section~\ref{sec:real}. 
Finally, Section~\ref{sec:conc} provides concluding remarks.

\section{{HLS data of Japan}}
\label{sec:data}
\begin{figure}[H]
    \centering
    \begin{tabular}{ccc}
        \includegraphics[scale=0.18]{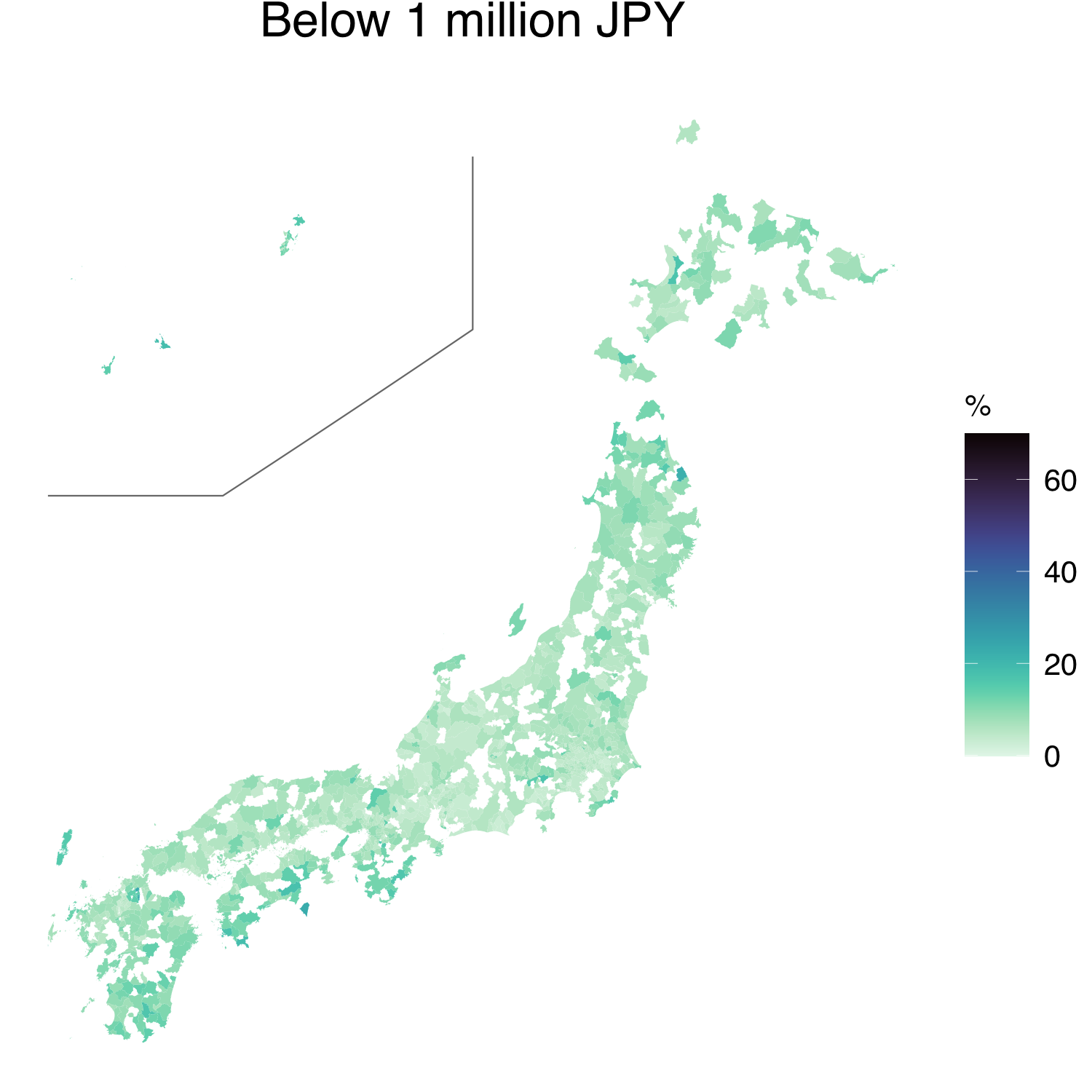}&
        \includegraphics[scale=0.18]{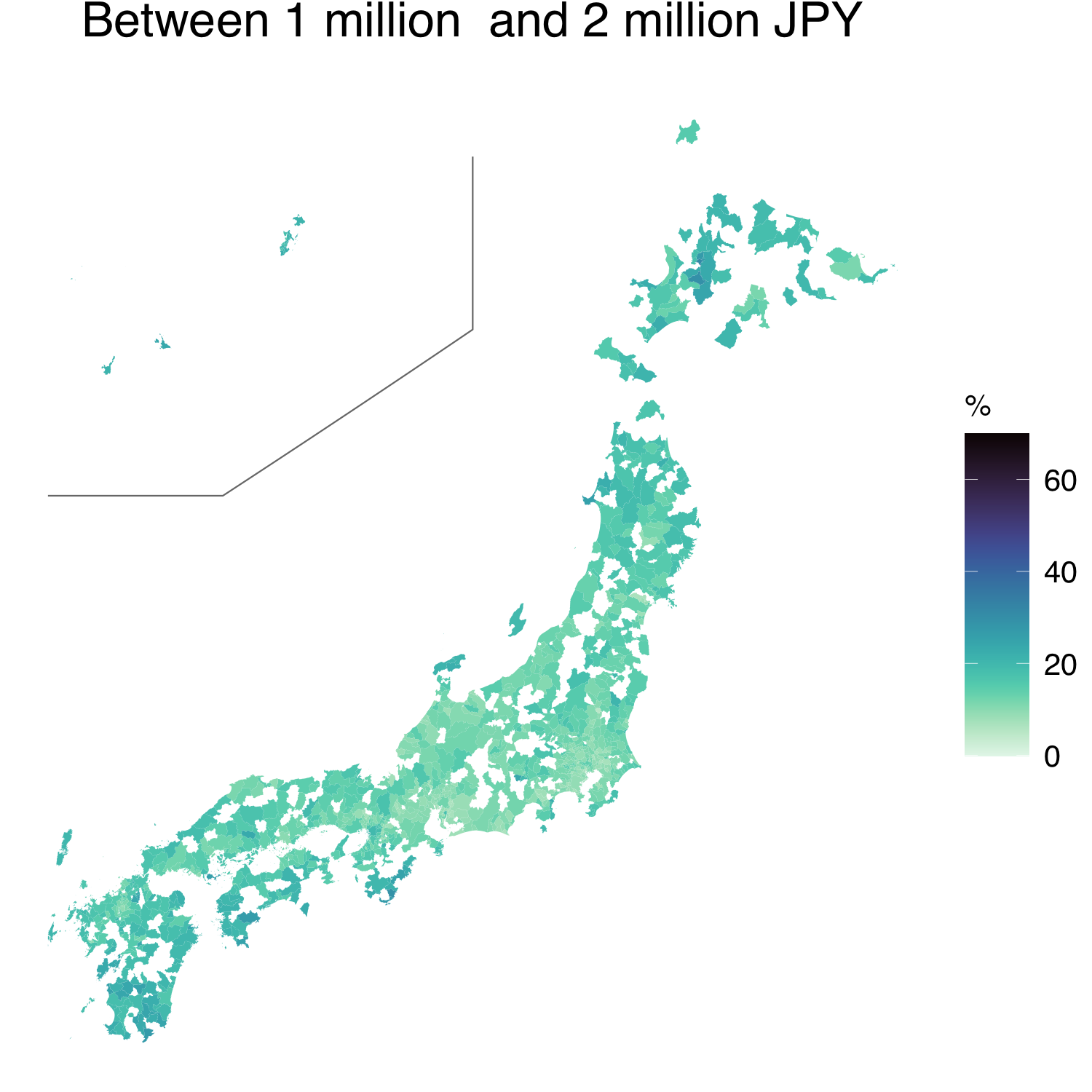}&
        \includegraphics[scale=0.18]{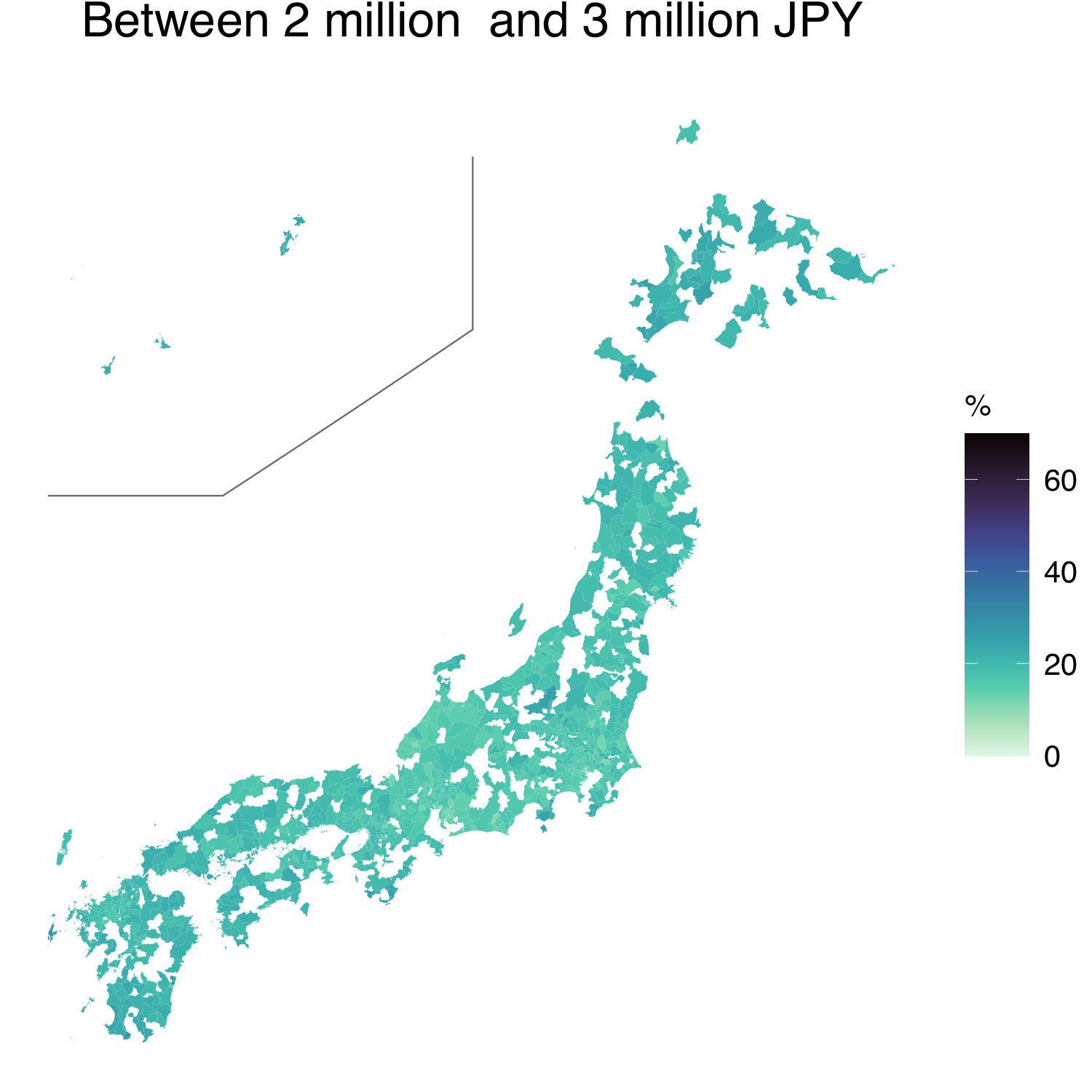}\\
        \includegraphics[scale=0.18]{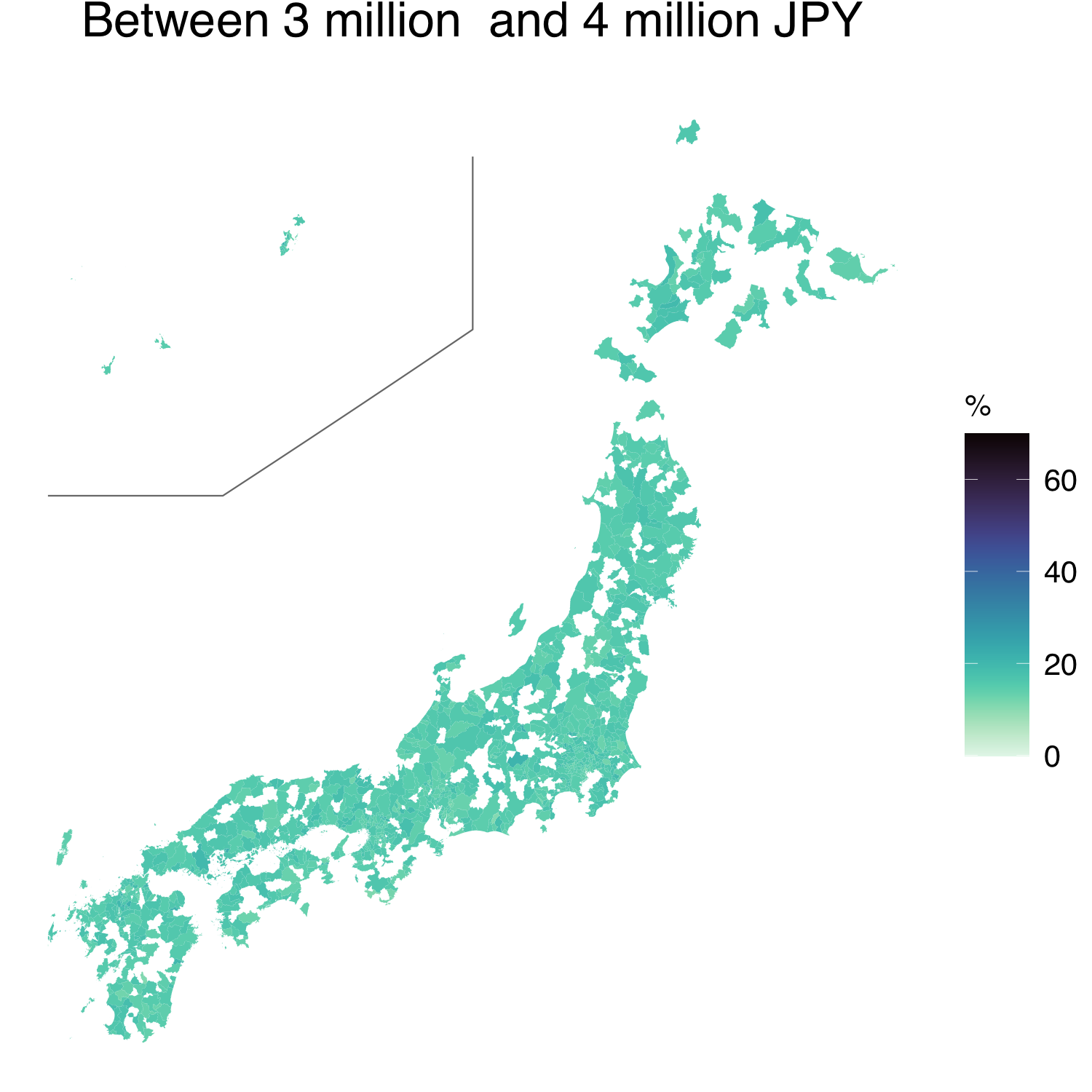}&
        \includegraphics[scale=0.18]{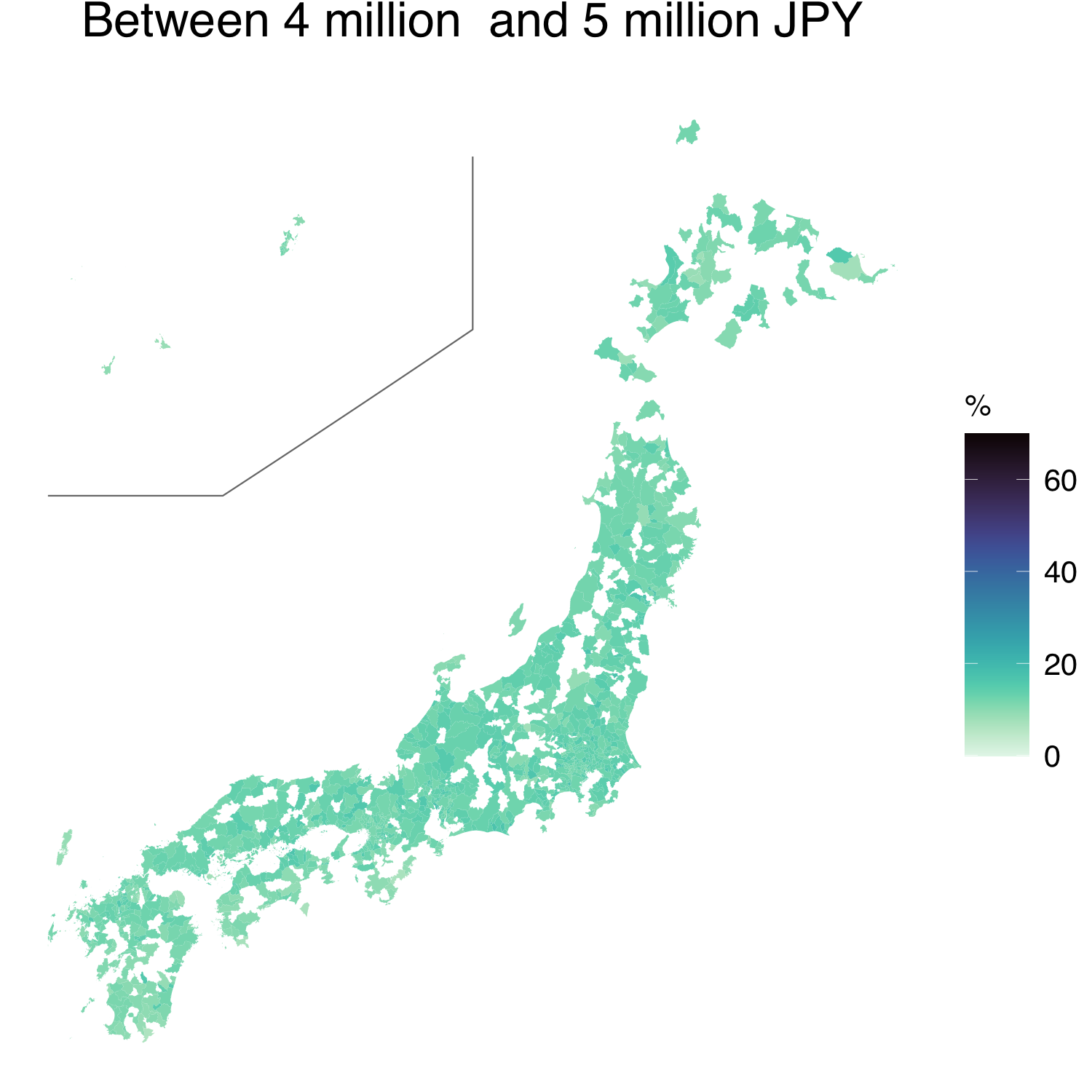}&
        \includegraphics[scale=0.18]{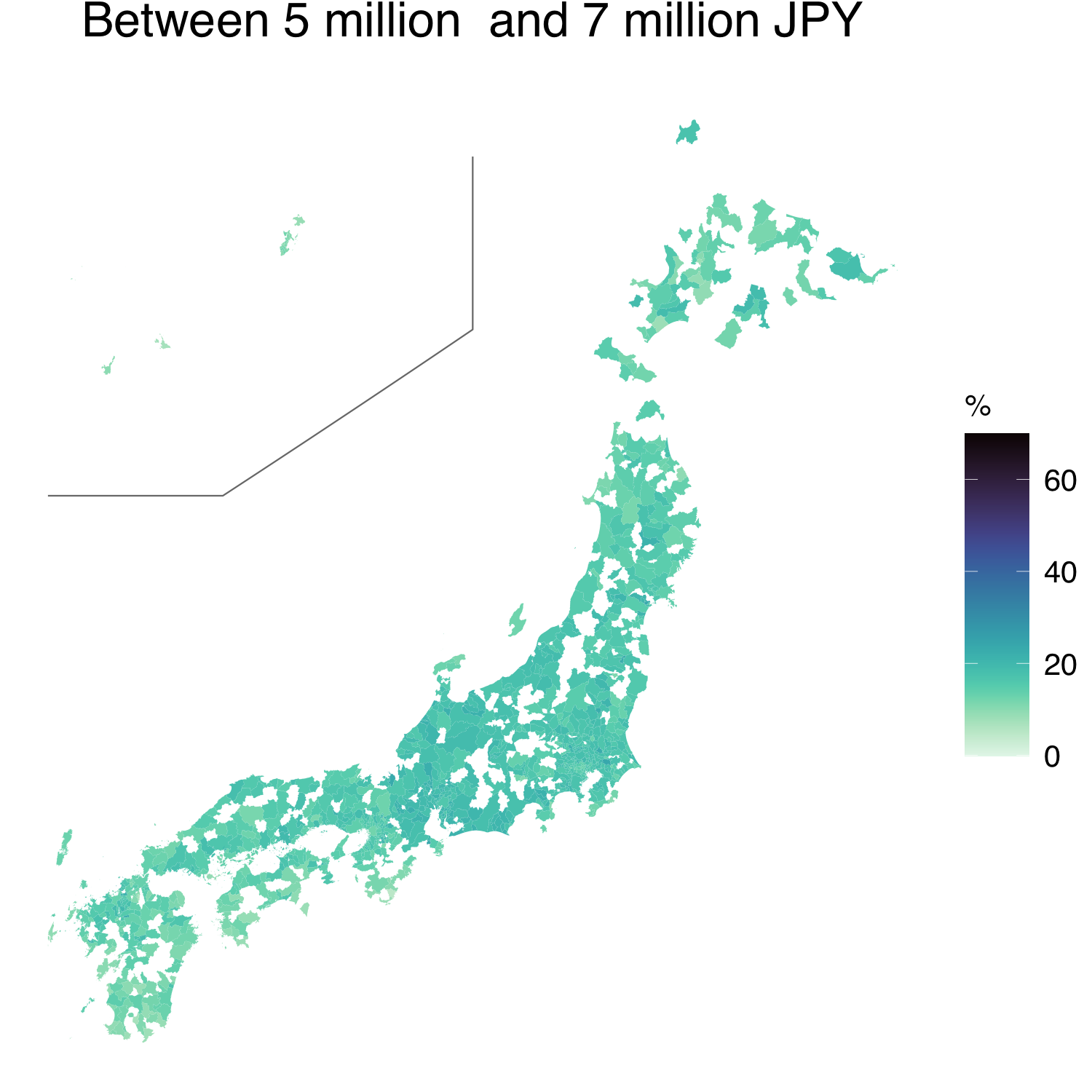}\\
       \includegraphics[scale=0.18]{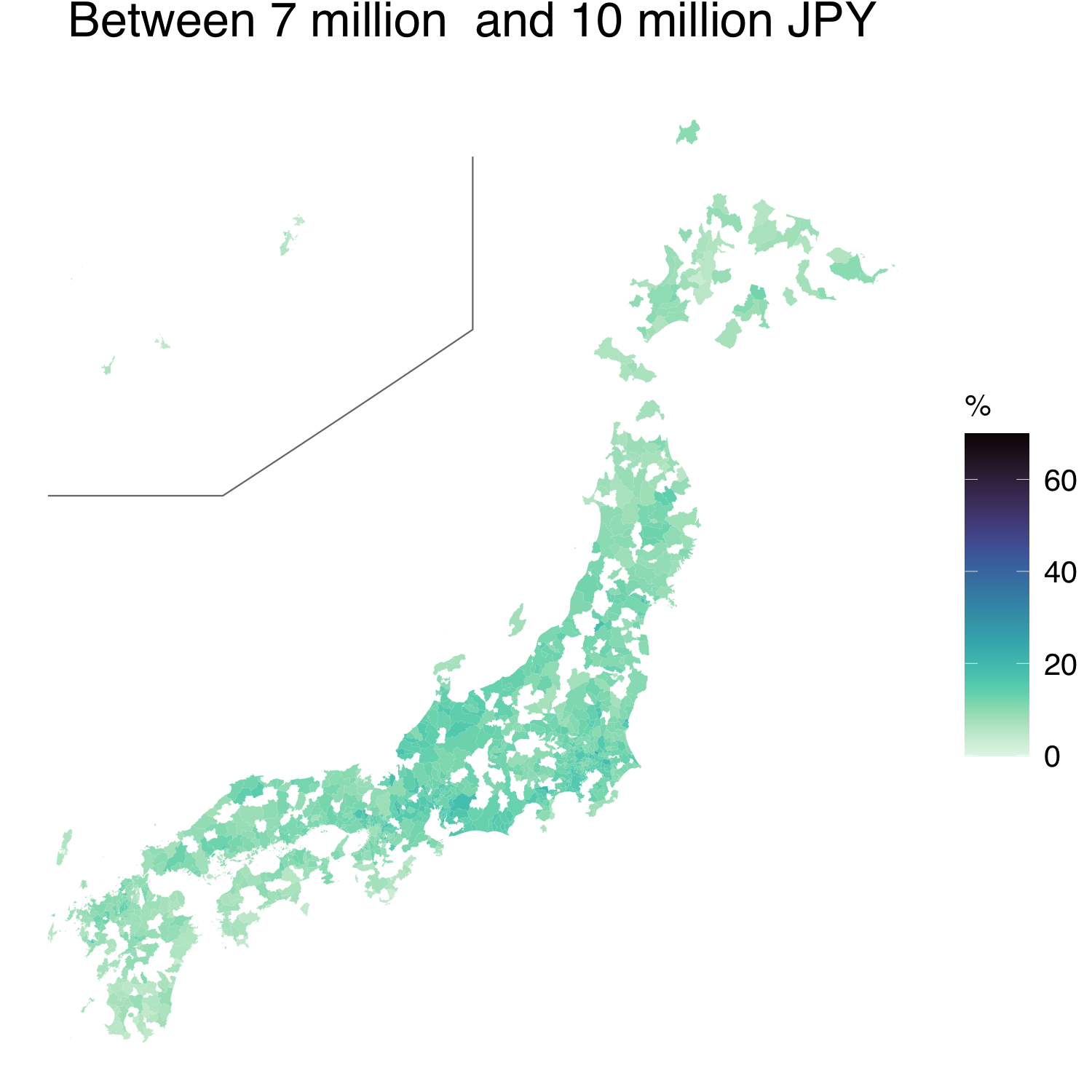}&
       \includegraphics[scale=0.18]{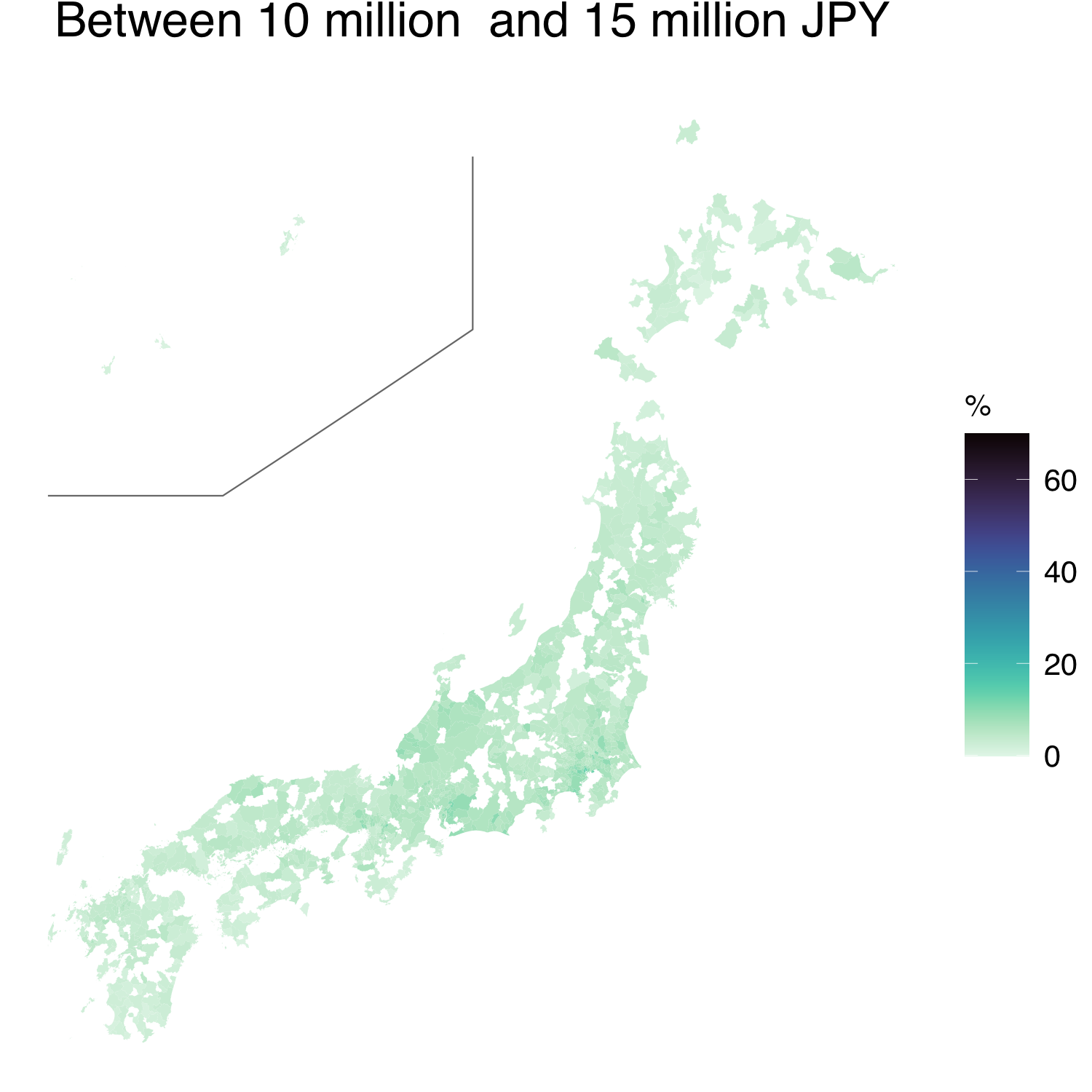}&
        \includegraphics[scale=0.18]{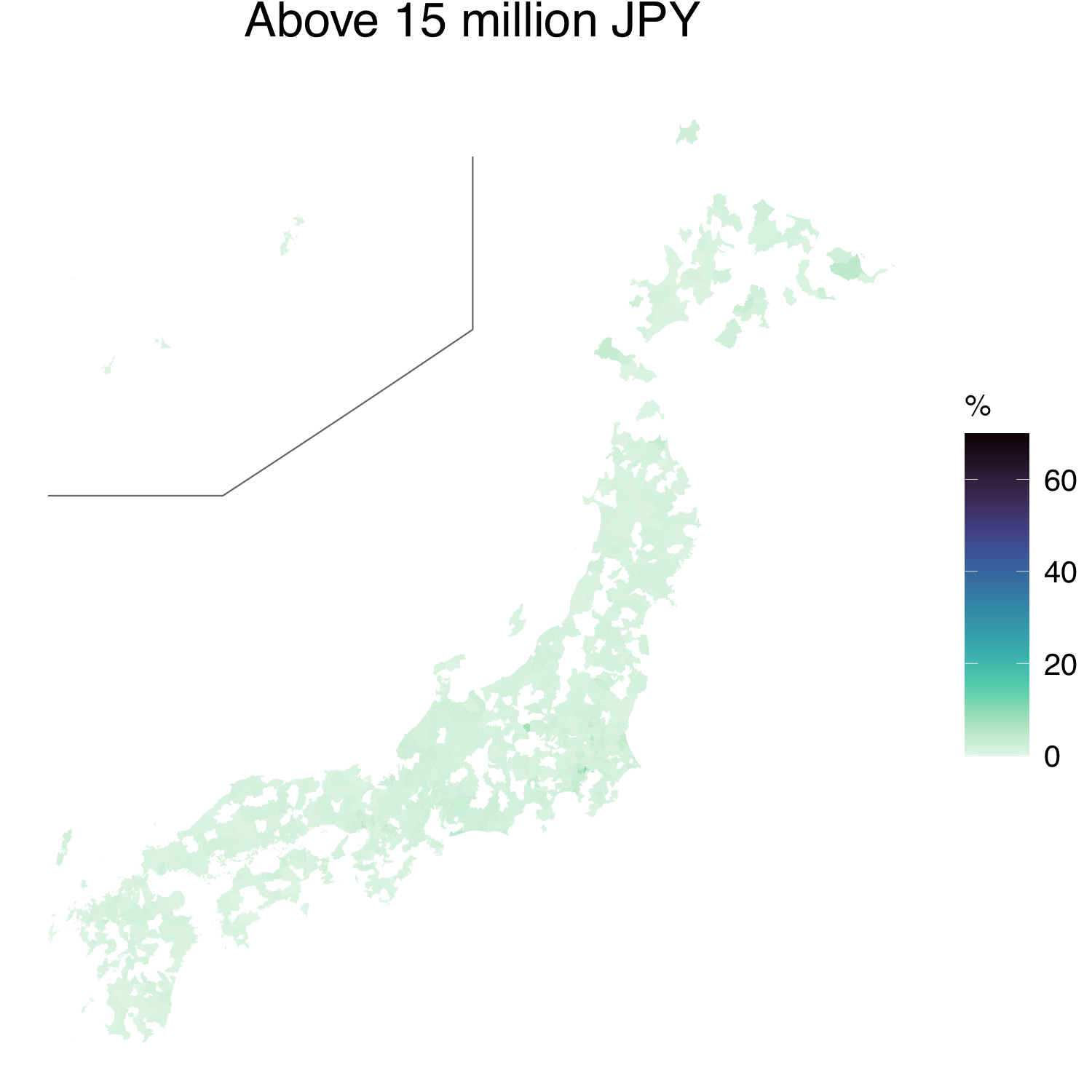}\\
    \end{tabular}
    \caption{Proportions of the nine income classes in the 2018 HLS data.}
    \label{fig:hls}
\end{figure}

HLS\footnote{The HLS data available at \url{ https://www.stat.go.jp/english/data/jyutaku/index.html}} is a large-scale household survey conducted by the Statistics Bureau of Japan every five years to monitor housing status and household wealth. 
The survey adopts a stratified two-stage sampling design.
Enumeration districts (EDs) are sampled first, followed by the sampling of dwelling units from enumeration unit districts (EUDs) formed within those sampled EDs.
Although the primary data are collected at the household level, the publicly available income data are municipality-level grouped data.
In these tables, the individual household income data are aggregated into frequency counts within predefined income classes to protect respondent privacy. 
Specifically, household income information is collected via questionnaires in which households are asked to indicate the class into which their annual pre-tax income falls, rather than reporting an exact figure. 
Consequently, the published municipality-level tables provide weighted design-based estimates of the number of households in these predefined income classes, rather than raw sample counts. 
The proportions shown in Figure~\ref{fig:hls} are derived from these published estimates. 

In our analysis, we utilise the reported household counts for each class. 
However, these published counts represent scaled population estimates rather than the actual number of households surveyed. 
Therefore, it is necessary to account for the underlying sample size for each municipality, yet this information is not publicly accessible. 
While \cite{kawakubo2019small} discussed the estimation of area-specific sample sizes or sample fractions for grouped survey data, their approach requires additional information on survey districts that cannot be obtained from the public HLS tables alone. 
As a practical solution, we assume a uniform sampling fraction of 1\% across all municipalities to estimate the required sample size, rather than treating the published counts as raw observed counts. 
Although a rigorous calculation of the effective sample size for each municipality would be ideal given the complex survey design of the HLS, our fixed fraction serves as a conservative and realistic proxy. 
Specifically, since the actual sampling fractions in most municipalities are expected to exceed 1\%, this specification provides a cautious estimation of uncertainty, thereby avoiding an overconfident assessment of the precision of the direct estimates.

Our analysis incorporates all available survey rounds from 1998 to 2018 ($T=5$). 
Table~\ref{tab:classes} presents the income classes defined for each round. 
During the two decades spanning the 1998 and 2018 surveys, municipal units underwent extensive restructuring, partly to enhance the efficiency of local taxation. 
Some municipalities were absorbed into larger entities, while others merged to form entirely new municipalities or were upgraded from villages to cities. 
We tracked all such administrative changes and compiled the grouped income data by aggregating the observations from the merged and absorbed units. 
Specifically, we standardised the spatial units to match the 1,741 municipalities existing in the 2018 round. 
The number of municipalities with observed income tabulations is also reported in Table~\ref{tab:classes}. 
Although data for more than half of the municipalities were missing in 1998, the coverage of the published tables has increased, with approximately 60\% available in recent rounds. 
Furthermore, while the data for 800 out of 1,741 municipalities (45.9\%) are available for all five rounds, 597 (34.3\%) are entirely missing across all rounds.

\begin{table}[H]
    \centering
    \caption{{Income classes of the HLS in million Japanese Yen, and the numbers and proportions of observed municipalities.}}
    \begin{tabular}{clrr}\hline
Year & Income classes & Observed & Proportion \\\hline
    1998 & $[0,2)$, $[2,3)$, $[3,4)$, $[4,5)$, $[5,7)$, $[7,10)$, $[10,15)$, $[15,\infty)$ & 814 & 46.8\% \\
    2003 & $[0,3)$, $[3,5)$, $[5,7)$, $[7,10)$, $[10,15)$, $[15,\infty)$ & 1040 & 59.7\% \\
    2008 & $[0,1)$, $[1,2)$, $[2,3)$, $[3,4)$, $[4,5)$, $[5,7)$, $[7,10)$, $[10,15)$, $[15,\infty)$ & 1117 & 64.2\% \\
    2013 & $[0,1)$, $[1,2)$, $[2,3)$, $[3,4)$, $[4,5)$, $[5,7)$, $[7,10)$, $[10,15)$, $[15,\infty)$ & 1110 & 63.8\% \\
    2018 & $[0,1)$, $[1,2)$, $[2,3)$, $[3,4)$, $[4,5)$, $[5,7)$, $[7,10)$, $[10,15)$, $[15,\infty)$ & 1086 & 62.3\% \\
    \hline
    \end{tabular}
    \label{tab:classes}
\end{table}

Figure~\ref{fig:hls} illustrates the proportions of the nine income classes in the 2018 HLS data. 
The figure reveals that many inland municipalities are missing. 
Due to the open ended income class and the lack of information on variability of incomes within classes, constructing direct estimates for income or poverty measures, such as average incomes, based on the counts published by HLS and, especially, associated uncertainty measures is not straightforward without strong distributional assumptions that may introduce arbitrary bias.

In most municipalities, the highest proportions of households fall within the third (between 2  million and 3 million JPY) and fourth (between 3 million and 4 million JPY) income classes. 
Municipalities situated far from urban centres, particularly those in the south-western and northern regions of Japan, tend to exhibit larger proportions in the first (below 1 million JPY) and second (between 1 million and 2 million JPY) income classes.

Finally, the HLS is conducted every five years, and the results do not become publicly available until at least one year after the survey's completion. 
To facilitate flexible and rapid policymaking with fine granularity in both space and time, it is desirable to have access to updated information regarding the income status of all municipalities on a more frequent basis, such as annually. 

In summary, although the HLS collects detailed data on household incomes across Japan, an analyst faces the following three challenges:
\begin{enumerate}
    \item
    the data are provided in a grouped format, from which income and poverty measures cannot be readily derived;
    \item
    the data contain a significant number of missing municipalities;
    \item
    the survey is conducted every five years, whereas the latest income status is often required on a more frequent basis (e.g. annually). 
\end{enumerate}
These challenges are addressed by the spatio-temporal mixture model proposed in the following section.

\section{Spatio-temporal mixture modelling }\label{sec:method}

\subsection{Model}\label{sec:model}
Suppose we are interested in the income distributions of $M$ areas over $T$ periods, denoted by $F_{it}(\cdot)$, with the corresponding density function $f_{it}(\cdot)$ for $i=1,\dots,M$ and $t=1,\dots,T$. 
{Let $[Z_{t0}, Z_{t1})$, $[Z_{t1}, Z_{t2}),\dots,[Z_{t,G_t-1}, Z_{tG_t})$ represent $G_t$ income classes in the $t$-th period, where $Z_{t0}$ and $Z_{tG_t}$ are typically set to $0$ and $\infty$, respectively. }

As in the case of the HLS data analysis in Section~\ref{sec:real}, we assume that not all of the $M$ areas are observed. 
Specifically, to ensure a streamlined presentation and to facilitate an evaluation of spatial interpolation performance, we assume that a consistent set of the first $m$ areas is observed across all time periods, while the remaining $m^*=M-m$ areas are unobserved for $t=1,\dots,T$. 

\label{page:D}{
Let $D_{itg}$ denote the published design-based estimate of the number of households in income class $g$ for municipality $i$ at time $t$, and let $D_{it}=\sum_{g=1}^{G_t}D_{itg}$.
Following the discussion in {Section~\ref{sec:data}}, to link these  grouped data to the multinomial likelihood introduced below, let $r_{it}$ denote the sample fraction for municipality $i$ at time $t$. 
We then define the corresponding observed counts for the likelihood as $N_{itg} = D_{itg}r_{it}$, with the total area-specific sample size given by $N_{it}=\sum_{g=1}^{G_t}N_{itg}=D_{it}r_{it}$. 
}

{
The multinomial likelihood is specified in terms of $N_{itg}$, ensuring that the sampling uncertainty is appropriately governed by the sample-size parameter.
In the actual data analysis, however, municipality-specific sampling fractions, sample sizes, or effective sample sizes are not publicly available.
We therefore assume a fixed working value for $r_{it}$, setting $r_{it}=r = 0.01$ for all $i$ and $t$. 
This yields $N_{itg} = 0.01D_{itg}$ and $N_{it} = 0.01D_{it}$, providing a conservative and practical quantification of the uncertainty associated with the design-based estimates. 
}

{
Let $y$ denote the unobserved household income}. \label{page:y}
To flexibly model the underlying spatio-temporal income distributions, we employ a finite mixture of log-normal distributions for $y$: %
\begin{equation}\label{MID}
f_{it}(y)=\sum_{k=1}^K\pi_{itk}\phi_{LN}(y;\x_{it}^{t}\bbe_k,\sigma_k^2), \quad i=1,\ldots,M,
\end{equation}
where $\x_{it} = (1,x_{it1},\dots,x_{itp})^t$ is a $(p+1)$-dimensional vector of covariates (e.g. consumption or tax levels) correlated with the income data, $\bbe_k=(\beta_{k0},\beta_{k1},\dots,\beta_{kp})^t$ and $\sigma_k^2$ are the coefficient and scale parameters of the $k$-th component, respectively. 
Here, $\phi_{LN}(\cdot;\mu,\sigma^2)$ denotes the density function of the log-normal distribution with parameters $\mu$ and $\sigma^2$. 

The inclusion of the covariate $\x_{it}$ in the component distribution in \eqref{MID} is essential. 
Regarding spatial interpolation of the unobserved areas, $\x_{it}$ must be observed for all spatial units. 
Furthermore, regarding temporal prediction, we assume that the most recent observations of $\x_{it^\ast}$ with $t^\ast>T$ become available more frequently than the HLS data, e.g. annually. 
In our HLS data analysis, this covariate is based on taxable income per taxpayer (see Section~\ref{sec:setup}). 

It is well known that a single log-normal distribution often provides a poor fit for income data. 
In contrast, a mixture of log-normal distributions offers superior flexibility and goodness-of-fit \citep{lubrano16}. 
Furthermore, by employing a data augmentation strategy for grouped data, the parameters of these component distributions facilitate a relatively straightforward posterior computation algorithm via a Gaussian framework (see Section~\ref{sec:mcmc}).

Furthermore, the continuous density in \eqref{MID} characterises the latent structure of unobserved household incomes. 
For the observed grouped data, the likelihood contribution of the $i$-th area at time $t$ is based on a multinomial distribution:
\begin{equation}\label{multi}
\prod_{g=1}^{G_t}\left\{F_{it}(Z_{tg})-F_{it}(Z_{t,g-1})\right\}^{N_{itg}},
\end{equation}
for $i=1,\dots,m$ and $t=1,\dots,T$, where $F_{it}(y)=\sum_{k=1}^K\pi_{itk}\Phi_{LN}(y;\x_{it}^t\bbe_k,\sigma^2_k)$ is the distribution function of the latent income in the $i$-th area at time $t$, and $\Phi_{LN}$ denotes the distribution function of the log-normal distribution.

The mixing proportions, $\pi_{itk}$, are modelled as
\begin{equation}\label{Prob}
\pi_{itk}=\frac{\exp(\mu_k+u_{ik}+\eta_{tk})}{\sum_{\ell=1}^K\exp(\mu_\ell+u_{i\ell}+\eta_{t\ell})},
\end{equation}
where $\mu_k$ denotes the baseline intercept determining  the overall proportion, $u_{ik}$ is the area-specific spatial effect, and $\eta_{tk}$ is the temporal effect. 
For identification, we set $\mu_1=u_{i1}=\eta_{t1}=0$. 
\label{page:Q}
{Let  $\tilde{\W}$ denote the adjacency matrix for all areas, where the diagonal elements are zero and the $(i,j)$-th element equals one if areas $i$ and $j$ share a boundary, and zero otherwise. 
Areas consisting solely of islands with no land borders are treated as isolated. 
We denote the row-standardised version of $\tilde{\W}$ by $\W$. }Using $\W$, we assume a simultaneous autoregressive (SAR) model for $\u_{{\rm all},k}=(u_{1k},\ldots,u_{Mk})^{t}$ independently for $k=2,\dots,K$. 
Specifically, $\u_{{\rm all},k}\sim N(\zero, \tau_k^2\V_{{\rm all},k})$, where  
$\V_{{\rm all},k}^{-1}=\Q_{{\rm all},k}=(\I_M-\rho_k\W)^t(\I_M-\rho_k\W)$, 
with variance parameter $\tau_k^2$ and spatial correlation parameter $\rho_k\in(0,1)$. 
For each $k$, let $\V_{{\rm all},k}$ be partitioned as
$
\V_{{\rm all},k}=
\begin{psmallmatrix} 
\V_{11k} & \V_{12k}\\ \V_{21k} & \V_{22k}
\end{psmallmatrix}
$,
where $\V_{11k}$ is the $m\times m$ submatrix corresponding to the observed areas. 
{
The marginal distribution of the spatial effects for these observed areas $\u_{k}=(u_{1k},\dots,u_{mk})^t$ is given by $N(\zero,\tau_k^2\V_{11k})$. 
The remaining blocks of $\V_{{\rm all},k}$ are utilised for the spatial interpolation of the unobserved areas (see{Section~\ref{sec:mcmc}}).} \label{page:V}

We assume that the temporal effect follows a random-walk process: $\eta_{tk}|\eta_{t-1,k}\sim N(\eta_{t-1,k},\alpha_k^2)$ for $t=1,\dots,T$. 
Additionally, to ensure the identifiability of each effect within the mixing proportions, we impose $\sum_{i=1}^m u_{ik}=0$ and $\sum_{t=1}^T\eta_{tk}=0$ for $k=2,\dots,K$. 
These sum-to-zero constraints are applied at each iteration of the Gibbs sampling algorithm, with the surplus being absorbed into $\mu_k$.  

\subsection{Prior distributions}\label{sec:prior}
We now specify the prior distributions of the model parameters. 
These prior distributions are either conditionally conjugate or nearly so after data augmentation, which facilitates the development of the Gibbs sampling algorithm. 
For the coefficient parameters of the component distributions, $\bbe_k$, we assume $\bbe_k\sim N(\zero,c_\beta \I_{p+1})$, $k=1,\dots,K$. 
Following \cite{Simpson}, the standard deviation parameters of the component distributions are assigned exponential priors with mean $1/a_\sigma$: $\sigma_k\sim Exp(a_\sigma)$, $k=1,\dots,K$. 
Since the component distributions are common across space and time, they can be estimated stably.
We therefore adopt the default hyperparameters given by $c_\beta=100$ and $a_\sigma=1$, such that these priors are diffuse. 
This choice for $\sigma_k$ is particularly suitable as it covers a wide range of Gini index values for the log-normal distribution. 

Regarding $\mu_k$, we assume normal priors, $\mu_k\sim N(0,c_\mu), k=2,\dots,K$, with a default value of $c_\mu=10$. 
Given the row-standardised adjacency matrix, the priors for the spatial-correlation parameters are set as $\rho_k\sim U(0,1), k=2,\dots,K$. 
For the standard deviations of the SAR model, $\tau_k$, we also assume exponential priors, $\tau_k\sim Exp(a_\tau), k=2,\dots,K$. 
The default value is $a_\tau=5$, giving the prior mean of $0.2$. 
This choice ensures that the prior does not favour excessively large values for the standard deviation while still covering a plausible range. 
For the initial states and standard deviations of the random walk processes, we similarly assume $\eta_{0k}\sim N(0,c_\eta)$ and $\alpha_k\sim Exp(a_\alpha)$ for $k=2,\dots,K$, with default  hyperparameters $c_\eta=10$ and $a_\alpha=5$.

\subsection{Posterior inference}\label{sec:mcmc}
The posterior inference for the proposed model is conducted using Markov chain Monte Carlo (MCMC) methods. 
While the full details are provided in the Supplementary Material, we briefly outline the strategy for constructing the Gibbs sampler here. 
First, to facilitate the sampling of the parameters in $\pi_{itk}$, we employ the P\'olya-gamma data augmentation of \cite{polson2013bayesian}, which allows these variables to be sampled from normal distributions. 
Second, we introduce the latent variables representing the number of households in the $g$-th income class that belong to the $k$-th mixture component. 
Third, the latent individual household incomes are introduced to simplify the sampling of $\bbe_k$ and $\sigma^2_k$. 
Conditionally on the data and the allocation of households across the mixture components, the logarithms of these individual household incomes are generated from truncated normal distributions. 
Finally, $\bbe_k$ and $\sigma_k^2$ are sampled from their full conditional distributions. 
Under conditional conjugacy, the full conditional distribution of $\bbe_k$ is normal.  
The variance $\sigma_k^2$ is sampled via a simple independent Metropolis--Hastings (MH) algorithm, which requires no algorithmic tuning. 

Based on the MCMC output, posterior inference regarding the quantities of interest can be performed. 
In the present context, we focus on the average income, ${\rm AI}_{it}=\sum_{k=1}^K\pi_{itk}\exp(\x_{it}^t\bbe_k+\sigma_k^2/2)$, median income ${\rm MI}_{it}$, and Gini index for $i=1,\dots,M$ and $t=1,\dots,T$. 
Under the mixture model, the median income and Gini index do not have closed-form expressions. 
The median income is obtained by numerically solving $\sum_{k=1}^K\pi_{itk}\Phi_{LN}(\text{MI}_{it};\x_{it}^t\bbe_k,\sigma_k^2)=0.5$. 
For the Gini index, defined as ${\rm Gini}_{it}={\rm AI}_{it}^{-1}\int_0^\infty F_{it}(z)(1-F_{it}(z))dz$, the integral is evaluated numerically following the approach of \cite{lubrano16}. 

{
To interpolate the unobserved areas, given the MCMC draws of the parameters and latent variables, the spatial effects for the unobserved areas, $\u^\ast_{k}=(u_{m+1,k},\dots,u_{Mk})^t$, are sampled from the conditional distribution of the SAR model given the spatial effects $\u_{k}$ for the observed areas: $N\big(\V_{21k}\V_{11k}^{-1}\u_{k}, \tau_k^2(\V_{22k}-\V_{21k}\V_{11k}^{-1}\V_{12k})\big)$. }
For temporal prediction, the future temporal effect $\eta_{T+1,k}$ is drawn from $N(\eta_{Tk},\alpha^2_k)$.

In practice, the number of components, $K$, is unknown, yet its choice significantly influences both the estimation performance and the interpretation of the results. 
To determine an appropriate $K$, we follow \cite{sugasawa2020estimation} and employ a criterion mimicking the posterior predictive loss (PPL) of \cite{gelfand98}, defined as: 
\[
\text{PPL}(K)=\frac{1}{mT}\sum_{i=1}^{m}\sum_{t=1}^T\sum_{g=1}^{G_t}V_{itg}^{(K)} + \frac{1}{mT+1}\sum_{i=1}^{m}\sum_{t=1}^T\sum_{g=1}^{G_t}\left(N_{itg}-E_{itg}^{(K)}\right)^2,
\]
where $E_{itg}^{(K)}$ and $V_{itg}^{(K)}$, respectively, denote the mean and variance of the posterior predictive distribution of $N_{itg}$ for the model with $K$ components. 
A model with a lower $\text{PPL}$ value is preferred.

\section{{Simulation study}}\label{sec:sim}
\subsection{{Set-up}}
The proposed method is first demonstrated using simulated data. 
We generate $M=1,700$ areal units in a two-dimensional space, mirroring the scale of the HLS, by sampling coordinates from $(d_{i1}, d_{i2})\sim U(-1,1)^2$. 
The neighbours of each area are defined as those within a Euclidean distance of $0.065$, resulting in an average number of neighbours of $5.3$. 
The income classes are set as $[0,1)$, $[1,2)$, $[2,3)$, $[3,4)$, $[4,5)$, $[5,7)$, $[7, 10)$, $[10,15)$, and $[15,\infty)$, giving $G_t=G=9$ for all periods. 
These classes corresponded to those utilised in the latter three rounds of the HLS (see also Table~\ref{tab:classes}). 

We first generated the data for the periods $t^\ast=1,\dots,T^\ast$, where $T^\ast=31$, under the two model specifications described below. 
These data were then down-sampled at three different frequencies (low, medium, and high) to investigate the impact of survey frequency on model inference. 
The models were subsequently fitted to the resulting down-sampled datasets. 
Specifically, we retained subsets of the original data for $t^\ast\in\left\{1,6,\dots,21,26\right\}$ in the low-frequency setting, $t^\ast\in\left\{1,4,\dots,25,28,31\right\}$ in the medium-frequency setting, and $t^\ast\in\left\{1,2,\dots,19,20,21\right\}$ in the high-frequency setting. 
Consequently, model estimation was basedon $T=5$, $10$ and $20$ periods for the low, medium, and high frequencies, respectively. 

For each setting, data from the $m=800$ areas across $T$ periods were used for model estimation. 
The data for the remaining $m^\ast=900$ areas in each period were reserved for spatial interpolation, while data for the subsequent period $t=T+1$ were reserved for temporal prediction. 
Note that the low-frequency setting mimics the situation of the HLS data analysed in this paper, where $M\approx 1700$ and $T=5$.

In this simulation study, the latent continuous data were first generated from the mixture model \eqref{MID} with $K=3$ and subsequently grouped into the classes defined above. 
The parameters of the component distributions are set as follows:  $\bbe_1=(0.5,0,1)^t$, $\bbe_2=(-0.5,1,0)^t$, $\bbe_3=(2,-1,-1)^t$, and $(\sigma_1,\sigma_2,\sigma_3)=(0.5,0.5,0.5)$. 
Each covariate was sampled from $U(0,1)$. 
For each area, $N_{it^\ast}$ was generated from $U(100,500)$.

Using the same component distributions, we consider the following two settings for the mixing proportions. 
In the first setting, the mixing proportions were generated according to \eqref{Prob} with
$(\mu_2,\mu_3)=(0,0)$, 
$(\rho_2,\rho_3)=(0.7,0.9)$, 
$(\tau_2,\tau_3)=(0.3,0.3)$, 
$(\alpha_2,\alpha_3)=(0.06, 0.06)$, 
$(\eta_{02}, \eta_{03})=(0,0)$. 
In the second setting, the spatial and temporal effects were determined by the following deterministic sequences: $u_{i2}=d_{i1}-d_{i2}$, $u_{i3}=d_{i2}-d_{i1}$, $\eta_{t^\ast2}=t^\ast/10-(T^\ast+1)/20$, $\eta_{t^\ast3}=t^\ast/20-T^\ast/20$. 
Furthermore, the two-dimensional space, $(-1,1)^2$, is partitioned into $7\times 7$ square blocks. 
Areas within the same block shared a common random effect on the mixing proportions, mimicking prefectural effects in Japan: $a_{h_ik}\sim N(0,0.3^2),\ h_i\in\{1,\dots,49\},\ i=1,\dots,M,\ k=2,3$. 
In this configuration, each area $i$ is a sub-area of the $h_i$-th block. 
Thus, the mixing proportions for the $i$-th area within the $h_i$-th block are proportional to $\exp(\mu_k+u_{ik}+\eta_{t^\ast k}+a_{h_ik})$ for $k=2,3$. 
The average number of sub-areas per block is $34.7$. 
For each combination of setting and frequency, a single dataset was generated.

\subsection{{Alternative models}}
We compare the proposed spatio-temporal mixture model with four alternative models. 
The first is a mixture model based on \eqref{MID} and \eqref{Prob}, where
the mixing proportions are modelled using two-way independent random effects to represent the spatial and temporal heterogeneity: $u_{ik}\sim N(0,\tau_k^2)$, $\eta_{tk}\sim N(0,\alpha^2_k)$. 
Since this model incorporates both spatial and temporal effects, it can be used for both spatial interpolation and temporal prediction. 
The second alternative model is a mixture of log-normal distributions where the mixing proportions include only $\mu_k$ and $\u_{k}$, the latter following the same SAR model as in the proposed model. 
This is a purely spatial model, which is independently fitted for each $t$ (for the low and medium frequencies) and is used solely for spatial interpolation. 
The third alternative model is a simple log-normal model for grouped data, estimated independently for each area and period within the Bayesian framework. 
The prior distribution for the mean parameter is $N(0,100)$, and that for the standard deviation parameter is $Exp(1)$, consistent with those in the proposed model. 
This model is estimated using an MH algorithm.

The fourth alternative is the small area estimation (SAE) approach. 
SAE has been widely applied to income and poverty measures to obtain more reliable estimates from limited sample sizes by considering linear mixed models for direct estimators or individual observations. 
However, these methods generally do not aim to infer the entire distribution, and developments specifically tailored for grouped data or incorporating complex spatio-temporal structures remain scarce.\footnote{Due to space limitations, existing small area methods are reviewed in the Supplementary Material.}

Settings~1 and 2 utilises different linear mixed models for the crude average income, which serves as the direct estimator. 
For Setting~1, we employ a modified version of the spatio-temporal model of \cite{marhuenda2013small}, incorporating spatial effects via a SAR model and temporal effects following a random walk process. 
Therefore, it can be used for both spatial interpolation and temporal prediction of average incomes. 
In Setting~2, we adopt the two-fold linear mixed model proposed by \cite{Torabi}, following a suggestion from an anonymous reviewer. 
Since this model accounts for a sub-area structure, it is well suited to the hierarchical configuration of Setting~2. 
However, as the model of \cite{Torabi} does not incorporate temporal effects, it is estimated independently for each period. 
These small area models are implemented within the Bayesian framework by assigning prior distributions to the unknown parameters.
Further details regarding the model specifications and the posterior computation are provided in the Supplementary Material.

\subsection{{Performance measures}}
The models are compared based on the root mean square errors (RMSE) and the coverage of the 95\% credible or prediction intervals. 
These measures are computed for the average incomes and the number of households in the income classes across in-sample estimation, spatial interpolation, and temporal prediction. 
For the average income ${\rm AI}_{it}$, the RMSE and coverage in the in-sample case are defined as 
\[
{\rm RMSE}_{\rm AI}=\sqrt{\frac{1}{mT}\sum_{i=1}^{m}\sum_{t=1}^T(\widehat{{\rm AI}}_{it}-{\rm AI}_{it})^2}, \quad {\rm Coverage}_{\rm AI}=\frac{1}{mT}\sum_{i=1}^{m}\sum_{t=1}^TI\left\{L_{it}\leq {\rm AI}_{it}\leq U_{it}\right\},
\]
where ${\rm AI}_{it}$ is the true value, $\widehat{\rm AI}_{it}$ is the posterior (predictive) mean, and $(L_{it},U_{it})$ denotes the 95\% credible interval. 
For spatial interpolation, the area indices range over $i=m+1,\dots,M$ and $t=1,\dots,T$. 
For temporal prediction, they range over $i=1,\dots,M$ at $t=T+1$.

Similarly, we evaluate the predictive performance for the number of households $N_{itg}$ based on the RMSE and the coverage. 
In the in-sample case, these are defined as 
\[
{\rm RMSE}_{\rm N}=\sqrt{\frac{1}{mTG}\sum_{t=1}^T\sum_{i=1}^m\sum_{g=1}^G(\hat{N}_{itg}-N_{itg})^2},\quad 
{\rm Coverage}_{\rm N}=\frac{1}{mTG}\sum_{i=1}^m\sum_{t=1}^T\sum_{g=1}^GI\left\{L_{itg}\leq N_{itg}\leq U_{itg}\right\},
\]
where $\hat{N}_{itg}$ is the posterior predictive mean of $N_{itg}$, and $(L_{itg}, U_{itg})$ denotes the 95\% prediction interval. 
These measures for spatial interpolation and temporal prediction are defined analogously.

\subsection{{Results}}
\begin{figure}[H]
\centering
\includegraphics[width=\linewidth]{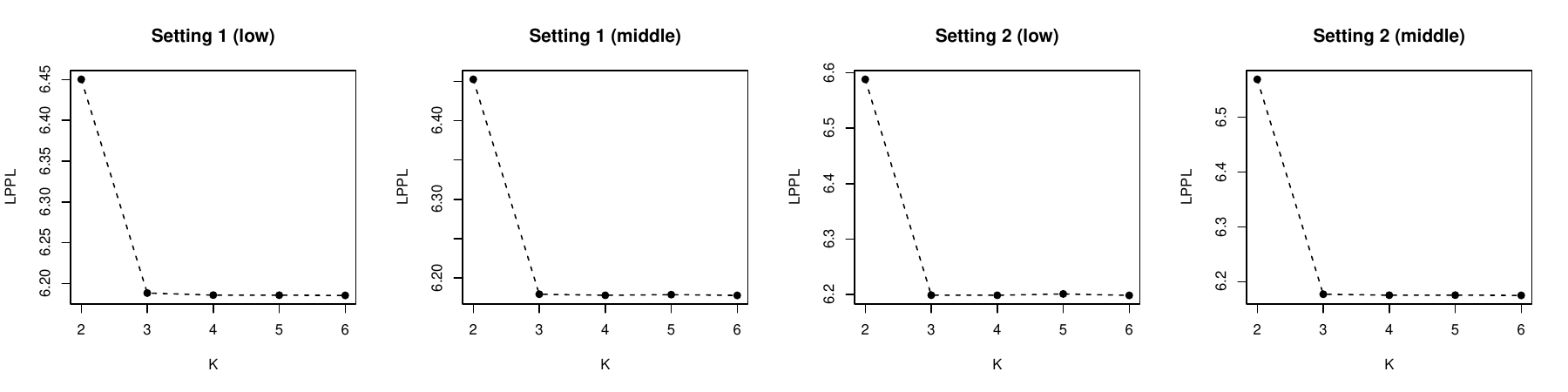}
\caption{Log posterior predictive loss (LPPL) for the simulation study.}
\label{fig:sim_K}
\end{figure}

First, we fitted the proposed spatio-temporal mixture model, as described in Section~\ref{sec:method}, with varying numbers of components: $K = 2, \dots, 6$. 
The Gibbs sampler was run for 30,000 iterations, with the initial 10,000 iterations discarded as burn-in. 
Figure~\ref{fig:sim_K} presents the log PPL (LPPL) for the mixture models for $K = 2, \dots, 6$ under the low- and medium-frequency settings. 
In all cases, the LPPL decreased substantially as $K$ increased from 2 to 3, the true number of components, beyond which no further improvements were observed. 
A similar phenomenon has been reported in the context of selecting the number of latent factors for multiple grouped count data \citep{Kobayashi25}. 
Since incorporating more components than necessary does not enhance predictive performance in terms of LPPL, particularly when analysing grouped data with limited information, these results suggest that monitoring the LPPL starting from small values of $K$ and selecting the value at the kink is a reasonable and practical approach.

Table~\ref{tab:sim_ai} presents the RMSE and coverage for the average income across the five models. 
Overall, the proposed model outperforms the alternatives in terms of RMSE. 
The coverage for the proposed model remains close to the nominal level of $0.95$ in most cases; however, some under-coverage occurs in spatial interpolation in Setting~2 due to model misspecification. 
Notably, the RMSE for in-sample estimation and temporal prediction decreases as the frequency increases in both settings. 
A similar, albeit more subtle, improvement is seen for spatial interpolation in Setting~1, whereas this is absent in Setting~2. 

Regarding the two-way model, although its in-sample RMSE is often the second smallest after that of the proposed model, it yields higher RMSE than the proposed or the spatial-only models in spatial interpolation and temporal prediction. 
While its in-sample performance improves with higher frequencies, no such gains are observed for spatial interpolation and temporal prediction. 
The spatial-only model produces the second smallest RMSE for spatial interpolation under both settings. 
However, it exhibits under-coverage that persists even as the frequency increases.  
These results highlight the efficacy of incorporating structured spatio-temporal effects; the proposed model demonstrates superior performance by effectively integrating both. 
Finally, estimating the grouped-data model for each area and period separately results in a high in-sample RMSE due to a lack of information borrowing across space and time.

The employed SAE methods are not specifically designed for grouped data, and the underlying model assumptions differ substantially from the data-generating processes considered in this study. 
Consequently, Table~\ref{tab:sim_ai} highlights the poor performance of these SAE methods, as evidenced by their large RMSEs in both settings and the low coverage in Setting~1. 

Finally, the proposed, two-way independent, and spatial-only models are compared based on the RMSE and coverage for the number of households. 
The results are presented in Table~\ref{tab:sim_yy}. 
The proposed model, which incorporates both spatial and temporal effects, performs robustly and achieves the smallest RMSE in most scenarios, with the exception of in-sample estimation. 
Furthermore, its coverage consistently remains close to the nominal level of $0.95$. 
Interestingly, unlike the results for the average income, the performance of the proposed model for household counts does not exhibit a clear trend of improvement with higher survey frequencies. 
The two-way independent model yields the largest RMSE for spatial interpolation and temporal prediction, whereas the spatial-only model produces smaller RMSE for spatial interpolation than the two-way model.

\begin{table}[H]
    \centering
    \caption{{RMSE and coverage of the 95\% intervals for the average income in in-sample estimation, spatial interpolation for the unobserved areas, and temporal prediction. Results are compared among five models: the proposed spatio-temporal model (Proposed), the two-way independent spatio-temporal model (Two-way), the spatial-only model (Spatial),  the small area estimation  (SAE), and the separate grouped log-normal model (LN).}}
    \label{tab:sim_ai}
\begin{tabular}{clllrrrrr}\toprule
    &Frequency&&& Proposed & Two-way & Spatial & SAE & LN \\\hline
Setting~1 & Low     & RMSE     & In-sample & 0.092 & 0.093 & 0.141 & 0.364 & 0.282\\
	      &         &          & Spatial   & 0.249 & 0.413 & 0.295 & 0.406 & ---\\
	      &         &          & Temporal  & 0.194 & 0.347 & ---   & 0.357 & --- \\
	      &         & Coverage & In-sample & 0.914 & 0.948 & 0.888 & 0.479 & 0.931 \\
	      &         &          & Spatial   & 0.937 & 0.937 & 0.860 & 0.812 & --- \\
	      &         &          & Temporal  & 0.986 & 0.946 & ---   & 0.994 & ---\\
          & Medium  & RMSE     & In-sample & 0.067 & 0.066 & 0.140 & 0.354 & 0.272\\
	      &         &          & Spatial   & 0.240 & 0.409 & 0.295 & 0.394 & ---\\
	      &         &          & Temporal  & 0.179 & 0.358 & ---   & 0.330  & --- \\
	      &         & Coverage & In-sample & 0.940 & 0.957 & 0.892 & 0.367 & 0.935 \\
	      &         &          & Spatial   & 0.956 & 0.947 & 0.870 & 0.780 & --- \\
	      &         &          & Temporal  & 0.984 & 0.869 & ---   & 0.987 & ---\\
          & High    & RMSE     & In-sample & 0.047 & 0.049 & ---   & 0.364 & 0.276\\
	      &         &          & Spatial   & 0.241 & 0.417 & ---   & 0.401 & ---\\
	      &         &          & Temporal  & 0.170 & 0.322 & ---   & 0.358 & --- \\
	      &         & Coverage & In-sample & 0.953 & 0.952 & ---   & 0.271 & 0.938 \\
	      &         &          & Spatial   & 0.932 & 0.945 & ---   & 0.747 & --- \\
	      &         &          & Temporal  & 0.954 & 0.899 & ---   & 0.949 & --- \\\hline
Setting~2 & Low     & RMSE     & In-sample & 0.076 & 0.088 & 0.126 & 0.199 & 0.259\\
	      &         &          & Spatial   & 0.218 & 0.623 & 0.239 & 0.441 & ---\\
	      &         &          & Temporal  & 0.266 & 0.795 & ---   & --- & --- \\
	      &         & Coverage & In-sample & 0.934 & 0.953 & 0.863 & 0.926 & 0.939 \\
	      &         &          & Spatial   & 0.856 & 0.905 & 0.816 & 0.946 & --- \\
	      &         &          & Temporal  & 0.981 & 0.769 & ---   & ---   & ---\\
          & Medium  & RMSE     & In-sample & 0.065 & 0.071 & 0.126 & 0.195 & 0.252\\
	      &         &          & Spatial   & 0.224 & 0.677 & 0.247 & 0.470 & ---\\
	      &         &          & Temporal  & 0.202 & 0.861 & ---   & ---   & --- \\
	      &         & Coverage & In-sample & 0.927 & 0.946 & 0.861 & 0.924 & 0.942 \\
	      &         &          & Spatial   & 0.870 & 0.923 & 0.811 & 0.937 & --- \\
	      &         &          & Temporal  & 0.931 & 0.534 & ---   & ---   & ---\\
          & High  & RMSE       & In-sample & 0.048 & 0.052 & ---   & 0.195 & 0.254\\
	      &         &          & Spatial   & 0.213 & 0.622 & ---   & 0.448 & ---\\
	      &         &          & Temporal  & 0.172 & 0.662 & ---   & ---   & --- \\
	      &         & Coverage & In-sample & 0.932 & 0.940 & ---   & 0.925 & 0.926 \\
	      &         &          & Spatial   & 0.864 & 0.902 & ---   & 0.942 & --- \\
	      &         &          & Temporal  & 0.931 & 0.855 & ---   & ---    & ---\\\bottomrule\end{tabular}
\end{table}

\begin{table}[H]
    \centering
    \caption{{RMSE and coverage of the 95\% prediction intervals for the number of households in in-sample estimation, spatial interpolation for the unobserved areas, and temporal prediction. Results are compared under the proposed spatio-temporal (Proposed), the two-way independent spatio-temporal (Two-way), and the spatial-only (Spatial) models.}}    \label{tab:sim_yy}
\begin{tabular}{clllrrrr}\toprule
    &&&& Proposed & Two-way & Spatial \\\hline
Setting~1 & Low     & RMSE     & In-sample & 5.058 & 5.023 & 4.935 \\
	      &         &          & Spatial   & 6.801 & 8.517 & 7.170 \\
	      &         &          & Temporal  & 6.127 & 7.760 & ---\\
	      &         & Coverage & In-sample & 0.969 & 0.971 & 0.974 \\
	      &         &          & Spatial   & 0.965 & 0.964 & 0.954 \\
	      &         &          & Temporal  & 0.974 & 0.965 & --- \\
          & Medium  & RMSE     & In-sample & 5.083 & 5.069 & 4.869 \\
	      &         &          & Spatial   & 6.808 & 8.578 & 7.318 \\
	      &         &          & Temporal  & 6.131 & 8.174 & --- \\
	      &         & Coverage & In-sample & 0.969 & 0.969 & 0.975 \\
	      &         &          & Spatial   & 0.968 & 0.963 & 0.952 \\
	      &         &          & Temporal  & 0.973 & 0.955 & ---\\
          & High    & RMSE     & In-sample & 5.136 & 5.133 & ---\\
	      &         &          & Spatial   & 6.780 & 8.665 & --- \\
	      &         &          & Temporal  & 6.095 & 7.416 & --- \\
	      &         & Coverage & In-sample & 0.967 & 0.967 & --- \\
	      &         &          & Spatial   & 0.962 & 0.963 & --- \\
	      &         &          & Temporal  & 0.963 & 0.959 & ---\\\hline
Setting~2 & Low     & RMSE     & In-sample & 5.137 & 5.085 & 5.090 \\
	      &         &          & Spatial   & 6.425 &12.072 & 6.604 \\
	      &         &          & Temporal  & 7.305 &16.464 & ---\\
	      &         & Coverage & In-sample & 0.968 & 0.970 & 0.970 \\
	      &         &          & Spatial   & 0.953 & 0.949 & 0.949 \\
	      &         &          & Temporal  & 0.975 & 0.904 & --- \\
          & Medium  & RMSE     & In-sample & 5.098 & 5.080 & 5.050 \\
	      &         &          & Spatial   & 6.244 &12.062 & 6.461 \\
	      &         &          & Temporal  & 6.921 &16.469 & --- \\
	      &         & Coverage & In-sample & 0.970 & 0.969 & 0.975 \\
	      &         &          & Spatial   & 0.956 & 0.957 & 0.953 \\
	      &         &          & Temporal  & 0.977 & 0.903 & ---\\
          & High    & RMSE     & In-sample & 5.197 & 5.192 & ---  \\
	      &         &          & Spatial   & 6.457 &12.064 & --- \\
	      &         &          & Temporal  & 5.917 &13.783 & --- \\
	      &         & Coverage & In-sample & 0.966 & 0.966 & ---  \\
	      &         &          & Spatial   & 0.953 & 0.947 & ---  \\
	      &         &          & Temporal  & 0.961 & 0.928 & --- \\\bottomrule
\end{tabular}
\end{table}

\section{Analysis  of the HLS data}\label{sec:real}
\subsection{Set-up}\label{sec:setup}
{In our analysis, we utilise the HLS data on the total annual income of general households, including sources such assalary and transfer incomes. 
For model estimation, we utilise data from $m=800$ municipalities where observations are consistently available across all survey rounds. 
The remaining 941 municipalities, including those with partially available data, are treated as unobserved during model estimation to facilitate for spatial interpolation and validation. 
Accordingly, the observations from the partially observed municipalities are reserved as a holdout set to evaluate and compare the spatial interpolation performance of the models. }
\label{page:unobs}

As the data span twenty years, the endpoints of the income classes provided in Table~\ref{tab:classes} are adjusted using the consumer price index, with 2018 as the base year. 
This adjustment ensures the temporal comparability of income distributions over the study period.

{Regarding the covariate, we employ the taxable income per taxpayer in millions of JPY, denoted by $\text{tax}_{it}$, setting $\x_{it}=(1,\text{tax}_{it},\text{tax}_{it}^2)^t$. 
We consider this quadratic specification to be more appropriate than a simple linear form, as the various tax deductions in the Japanese progressive tax system exhibit nonlinear behaviour relative to income levels.
Specifically, while the deduction rate is relatively high for low-income households, its marginal benefit diminishes as income rises, eventually reaching a statutory cap at an annual salary of 8.5 million JPY. 
Furthermore, for lower-income groups, social insurance contributions, which increase proportionally with earnings, act as a further constraint on the growth of taxable income, reinforcing the nonlinear relationship.  }
\label{page:tax}

The covariate data are obtained from the tax survey conducted by the Ministry of Internal Affairs and Communications of Japan.\footnote{The tax survey data are available at \url{http://www5.cao.go.jp/keizai-shimon/kaigi/special/future/keizai-jinkou_data/csv/file11.csv}} 
These data are available annually for all municipalities in Japan and can reasonably predict the income distributions observed in the HLS. 
Consistent with our treatment of the income classes, taxable income is adjusted by the consumer price index. 
Notably, while this auxiliary information is available annually at the municipal level, it is insufficient to infer full income distributions by itself. 
Incorporating such vital information as covariates in our proposed model is therefore essential for achieving accurate spatial interpolation and temporal prediction.

\subsection{Model comparison}\label{sec:comparison}
The proposed Gibbs sampler was implemented using four independent parallel chains. 
Each chain was initialised at randomly chosen starting values and run for 40,000 iterations, with the first 10,000 iterations discarded as burn-in. 
We retained every 15th draw from the subsequent 30,000 iterations, yielding a total of 8,000 pooled samples from the four chains for posterior inference. 
To ensure model identifiability, we imposed an order constraint on the intercept term, $\beta_{1k}$. 
Convergence was assessed by comparing the posterior samples across the four chains. 
As evidenced by the trace plots and posterior distribution summaries in the Supplementary Material, the chains converged to a common stationary distribution with substantial overlap.

\begin{figure}[H]
    \centering
    \includegraphics[scale=0.4]{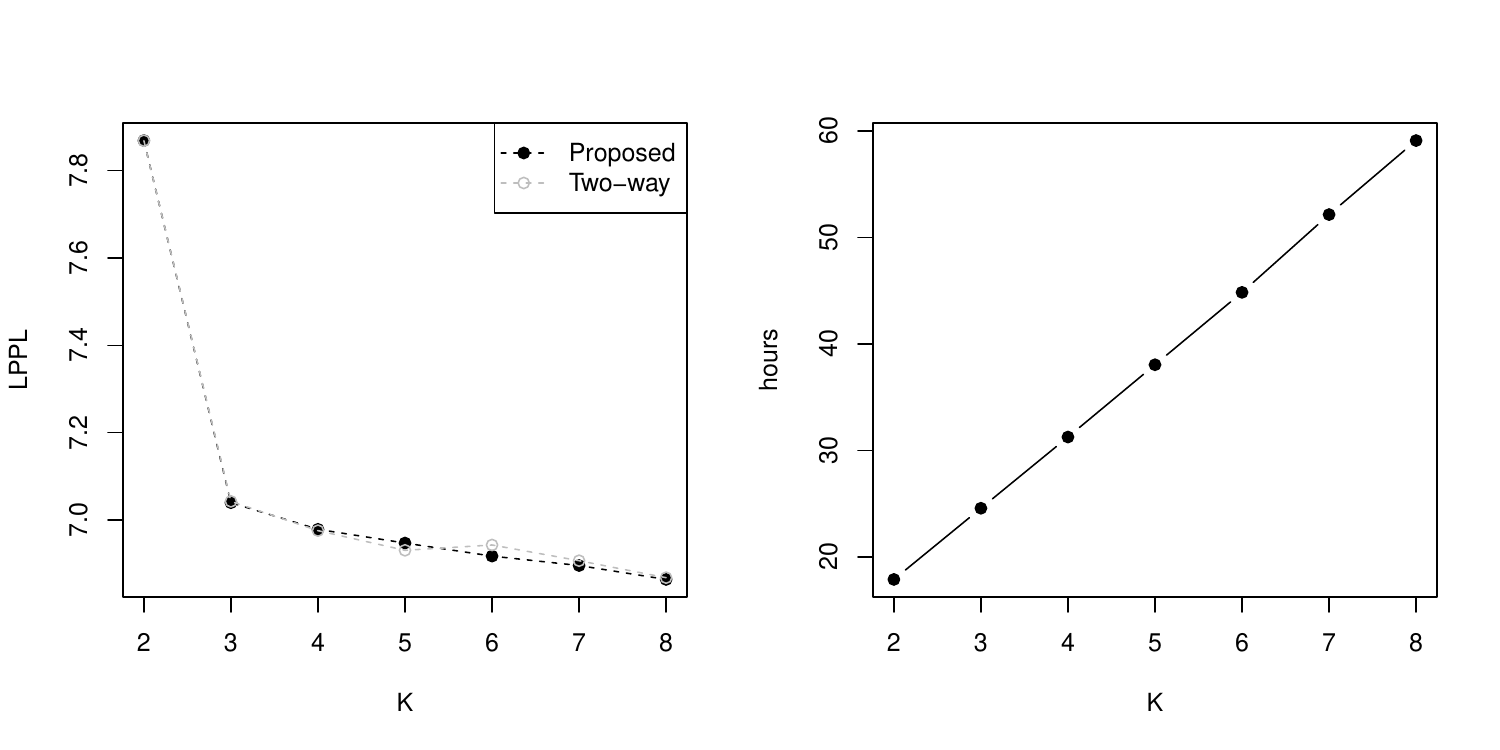}
    \caption{Log posterior predictive loss (LPPL) for the proposed and two-way independent models for the HLS data (left panel) and the computing time for the proposed model across varying $K$ (right panel).}
    \label{fig:k}
\end{figure}

Figure~\ref{fig:k} shows the LPPL for different numbers of mixture components $K$ ranging from $2$ to $8$, for both the proposed and two-way independent models. 
Consistent with the simulation study, the LPPL for both models declines sharply as $K$ increases from $2$ to $3$. 
Although the LPPL continues to decline gradually for larger values of $K$, we select $K=3$ at the elbow as a reasonable and parsimonious number of components for our analysis. 

{
We have also estimated a simple linear model and a model incorporating penalised splines (P-splines) to provide a flexible alternative for the effect of taxable income. 
For the linear model, the same prior was assigned to the regression coefficients as in the quadratic model.  
For the P-spline model, we employed a first-order random walk prior for the B-spline coefficients, $\bbe_k\sim N(\zero, \phi^2 \D^{-1})$, where $\D$ is the penalty matrix and $\phi^2 \sim IG(1,0.005)$ following {\cite{Lang01032004}}. 
Based on the LPPL, we selected $K=3$ for both the linear and P-spline models, which yielded values of $7.064$ and $7.152$, respectively. 
As the LPPL for the quadratic model with $K=3$ is $7.040$, we conclude that the quadratic specification provides the best fit among the candidate models. }
\label{page:tax_lppl}

{
{Figure~\ref{fig:k}} also presents the computing time for the proposed model across $K=2,\dots,8$. 
The results indicate that the computing time increases linearly with $K$. 
For the selected value of $K=3$, the MCMC algorithm required approximately 24 hours to complete using a non-optimised R code on an Apple M2 Ultra chip without any parallelisation. }
\label{page:time}

The number of households in each income class of the holdout municipalities is predicted using spatial interpolation. 
The proposed model is compared with the two-way independent and parametric spatial income models for grouped data proposed by \cite{sugasawa2020estimation}. 
In the approach of \cite{sugasawa2020estimation}, the parameters of a parametric income distribution vary over space through an appropriate transformation of spatial random effects. 
We consider the Singh-Maddala and Dagum distributions as the parametric models, as they are frequently employed in income modelling \citep{kleiber03book}. 

The performance of the four models is evaluated based on the coverage of the 95\% prediction interval, as well as the $0.5$, $0.9$, and $0.99$ quantiles of the relative absolute error (RAE) for the household counts. 
The RAE is defined as $|\hat{N}_{itg}-N_{itg}|/(N_{itg}+1)$, where $\hat{N}_{itg}$ denotes the posterior predictive mean for the holdout municipality. 
Table~\ref{tab:coverage} reveals that the proposed and two-way models achieve coverage rates close to the nominal level of $0.95$. 
In contrast, the Singh-Maddala and, particularly, the Dagum models result in under-coverage. 
While the proposed, Dagum and Singh-Maddala models yield comparable median RAE, the $0.9$ and $0.99$ quantiles of the RAE under the latter two models are substantially larger than those under the proposed model.

\begin{table}[H]
    \centering
    \caption{Coverage rates of the 95\% prediction intervals and the 0.5, 0.9, and 0.99 quantiles of the RAE for the holdout municipalities under the proposed spatio-temporal model (Proposed), two-way independent spatio-temporal model (Two-way), and Dagum and Singh-Maddala models. }
    
    \begin{tabular}{crrrr}\toprule
    & &\multicolumn{3}{c}{Quantiles of RAE} \\
    \cmidrule(lr){3-5}
    Model & \multicolumn{1}{c}{Coverage} & \multicolumn{1}{c}{$0.5$th} & \multicolumn{1}{c}{$0.9$th} & \multicolumn{1}{c}{$0.99$th}
    \\\hline
     Proposed        & 0.945 & 0.139 & 0.404 & 1.044 \\
     Two-way         & 0.939 & 0.158 & 0.471 & 1.201 \\
     Dagum           & 0.627 & 0.131 & 0.454 & 1.469 \\
     Singh-Maddala   & 0.886 & 0.134 & 0.467 & 1.415 \\\bottomrule
    \end{tabular}
    \label{tab:coverage}
\end{table}

\subsection{Income maps of Japan}\label{sec:map}
Based on the results presented in Figure~\ref{fig:k} and Table~\ref{tab:coverage}, we adopt $K=3$ for the proposed spatio-temporal mixture model in the subsequent analysis.
Using this model, we constructed the complete income maps of Japan for all HLS survey years as well as for 2020, a year in which HLS was not conducted. 

For the survey years, the maps integrate in-sample estimates for the observed municipalities with spatial interpolations for the unobserved municipalities. 
For 2020, the values for all municipalities were obtained through temporal prediction.
Specifically, the temporal effect for 2020 is generated from $N(\eta_{Tk},\frac{2}{5}\alpha^2_k)$, where the variance is scaled because the one-unit increment in $t$ in our model corresponds to a five-year period in the HLS rounds.

\begin{figure}[H]
    \centering
    \begin{tabular}{ccc}
    \includegraphics[scale=0.18]{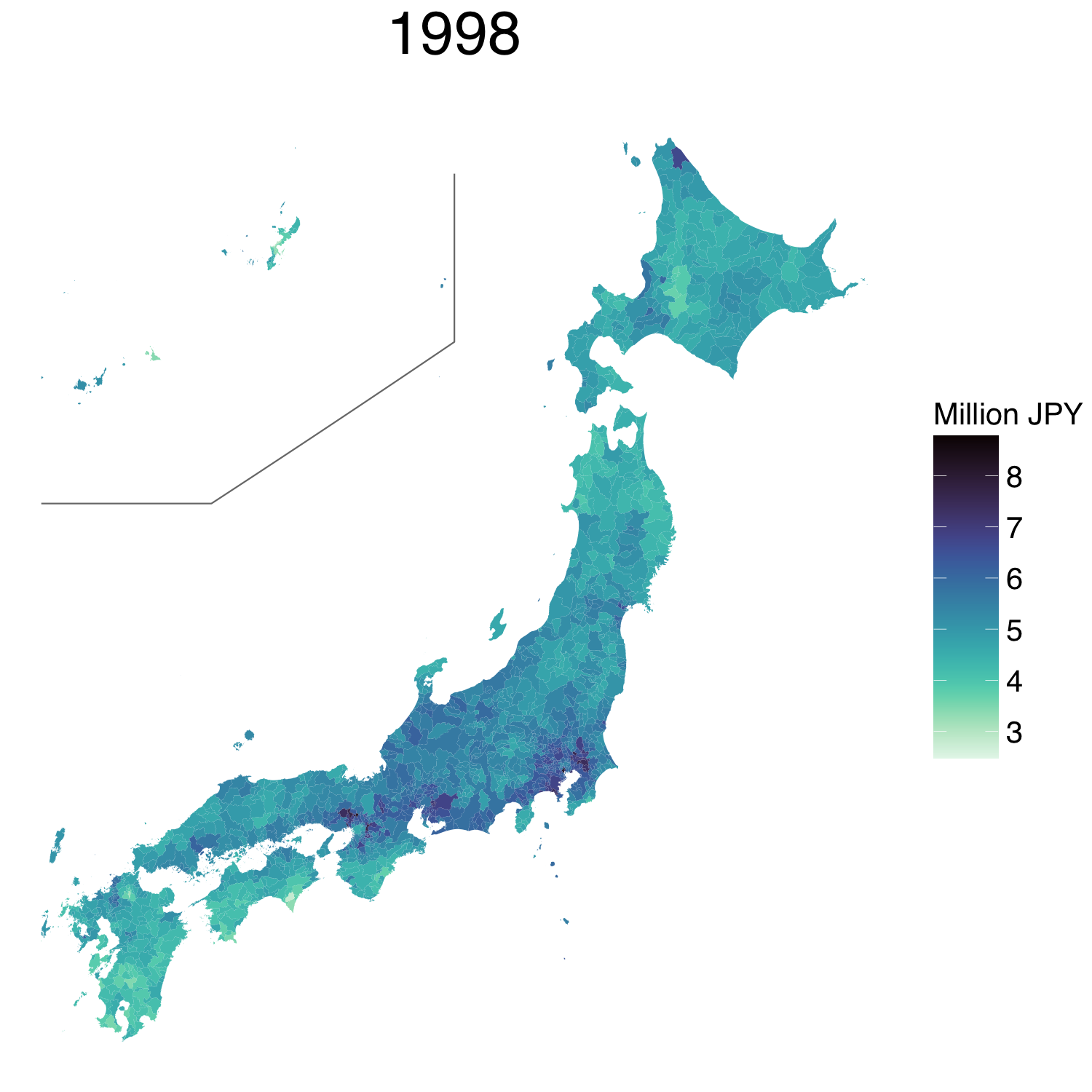}&
    \includegraphics[scale=0.18]{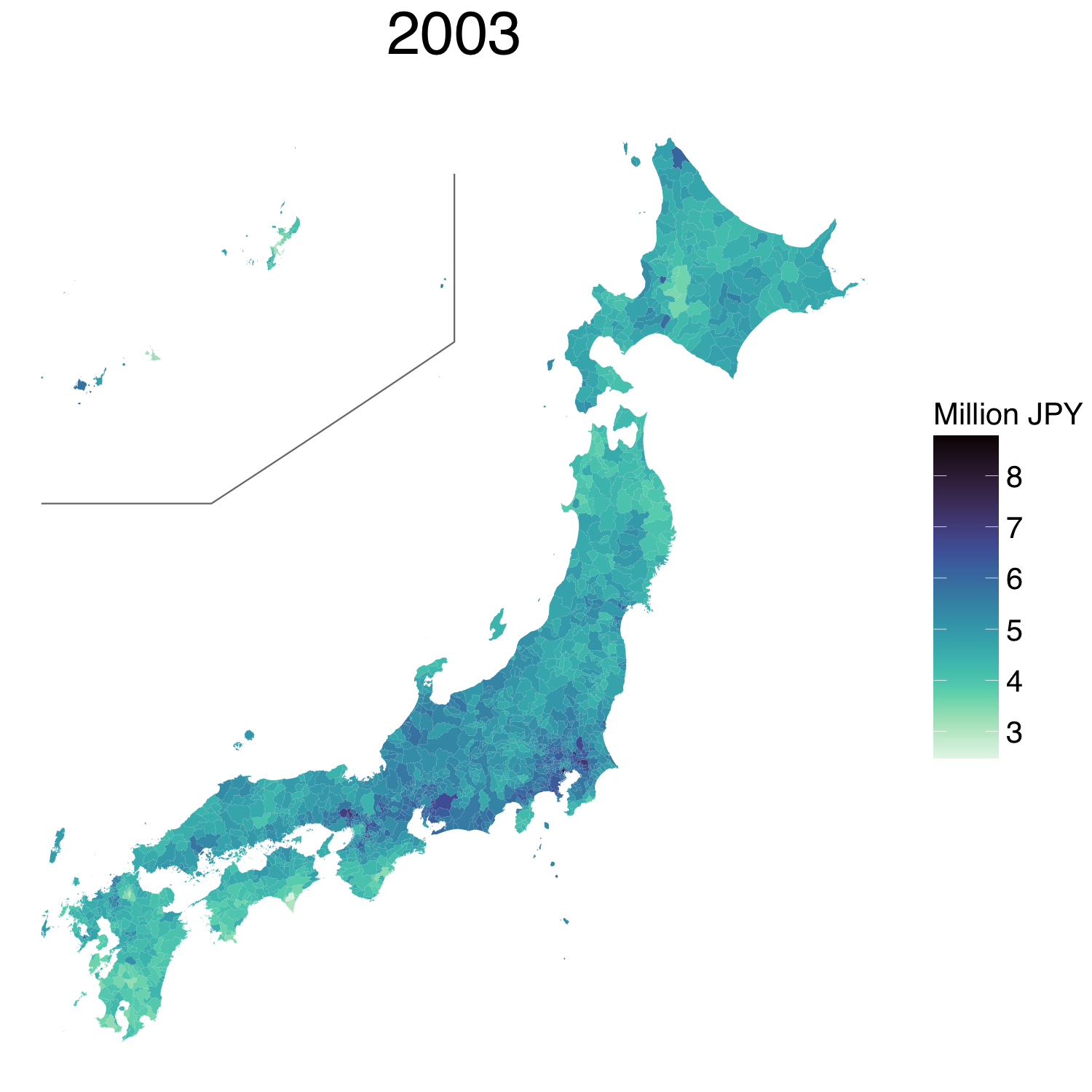}&
    \includegraphics[scale=0.18]{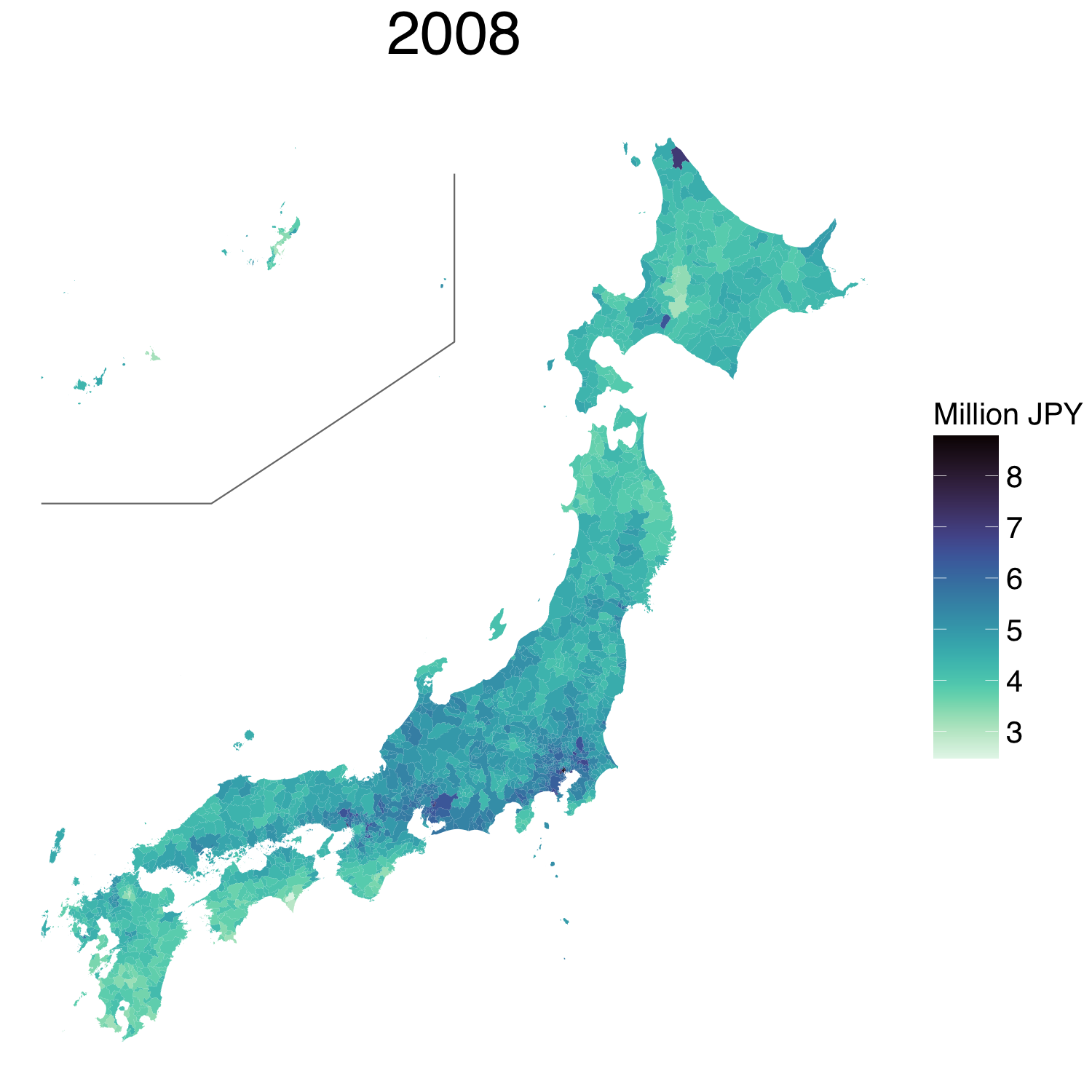}\\
    \includegraphics[scale=0.18]{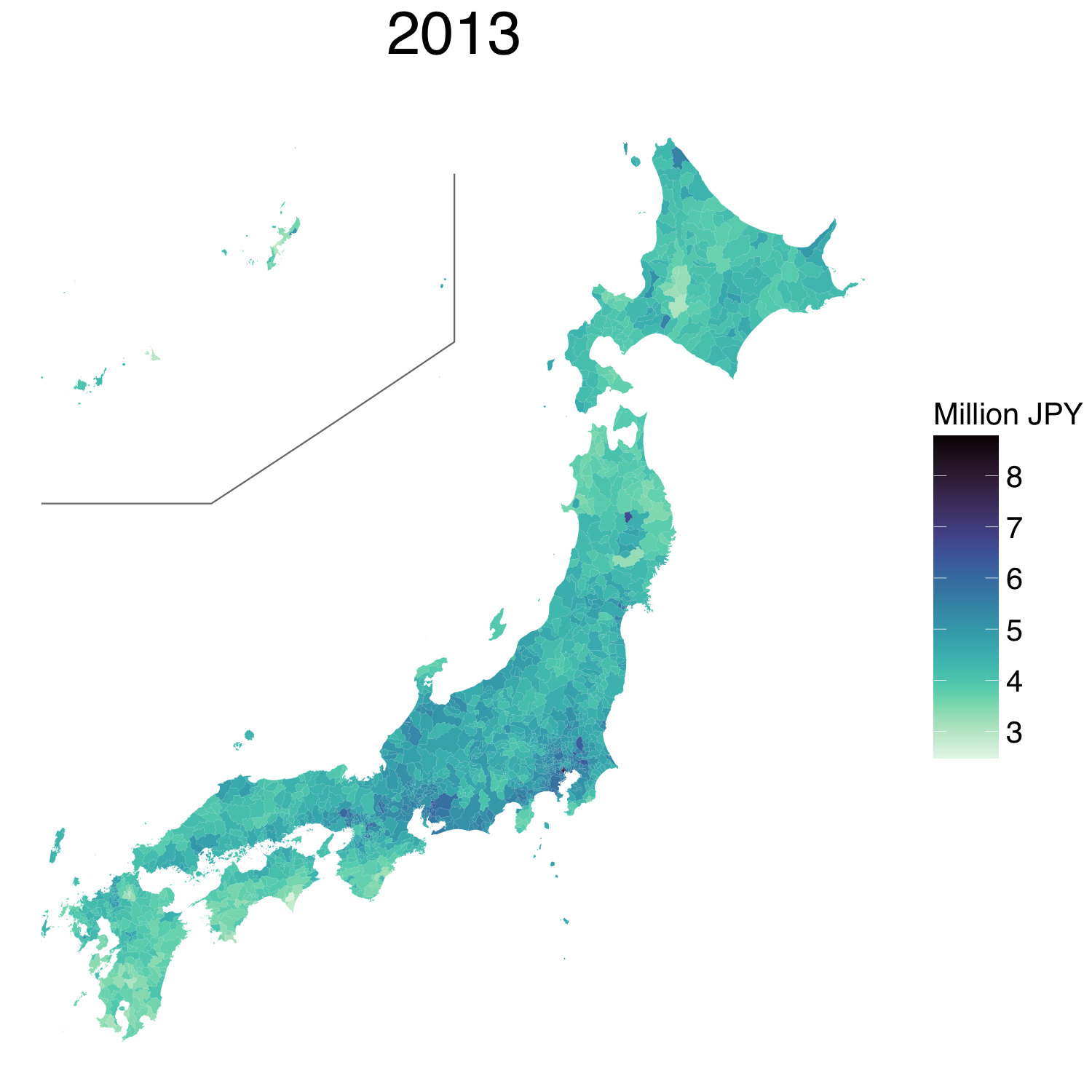}&
    \includegraphics[scale=0.18]{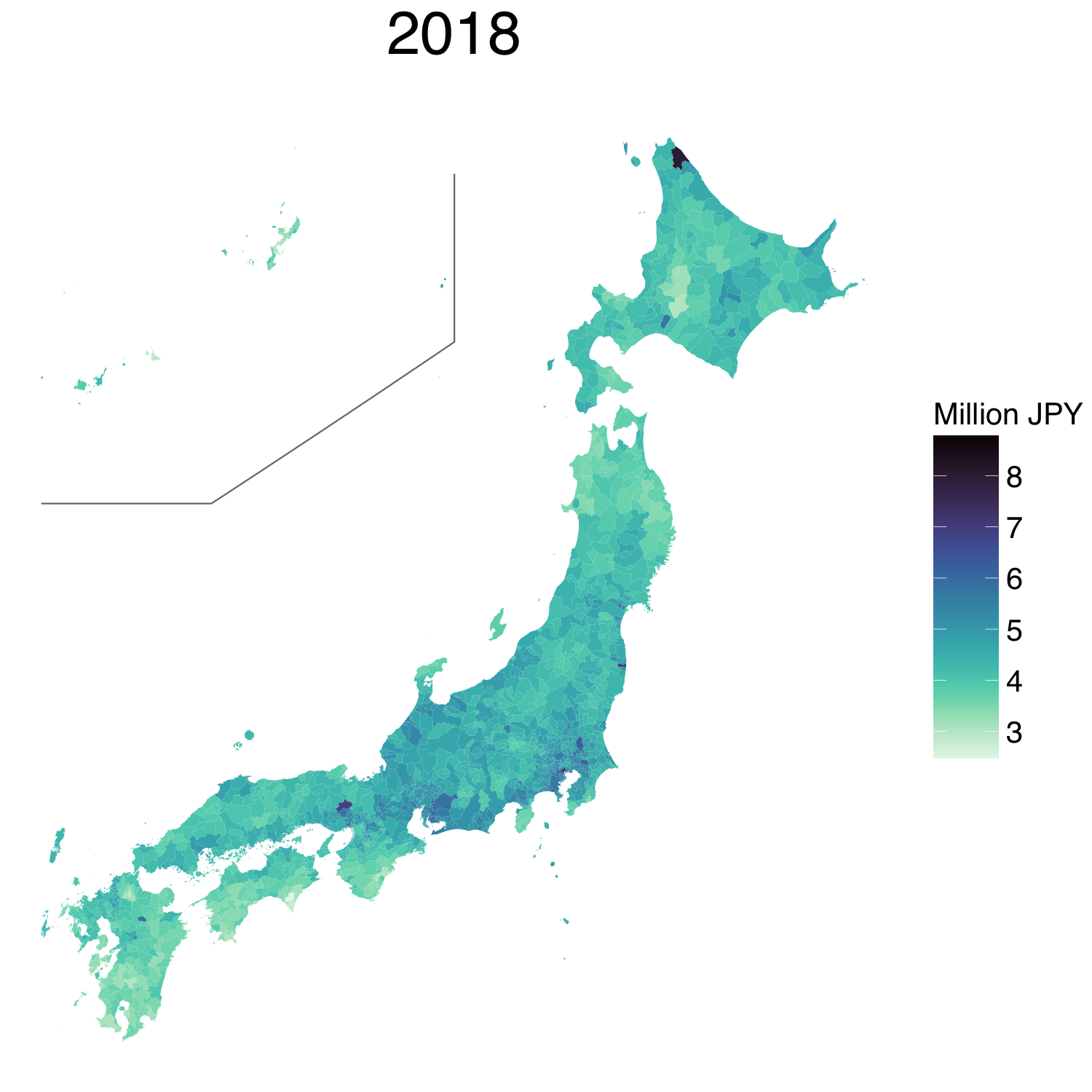}&
    \includegraphics[scale=0.18]{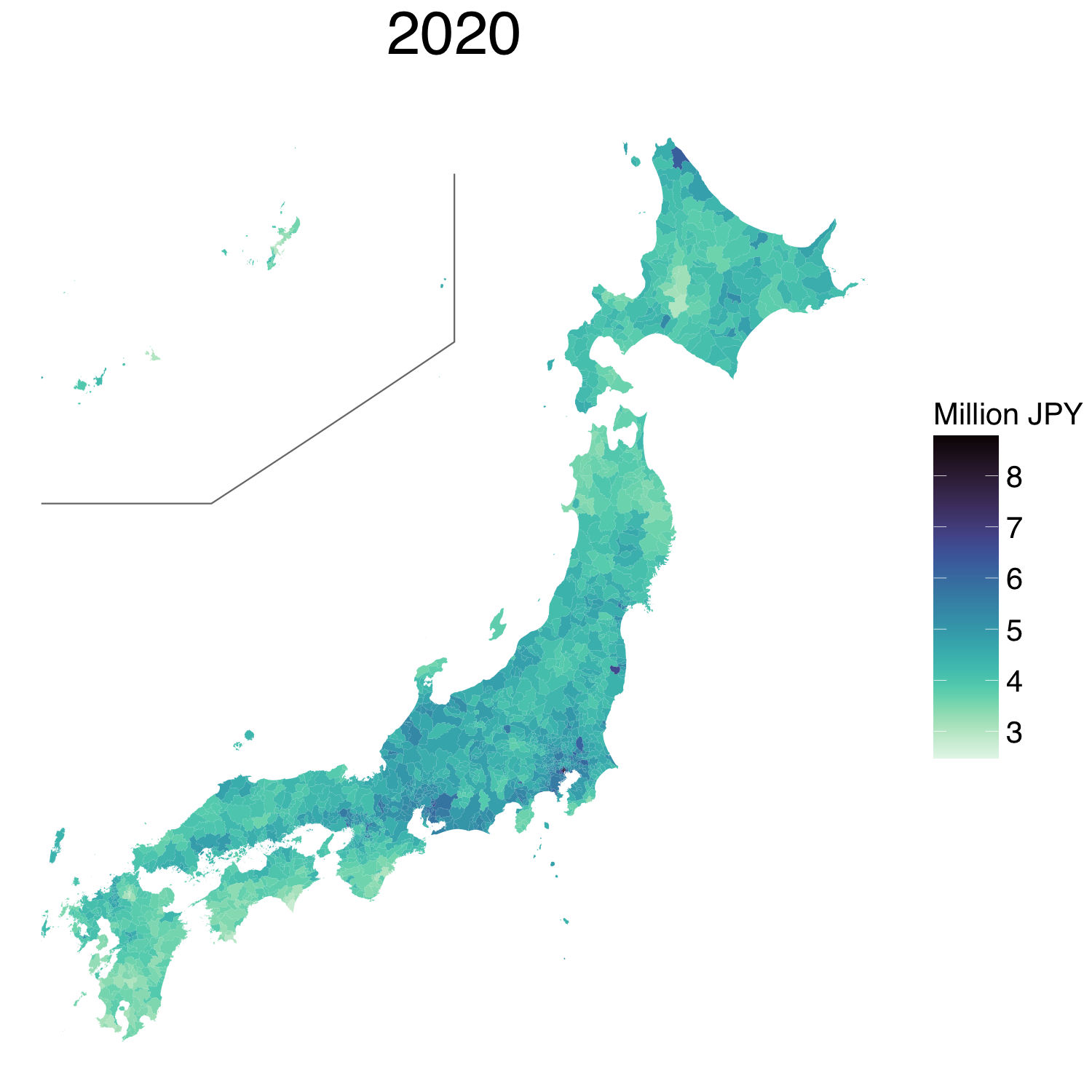}
    \end{tabular}
    \caption{Complete maps of average income.}
    \label{fig:map_ai}
\end{figure}

Figure~\ref{fig:map_ai} shows the complete maps of average income across Japan. 
Municipalities with darker colours indicate areas with higher income levels. 
A notable observation is that the overall colouration of the maps tends to lighten over time. 
In 1998, for example, municipalities along the coastline of central Honshu exhibit much darker colours than those on the south-western islands of Shikoku and Kyushu. 
By 2018, however, high-income municipalities are increasingly concentrated within metropolitan areas, reflecting a growing disparity between urban and rural regions. 
The predicted average income for 2020 exhibits spatial patterns similar to those observed in 2018. 
A consistent trend is also found in median income, the results of which are presented in the Supplementary Material. 

{
{Figure~\ref{fig:ci_ai}} presents the posterior (predictive) means along with the 95\% credible and prediction intervals of the average income of the municipalities. 
The observed municipalities are plotted in black, whereas the unobserved ones are shown in grey. 
The figure shows that the intervals for the observed municipalities are notably narrower than those for the unobserved ones, indicating that the uncertainty inherent in spatial interpolation is appropriately captured by the model. 
}

\begin{figure}[H]
    \centering
    \begin{tabular}{ccc}
    \includegraphics[scale=0.18]{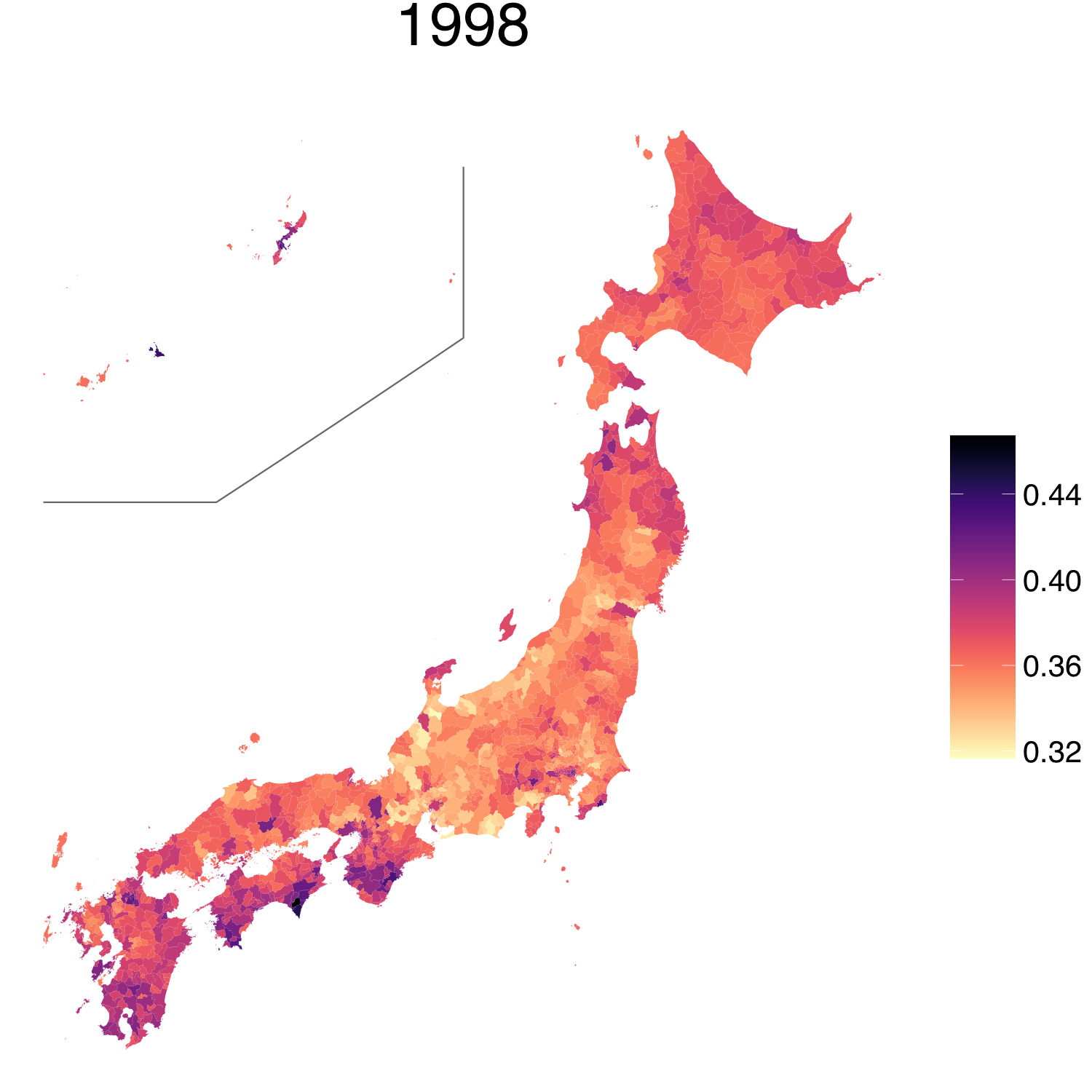}&
    \includegraphics[scale=0.18]{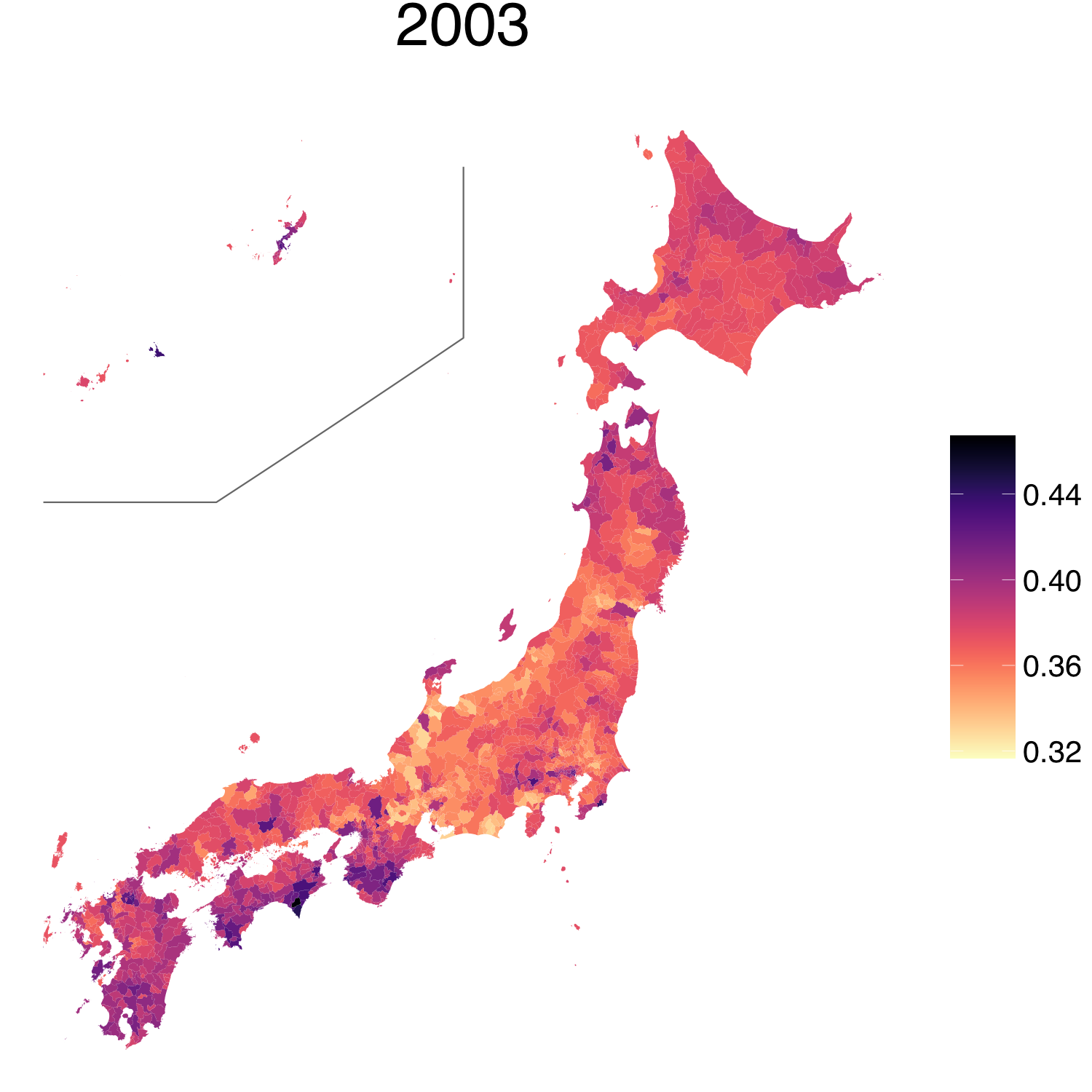}&
    \includegraphics[scale=0.18]{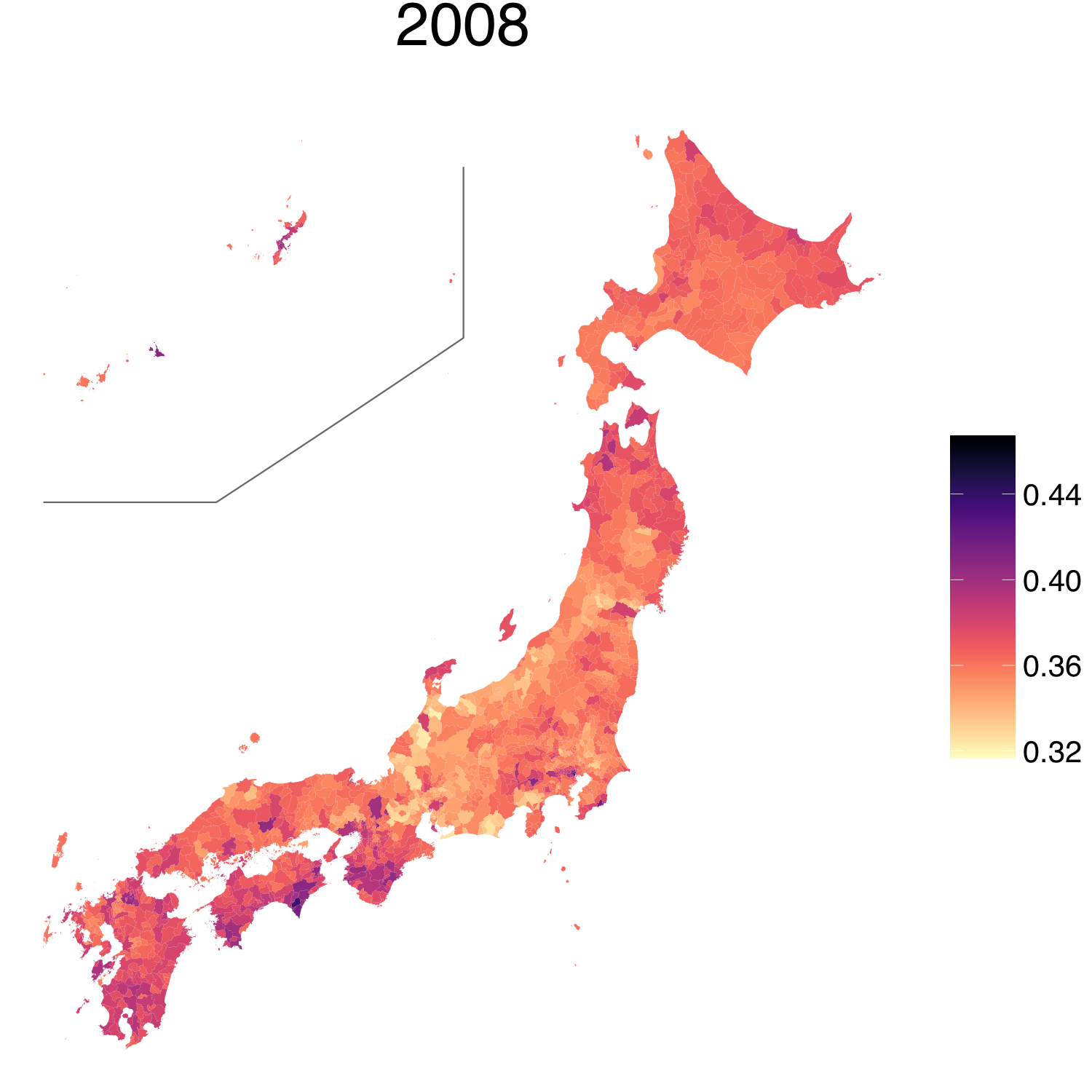}\\
    \includegraphics[scale=0.18]{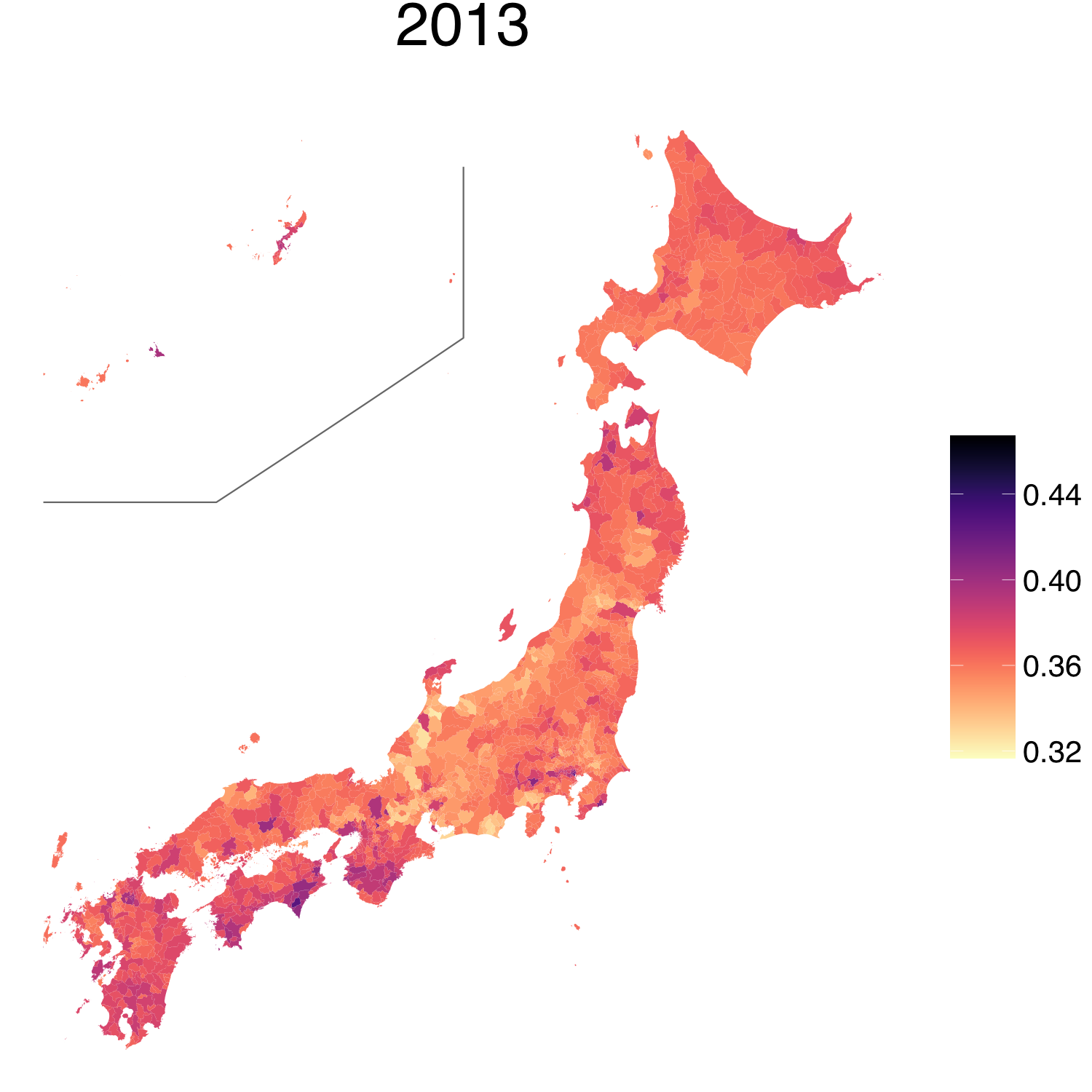}&
    \includegraphics[scale=0.18]{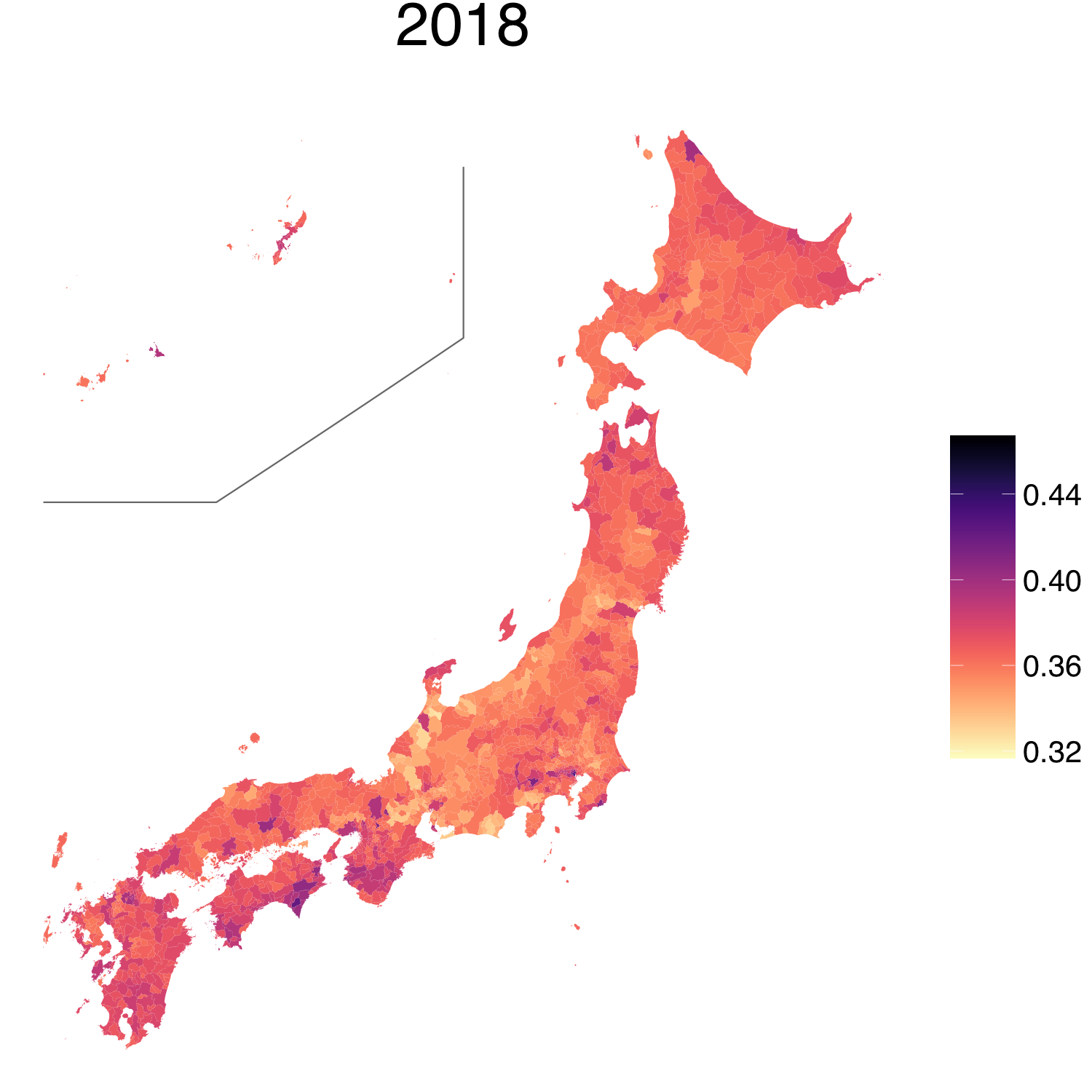}&
    \includegraphics[scale=0.18]{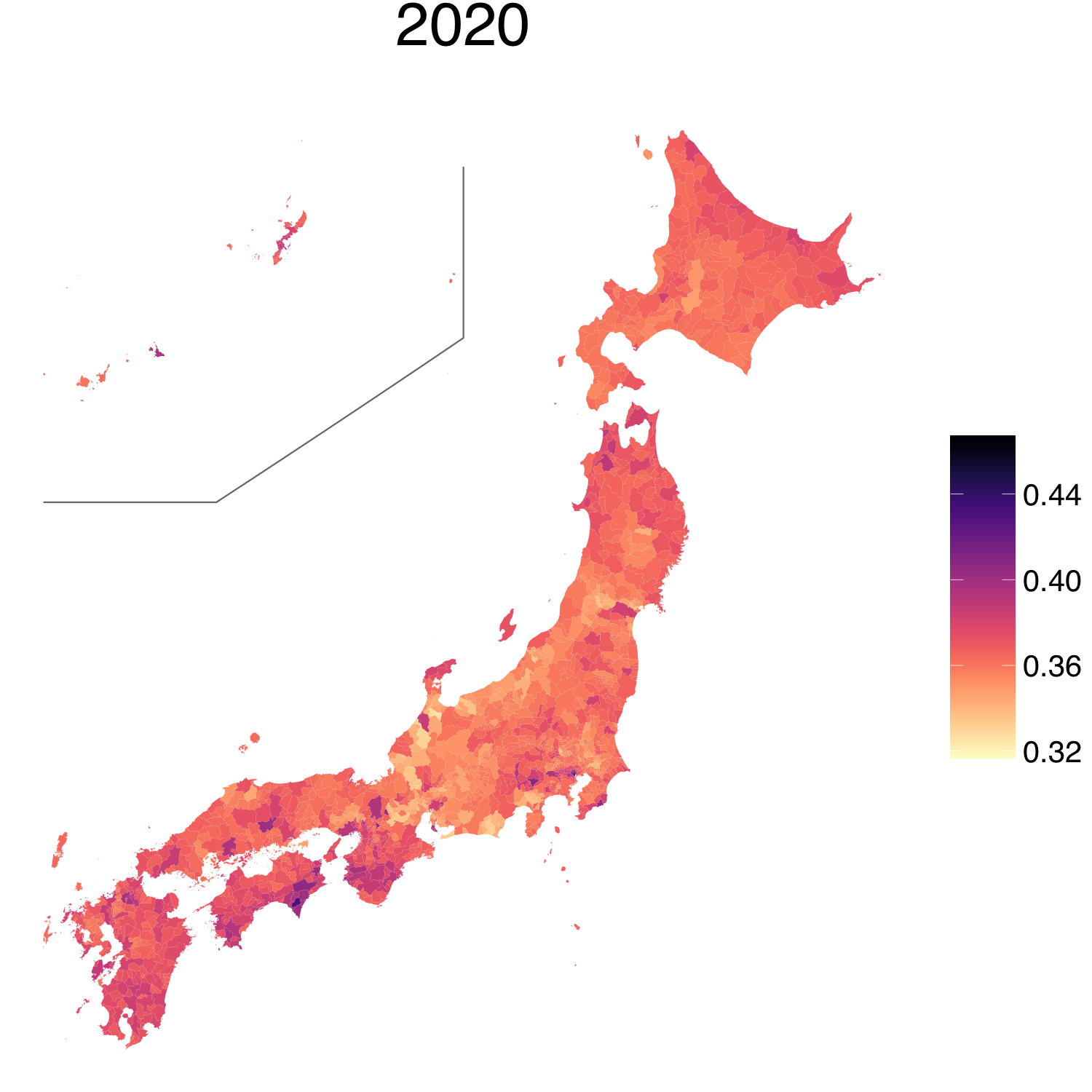}
    \end{tabular}
    \caption{Complete maps of the Gini index.}
    \label{fig:map_gini}
\end{figure}

Figure~\ref{fig:map_gini} presents the posterior (predictive) means of the Gini index across Japan. 
Darker shading indicates municipalities with higher Gini indices, representing greater income inequality. 
The results suggest that the spatial variation in the Gini index across the country has diminished over the 20-year study period. 
In 1998, for example, municipalities in northern and particularly south-western Japan are characterised by dark shading, whereas those in central Honshu exhibit relatively light shading. 
The figure also highlights that the Gini index in 2003 appears uniformly higher than in other periods across Japan. 
By 2018, however, although municipalities in western and northern  Japan still exhibit relatively high inequality, the overall colouration of the map appears more uniform, indicating a degree of spatial convergence in inequality levels. 

\begin{figure}[H]
    \centering
    \includegraphics[width=\linewidth]{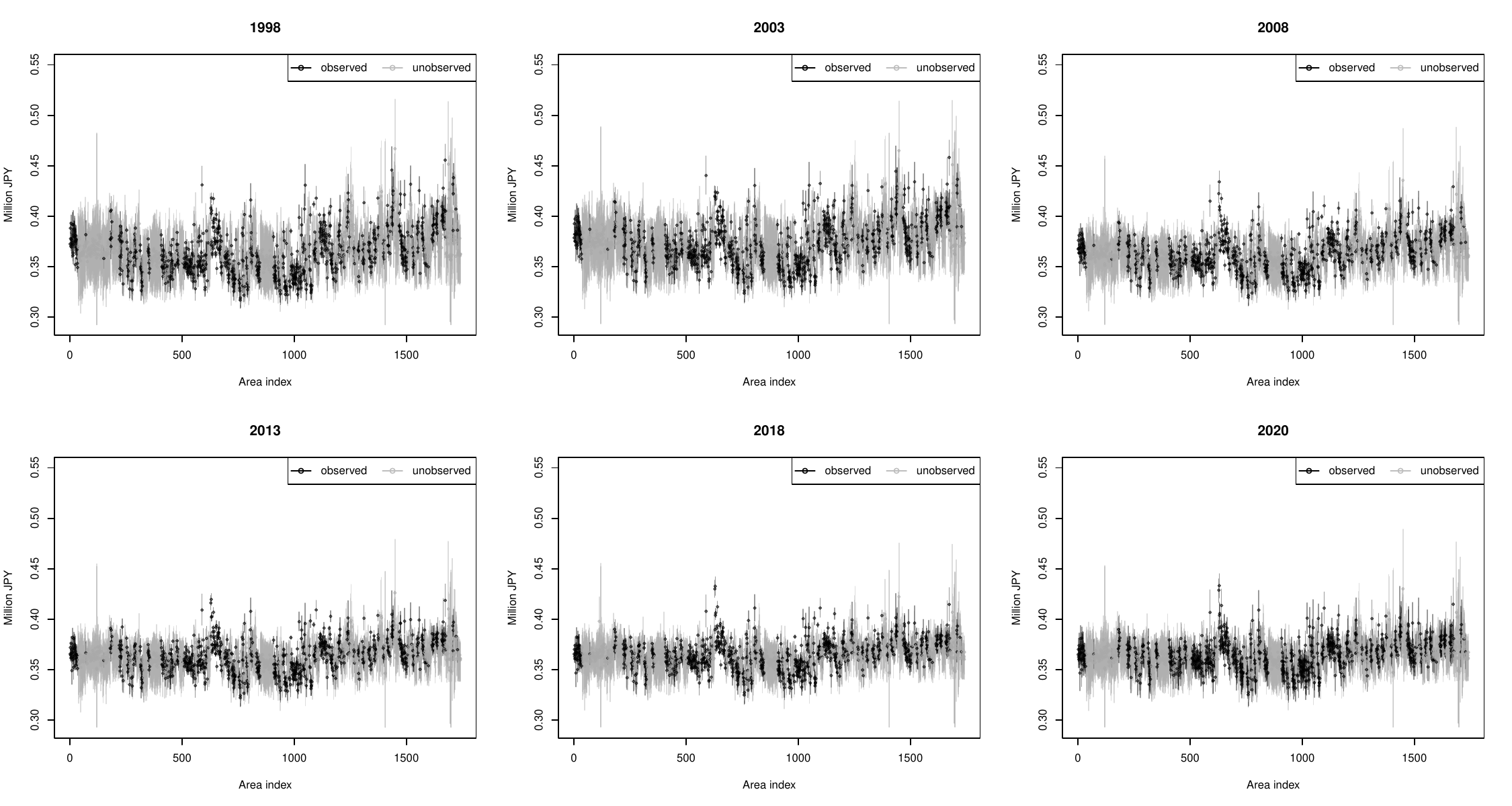}
    \caption{Posterior and posterior predictive means (points) and 95\% credible and prediction intervals (line segments) for the Gini index of the observed (black) and unobserved  (grey) municipalities. }
    \label{fig:ci_gini}
\end{figure}

{
{Figure~\ref{fig:ci_gini}} presents the posterior (predictive) means and 95\% credible and prediction intervals for the Gini index of both observed and unobserved municipalities. 
The patterns of uncertainty are similar to those observed for average income. 
Notably, the overall widths of the 95\% intervals have shrunk over time. 
}

\begin{figure}[H]
    \centering
    \includegraphics[width=\textwidth]{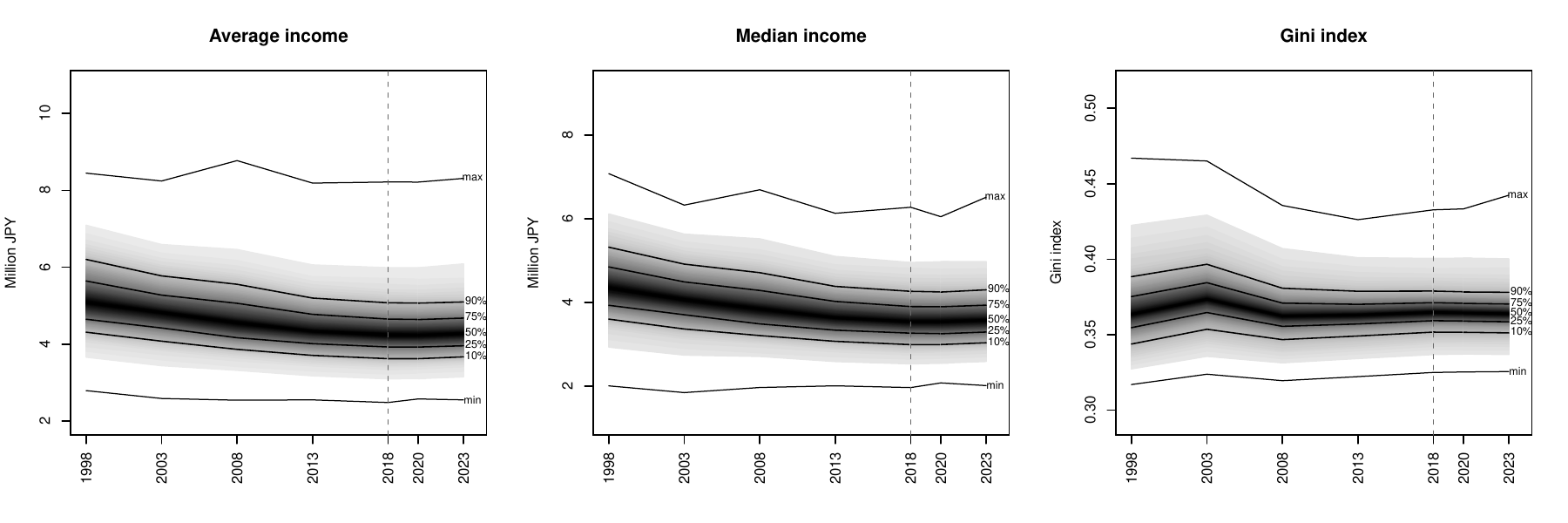}
    \caption{Temporal changes in the distributions of average income, median income, and the Gini index across Japan. The dashed vertical lines indicate that the final HLS in 2018, with subsequent values representing temporal predictions. }
    \label{fig:japan}
\end{figure}

{
{Figure~\ref{fig:japan}} illustrates the temporal evolution of the distributions of average income, median income, and the Gini index across Japan from 1998 to 2023. 
The overall average and median income levels exhibit a clear and persistent downward trend over the study period. 
Furthermore, the figure reveals that the variation in the Gini index has clearly diminished over time, suggesting a trend the towards geographical homogenisation in income inequality across Japanese municipalities. 
}

\begin{figure}[H]
    \centering
    \includegraphics[width=\linewidth]{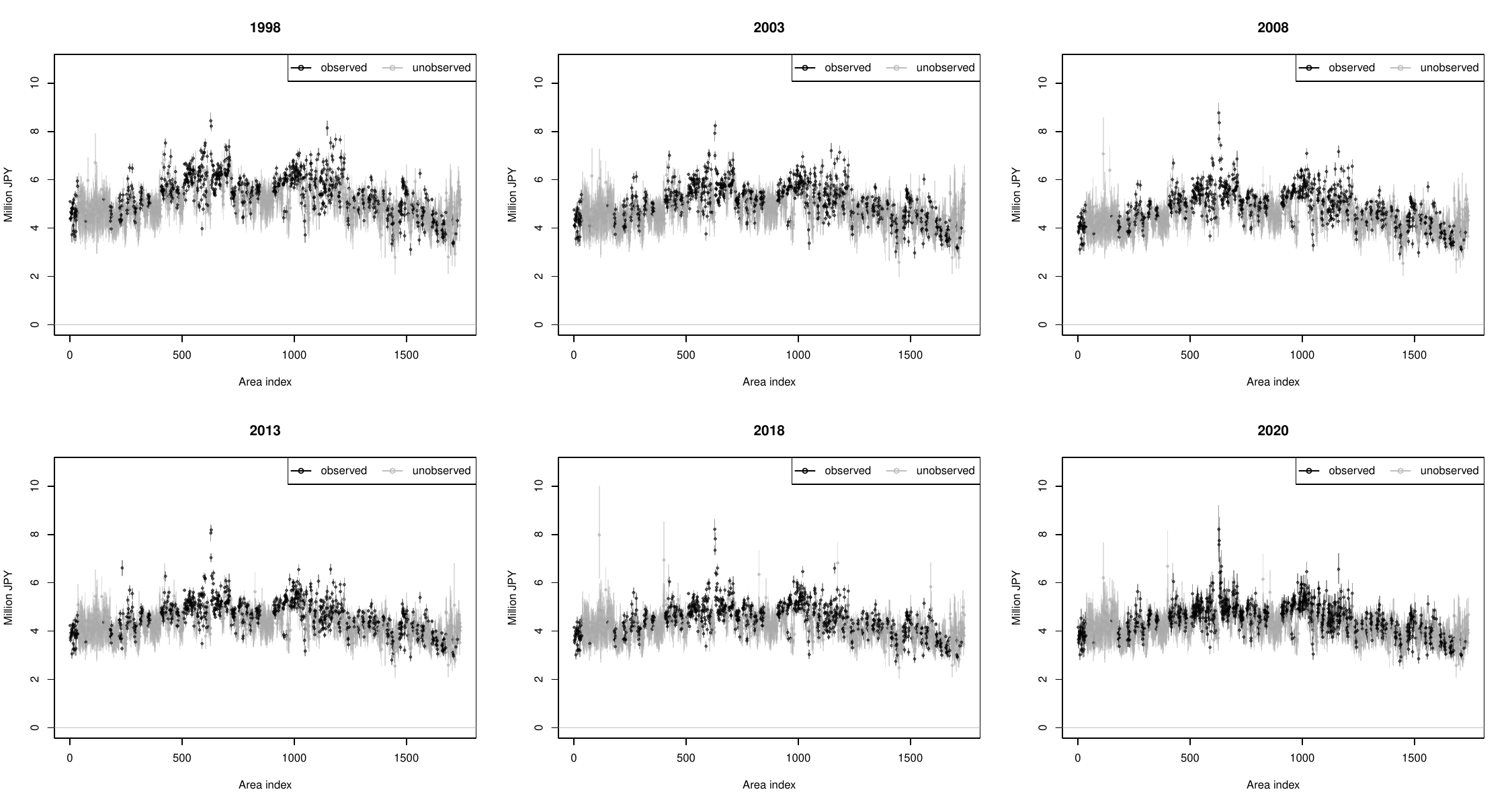}
    \caption{Posterior and posterior predictive means (points) along with 95\% credible and prediction intervals (line segments) for the average income for the observed (black) and unobserved  (grey) municipalities. }
    \label{fig:ci_ai}
\end{figure}

\subsection{{Comparison with crude average incomes}}
\label{sec:crude}
In the context of small area estimation, it is  standard practice to compare model-based estimates with direct estimators to demonstrate the reduction in uncertainty. 
However, in our application, the HLS does not release any official direct estimates or associated uncertainty measures for income indices at the municipality level.  
Furthermore, as discussed in Section~\ref{sec:data}, constructing direct estimates based on the counts published by the HLS and, more crucially, determining their associated uncertainty are far from straightforward. 
This difficulty arises from the presence of open-ended income classes and the lack of information regarding within-class variability. 
Addressing these issues typically requires strong distributional assumptions, which may, in turn, introduce arbitrary bias. 

\begin{figure}[H]
    \centering
    \includegraphics[scale=0.45]{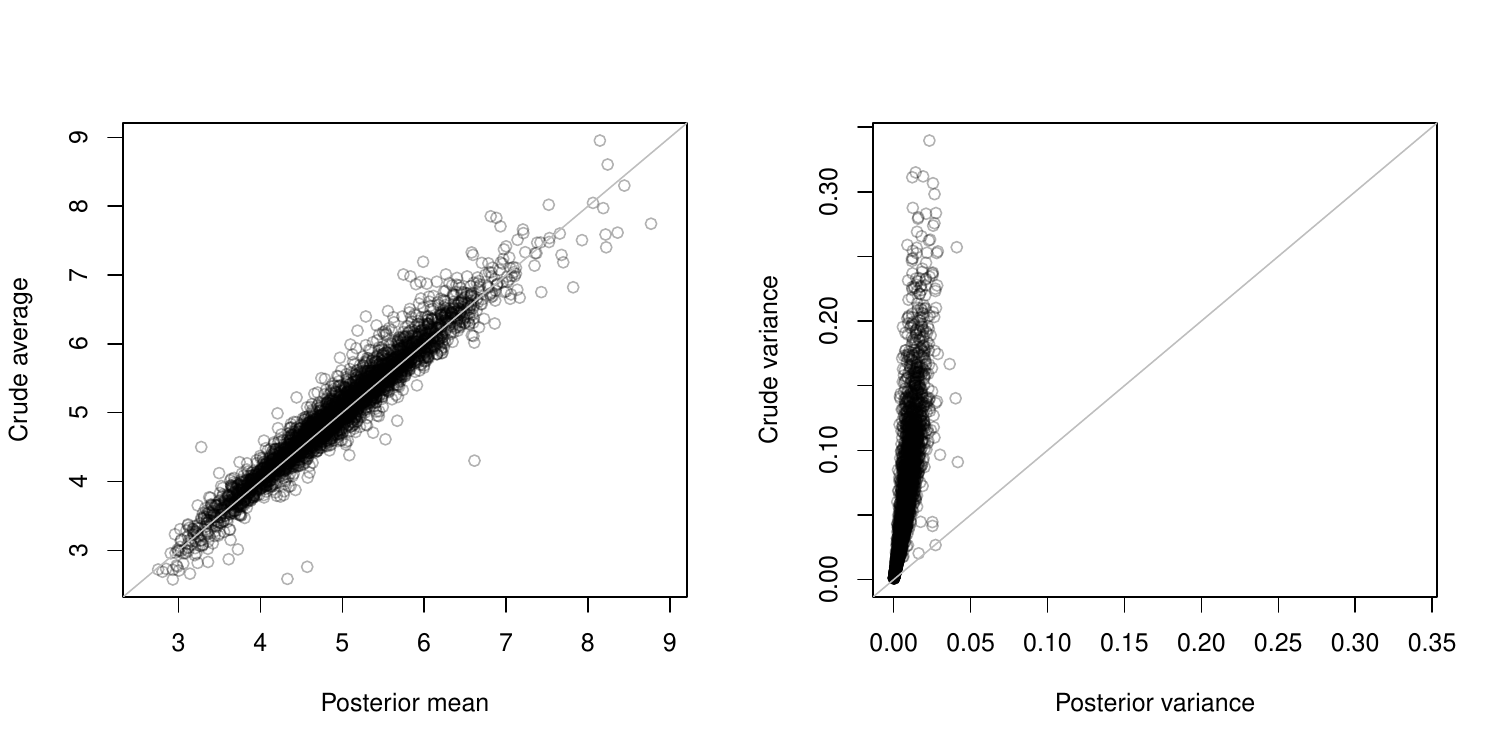}
    \caption{Comparison between the posterior means and the crude estimates (left panel), and between the posterior variances and  the crude sampling variances (right panel), for the average income in the observed municipalities.}
    \label{fig:crude}
\end{figure}

Nonetheless, to provide a comparative benchmark for the observed municipalities, we compare the posterior means and variances of average income with the ``crude'' average incomes and their sampling variances, computed using the same procedure as the SAE models in the simulation study (see the Supplementary Material for details). 
The results are presented in Figure~\ref{fig:crude}. 
While the posterior means and the crude estimates are mostly aligned along the 45-degree line, there are instances where the crude estimates are noticeably smaller. 
Most importantly, the posterior variances are markedly and consistently smaller than the crude variances. 
This result highlights the substantial reduction in uncertainty achieved through the proposed model framework, which effectively borrows strength across space and time.

\subsection{Area-wise income distributions}\label{sec:area-wise}
In this section, we examine the income distributions for eight selected municipalities. 
They include four observed municipalities and four unobserved municipalities, chosen to represent diverse geographical and economic contexts. 
The locations of these selected municipalities are provided in the Supplementary Material. 

Figure~\ref{fig:area_dist} presents the posterior means, along with the 95\% credible and prediction intervals, of the income distributions for the survey years, as well as for 2020 and 2023.  
The top and bottom panels of the figure correspond to the observed and unobserved municipalities, respectively. 
The results demonstrate that the proposed mixture model effectively captures a diverse range of distributional shapes, reflecting the spatial and temporal heterogeneity across different regions and periods. 

For example, Minato, one of the wealthiest districts in Tokyo, exhibits markedly long right tails than other municipalities. 
In contrast, municipalities such as Ninohe in northern Honshu, Nichinan in western Honshu, and Nakatosa in southern Shikoku exhibit much shorter right tails. 
The bulk of these distributions is heavily concentrated below the annual income of 5 million JPY, reflecting the stark contrast in income structures between metropolitan and rural areas. 

Interestingly, the income distributions of Oshino exhibit longer right tails than those of other municipalities, with the exception of Minato. 
Although Oshino is a small municipality that was not observed in the HLS, it hosts the headquarters and factories of a major electric machinery manufacturer, which likely accounts for its high income levels.
This result demonstrates the capacity of the proposed model to infer the income distribution of such unique areas by leveraging auxiliary information through covariates and spatio-temporal mixing proportions. 

For all municipalities shown in the figure, with the exception of Oshino, the right tails of the income distributions have shortened slightly, and the density in low-income regions has increased over the past twenty years. 

\begin{figure}[H]
    \centering
    \includegraphics[width=\textwidth]{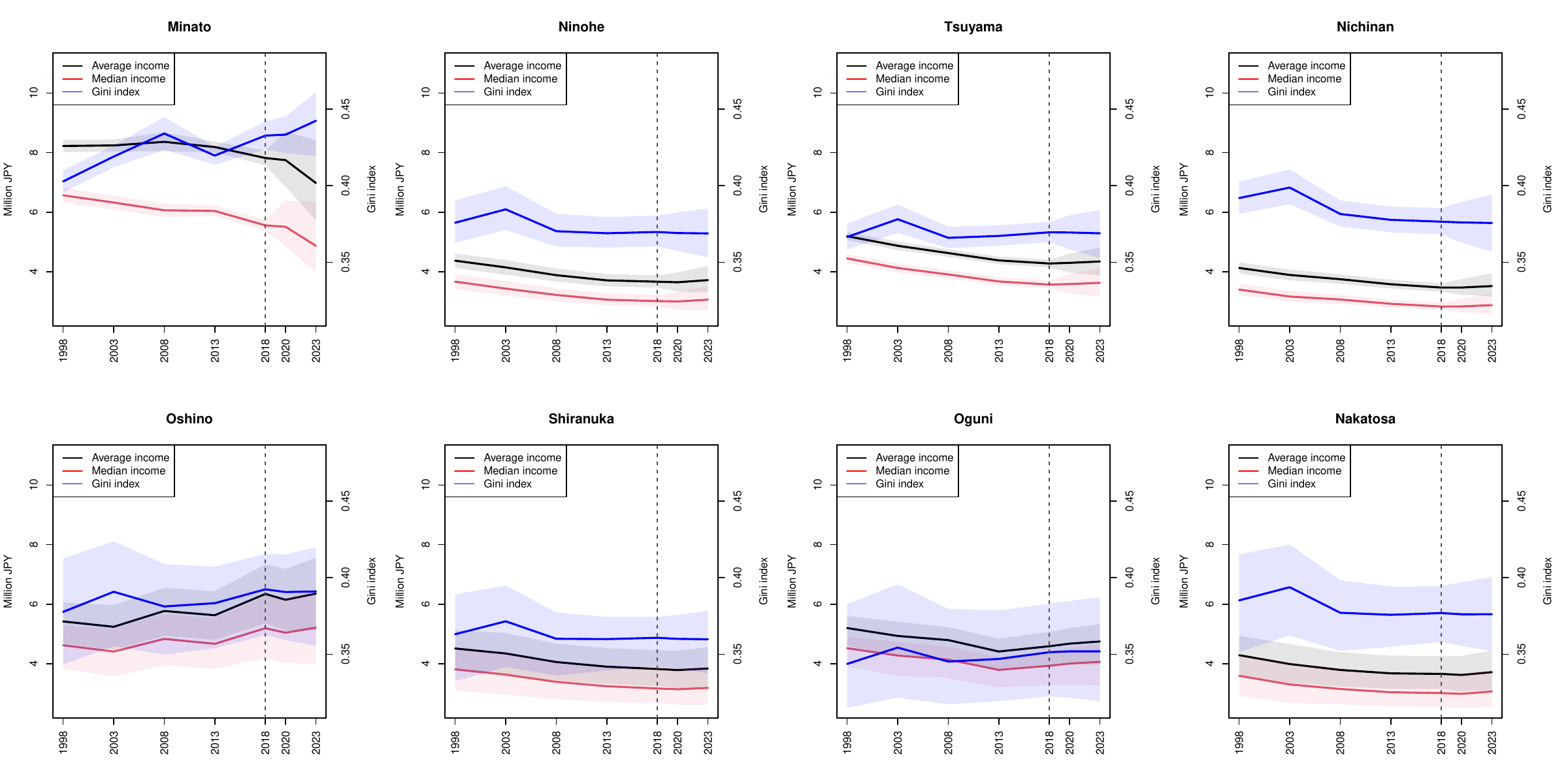}
    \caption{Temporal changes in average and median incomes (left  axis) and the Gini index (right axis) for the selected observed (top panels) and unobserved (bottom panles) municipalities. 
    The shaded areas represent the 95\% intervals, and the vertical dashed lines indicate that the final HLS round in 2018, with the subsequent values representing temporal predictions.}
    \label{fig:area}
\end{figure}

{
{Figure~\ref{fig:area}} illustrates the temporal changes in the posterior  (predictive) means, along with 95\% credible and prediction intervals for the average income, median income, and the Gini index of the selected municipalities. 
The figure clearly shows that the average and median income levels in all  municipalities, with the exception of Oshino, have decreased over time. 
Notably, the Gini index for Minato has remained above $0.4$ and exhibits an increasing trend. 
Similarly, the Gini index for Oshino has shown a slight increase. 
For the remaining municipalities, the variation in the Gini index has diminished over the past twenty years, consistent with the nationwide trend. 
The figure also illustrates that the 95\% intervals for the income measures are substantially wider for the unobserved municipalities, particularly for Oguni, than for the observed municipalities. 
}

\begin{figure}[H]
    \centering
    \includegraphics[width=\textwidth]{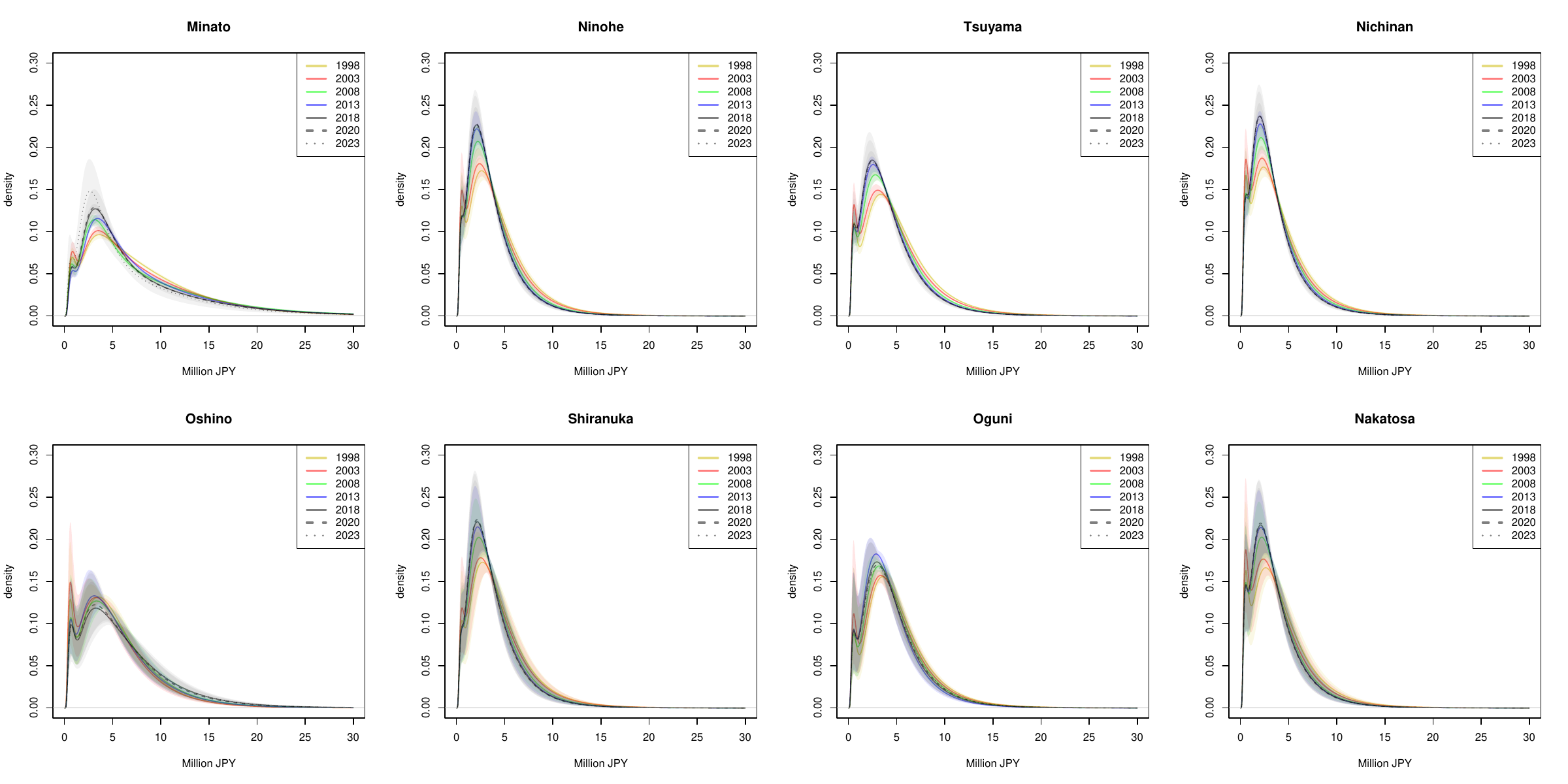}
    \caption{Posterior and posterior predictive means and 95\% credible and prediction intervals for the income distributions of the selected observed (top row) and unobserved (bottom row) municipalities. The shaded areas represent the 95\% intervals. }
    \label{fig:area_dist}
\end{figure}

\subsection{Prior sensitivity check}\label{sec:alt}
\begin{figure}[H]
    \centering
    \includegraphics[width=\textwidth]{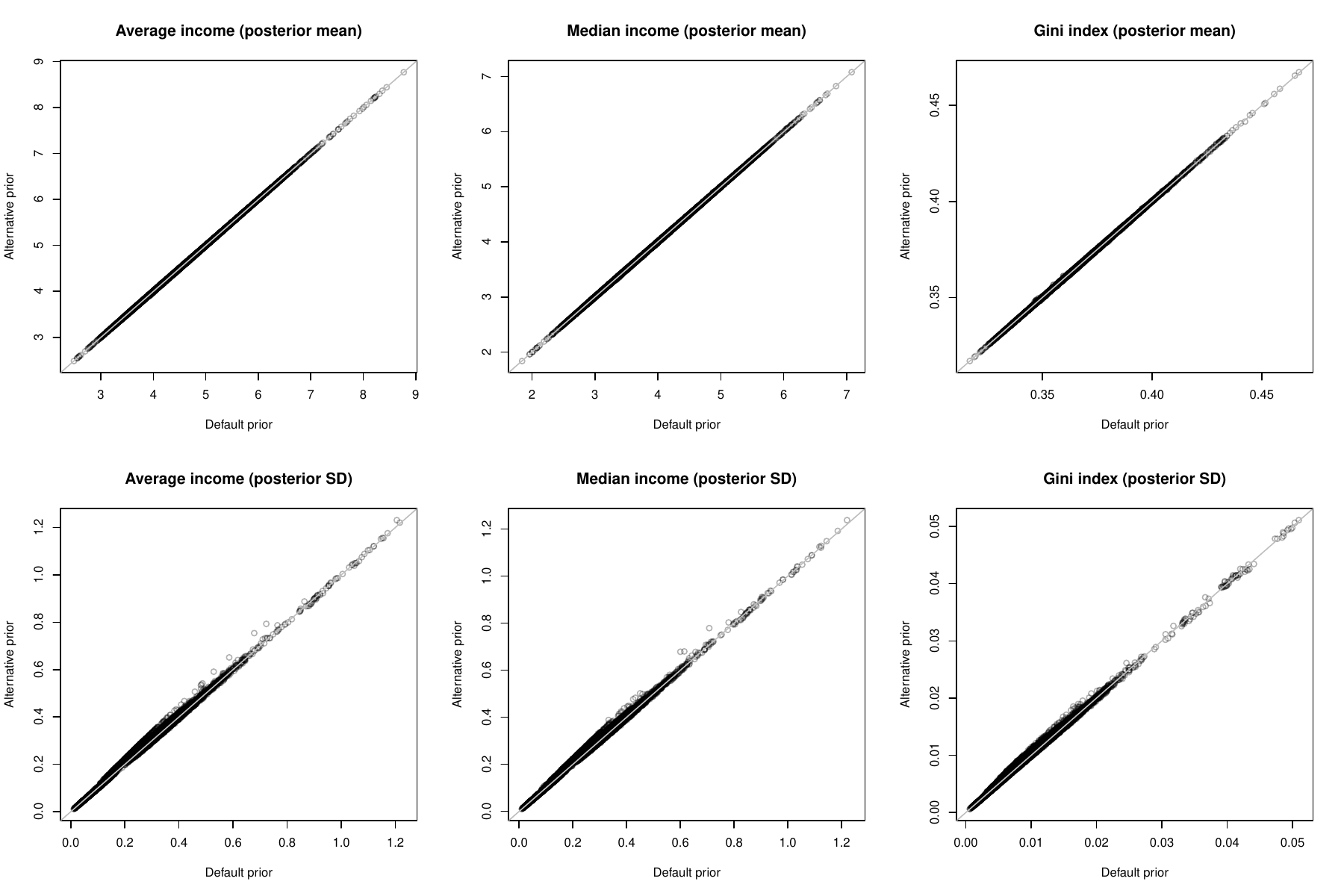}
    \caption{Scatter plots of the posterior means (top panels) and posterior standard deviations (bottom panels) for average income, median income and the Gini index across all municipalities, under the default and alternative prior settings.} 
    \label{fig:prior}
\end{figure}

Finally, we assess the sensitivity of our results to the prior specifications for the standard deviation parameters.  
In the alternative prior setting, we double the prior means as $a_\sigma=0.5$ and $a_\tau=a_\alpha=2.5$, thereby allowing for  standard deviation larger values. 
Figure~\ref{fig:prior} presents the scatter plots comparing the posterior means and standard deviations for average income, median income and the Gini index under the default and alternative exponential priors. 
The figure demonstrates that the posterior distributions of the quantities of interest are highly robust to the prior specifications, as the posterior means and standard deviations are closely aligned along the 45-degree line. 
Further comparisons of the posterior distributions of the standard deviation parameters under these different prior settings are provided in the Supplementary Material.

\section{Discussion}\label{sec:conc}
In this study, we proposed a novel spatio-temporal mixture model for analysing income distribution based on grouped data. 
By incorporating covariate information and accounting for spatio-temporal heterogeneous effects, the proposed model enables the inference and prediction of income distributions for both observed and unobserved areas at any given time point. 
Particularly, when an appropriate covariate is available at a higher frequency than the grouped income data itself, these predictions can provide significant insights into the latest income status. 
By applying the proposed method to the HLS data, we produced the complete maps of average income, median income, and the Gini index across Japan. 
Our findings revealed that the overall average and median income levels have decreased over the past twenty years, while the spatial variation in Gini indices has also diminished in the same period. 

Furthermore, we demonstrated that the inclusion of appropriate covariates is essential for accurately revealing the income status of specific areas. 
This is especially true for unobserved municipalities, where covariate information often serves as the primary source of information for inferring income distributions. 
In our analysis, we utilised the taxable income as a key covariate for predicting income distributions in the year 2020, in which the HLS was not conducted, owing to its annual availability for all municipalities. 
While it is possible to incorporate other sources to further enhance the model performance, such as the number of business sites or the proportions of workers across different industries, these large-scale surveys (e.g., the Economic Census or Population Census) are typically conducted only every few years. 
This periodicity limits their utility for high-frequency temporal prediction. 
Nevertheless, as incorporating diverse information sources would yield valuable economic implications, exploring methods to integrate such multi-frequency data remains a promising avenue for future research.

The proposed method can reveal the income distributions and associated poverty measures of any area at any given period, provided that relevant covariate information is available. 
It is also possible to estimate the proportion of the population in each area whose income falls below a specific threshold. 
Based on the results of the proposed method, policymakers can identify regions requiring targeted intervention to mitigate poverty and inequality across the nation, thereby facilitating efficient policy implementation and resource allocation.  
Such interventions can be informed by the most recent information on income distributions, as elucidated by the spatial interpolation and temporal prediction capabilities of the proposed model.

This study adopts a full MCMC approach to the posterior inference. 
A key advantage of the proposed Gibbs sampler is that it does not require algorithmic tuning.  
While the inclusion of many latent variables can lead to computational inefficiencies, we considered the necessity of alternative strategies. 
In our preliminary analysis, we attempted to implement a standard MH algorithm and its adaptive version to sample $\bbe_k$ and $\sigma^2_k$ directly, without augmenting individual household incomes. 
In this alternative approach, \eqref{multi} was directly used in the likelihood function. 
However, the MH algorithms failed to explore the parameter space effectively and often remained stuck near the initial values. 

{
Alternative high-speed computation methods, such as integrated nested Laplace approximation {\citep{rue2009approximate}} and  variational Bayes approximation {\citep{blei2017variational}}, remain appealing candidates for future work.  
In our application to the HLS data, the computing time for $K=3$ was approximately 24 hours. 
Given that our priority was to ensure the quality of posterior inference and that the computing time was not a prohibitive constraint in the present context, we consider our computational approach to be appropriate. 
Nevertheless, developing more efficient algorithms remains an important direction for further research.
}


\section*{Acknowledgements}
This work was supported by JSPS KAKENHI (\#24K00244, \#25H00546,  \#22K01421, \#21K01421, \#21H00699, \#20H00080 and \#18H03628) and Japan Center for Economic Research. 

\newpage
\appendix

\renewcommand{\thesection}{S.\arabic{section}}
\setcounter{section}{0}

\renewcommand{\theequation}{S.\arabic{equation}}
\renewcommand{\thefigure}{S.\arabic{figure}}
\renewcommand{\thetable}{S.\arabic{table}}

\setcounter{equation}{0}
\setcounter{figure}{0}
\setcounter{table}{0}

\section{Additional literature reviews}\label{sup:review}
\subsection{Income distribution estimation for a single area and period}
Existing methods (and the countries to which they have been applied) include the maximum-likelihood estimation for the flexible five-parameter beta distribution \citep{mcdonald95} (the United States), generalised method of moments estimation for grouped income data \citep{haja12, grif15} (China, India, Russia, Pakistan, Poland, and Brazil) and its generalisation \citep{chen18} (the United States and China), Bayesian estimation of parametric distributions for grouped income data \citep{kakamu, eckern21} (Japan, India, Peru, Ethiopia, and Iraq), maximum-likelihood and Bayesian estimations of a parametric Lorenz curve based on the Dirichlet likelihood \citep{choti02, choti05} (Sweden and Brazil) and finite mixtures of log-normal distributions for individual incomes \citep{flach,lubrano16} (the United Kingdom), grouped-data-based estimation of the parametric Lorenz curve via approximate Bayesian computation \citep{kobakakakmu} (Japan), and finite mixture of the beta distribution of the second kind \citep{choti07} (Hong Kong, Japan, Malaysia, Philippines, Singapore, Korea, Taiwan, and Thailand). 

\subsection{Small area estimation}
\cite{marhuenda2013small} explored the Fay--Herriot model incorporating a spatio-temporal correlation structure to analyse Italian income data. 
However, as the Fay--Herriot model is designed for area-level direct estimators, such as median income, it cannot be used to treat grouped data directly. 
\cite{kawakubo2019small} considered a version of the area-level linear mixed model, while \cite{walter} proposed a unit-level small area model for grouped data. 
These methods were applied to Japanese and Mexican data, respectively.
Although these approaches can predict general small area parameters in unobserved areas, they have neither spatial nor temporal considerations. 
Furthermore, as the method of \cite{walter} requires individual-level observations within the grouped categories, it cannot be applied to the HLS data. 
\cite{gardini} considered a finite mixture of log-normal distributions for a unit-level small area model to stably estimate income and poverty measures for Italy.

\section{Gibbs sampler for the proposed model}\label{sup:algo}
\noindent\paragraph*{Sampling step for the component allocation:}
To facilitate the sampling of the parameters for the component distributions, $\bbe_k$ and $\sigma^2_k$, several latent variables are introduced. 
First, we define a latent indicator of the mixture component representing the component to which the $j$-th household in the $g$-th group belongs.
This latent indicator is denoted by $h_{itgjk}$, which takes the value $1$ if the household belongs to the $k$-th component and $0$ otherwise, for $j=1,\dots,N_{itg}$ and $k=1,\dots,K$. 
Consequently, the likelihood contribution for the $g$-th group in  the $i$-th area at time $t$ is given by: 
\begin{eqnarray*}
\prod_{k=1}^K\prod_{j=1}^{N_{itg}}\left[\pi_{itk}\left\{\Phi_{LN}(Z_{tg};\x_{it}^t\bbe_k,\sigma^2_k)-\Phi_{LN}(Z_{t,g-1};\x_{it}^t\bbe_k,\sigma^2_k)\right\}\right]^{h_{itgjk}}
\propto\prod_{k=1}^K p_{itgk}^{\sum_{j=1}^{N_{itg}}h_{itgjk}}, 
\end{eqnarray*}
where $p_{itgk} \propto \pi_{itk}\left\{\Phi_{LN}(Z_{tg};\x_{it}^t\bbe_k,\sigma^2_k)-\Phi_{LN}(Z_{t,g-1};\x_{it}^t\bbe_k,\sigma^2_k)\right\}$.
The full conditional distribution of each $h_{itgjk}$ is the multinomial distribution $M(1,\p_{itg})$, where  $\p_{itg}=(p_{itg1},\dots,p_{itgK})^t$. 
Let $\s_{itg}=(s_{itg1},\dots,s_{itgK})^t$ denote a random vector where each element  $s_{itgk}=\sum_{j=1}^{N_{itg}}h_{itgjk}$ represents the unobserved number of households belonging to the $k$-th component of the mixture within the $g$-th income class. 
By definition, $\sum_{k=1}^Ks_{itgk}=N_{itg}$. 
Rather than sampling the individual mixture memberships, $h_{itgjk}$, for each $j$, we sample the aggregate $\s_{itg}$ for $i=1,\dots,m$, $t=1,\dots,T$, $g=1,\dots,G_t$. 
From the above expression, $\s_{itg}$ is sampled from $M(N_{itg}, \p_{itg})$. 

\noindent\paragraph*{Sampling steps for $\pi_{itk}$:}
To facilitate the sampling of the parameters and variables included in $\pi_{itk}$, P\'olya-gamma data augmentation \citep{polson2013bayesian} is utilised.
The full conditional distributions of $\mu_k$, $\u_k$ and $\bet_k$ are proportional to %
$$
\pi(\mu_k)\pi(\bet_k)\pi(\u_k)\prod_{i=1}^{m}\prod_{t=1}^T\frac{\exp(\mu_k+u_{ik}+\eta_{tk}-C_{itk})^{s_{itk}}}{\{1+\exp(\mu_k+u_{ik}+\eta_{tk}-C_{itk})\}^{N_{it}}},
$$
where $\pi(\mu_k)$, $\pi(\bet_k)$ and $\pi(\u_k)$ are the prior densities of $\mu_k$, $\bet_k=(\eta_{1k},\dots,\eta_{Tk})^t$ and $\u_k$, respectively, $C_{itk}=\log\{\sum_{\ell\neq k} \exp(\mu_\ell+ u_{i\ell} +\eta_{t\ell})\}$ and $s_{itk}=\sum_{g=1}^{G_t} s_{itgk}$.
This can be rewritten as follows: 
\begin{align*}
\pi(\mu_k)\pi(\bet_k)&\pi(\u_k)\prod_{i=1}^{m}\prod_{t=1}^T\exp\left\{(\mu_k+u_{ik}+\eta_{tk}-C_{itk})\left(s_{itk}-\frac{N_{it}}{2}\right)\right\}\\
&\times\int_{0}^{\infty}\exp\left\{-\frac{1}{2}(\mu_k+u_{ik}+\eta_{tk}-C_{itk})^2\omega_{itk}\right\}p_{PG}(\omega_{itk}; N_{it}, 0)d\omega_{itk},
\end{align*}
where $p_{PG}(\cdot; b,c)$ denotes the density function of the P\'olya-gamma distribution with the parameters $b$ and $c$, denoted by $PG(b,c)$.
As $\pi(\mu_k)$, $\pi(\bet_k)$ and $\pi(\u_k)$ are all Gaussian, their full conditional distributions are also Gaussian given the additional latent variables, $\omega_{itk}$. 
This part of the Gibbs sampler involves sampling $\omega_{itk}$, $\mu_k$, $\u_{k}$ and $\eta_{tk}$. 
\begin{itemize}
\item {Sampling $\omega_{itk}$:}
For $i=1,\dots,m$, $t=1,\dots,T$, $k=2,\dots,K$, $\omega_{itk}$ is drawn from $PG(N_{it}, \mu_k+u_{ik}+\eta_{tk}-C_{itk})$.

\item{Sampling $\u_k$:}
For $k=2,\dots,K$, the unrestricted $\u^*_k$ is sampled from $N(\m_{u,k},\V_{u,k})$, where
$$
\V_{u,k} = \left[\sum_{t=1}^T\bOmega_{tk}+\frac{1}{\tau_k^2}\V_{11k}^{-1}\right]^{-1},\quad
\m_{u,k} = \V_{u,k}\left[\sum_{t=1}^T\left\{\d_{tk} - \bOmega_{tk}((\mu_k+\eta_{tk})\biota_m - \C_{tk})\right\} \right],
$$
$\bOmega_{tk}=\diag(\omega_{1tk},\dots,\omega_{mtk})$, $\d_{tk}=(s_{1tk}-N_{1t}/2,\dots,s_{mtk}-N_{mt}/2)^t$, $\C_{tk}=(C_{1tk},\dots,C_{mtk})^t$ and $\biota_m$ is the $m$-dimensional vector of ones. 
To impose the sum-to-zero constraint to $\u_k$, we set $\u_k\leftarrow \u_k^*-\bar{u}_k^*\biota_m$ and $\mu_k\leftarrow \mu_k+\bar{u}_k^*$ where $\bar{u}_k^*=m^{-1}\sum_{i=1}^mu_{ik}^*$. 

\item{Sampling $\eta_{tk}$:}
For $t=1,\dots,T$, the unrestricted $\eta^*_{tk}$ is sampled from $N(m_{\eta,tk},V_{\eta,tk})$, where 
\begin{eqnarray*}
V_{\eta,tk}&=&\left\{
\begin{array}{ll}
\left(\sum_{i=1}^m\omega_{itk} + \frac{2}{\alpha_k^2}\right)^{-1}    &  \text{if}\quad t\neq T,\\
\left(\sum_{i=1}^m\omega_{itk} + \frac{1}{\alpha_k^2}\right)^{-1}    &  \text{if}\quad t= T
\end{array}
\right.\\
m_{\eta,tk}&=&V_{\eta,tk}\left[\sum_{i=1}^m \left\{d_{itk}-\omega_{itk}(\mu_k+u_{ik}-C_{itk})\right\}+\frac{e_{tk}}{\alpha_k^2}\right],\\
e_{tk}&=&\left\{
\begin{array}{ll}
\eta_{t-1,k}    &  \text{if}\quad t=T,\\
\eta_{t-1,k}+\eta_{t+1,k}    &  \text{otherwise}.
\end{array}
\right.\\
\end{eqnarray*}
Then, similar to $\u_k$, we set $\eta_{tk}\leftarrow \eta_{tk}^*-\bar{\eta}_k^*$ and $\mu_k\leftarrow \mu_k+\bar{\eta}_k^*$ where $\bar{\eta}_k^*=T^{-1}\sum_{t=1}^T \eta_{tk}^*$.

\item
{Sampling $\mu_k$:}
For $k=2,\dots,K$, $\mu_k$ is drawn from $N(m_{\mu,k},V_{\mu,k})$, where
\[
V_{\mu,k} = \left(\sum_{i=1}^m\sum_{t=1}^T\omega_{itk}+1/c_\mu\right)^{-1}, \quad 
m_{\mu,k}=V_{\mu,k}\left[\sum_{i=1}^m\sum_{t=1}^T\left\{d_{itk}-\omega_{itk}\left((u_{ik}+\eta_{tk})-C_{itk}\right)\right\}\right]. 
\]

\item
{Sampling $\tau_k^2$:}
Since $\tau_k\sim Exp(a_\tau)$, $\pi(\tau_k^2)\propto \tau_k^{-1}\exp\left\{-a_\tau\tau_k\right\}$.
The full conditional distribution of $\tau_k^2$ is given by
\[
\begin{split}
\pi(\tau_k^2|-)
&\propto \left(\frac{1}{\tau_k^2}\right)^{m/2}\exp\left\{-\frac{1}{2\tau_k^2}\u_k^t\V_{11k}^{-1}\u_k\right\}\left(\frac{1}{\tau_k^2}\right)^{1/2}\exp\left\{-a_\tau \tau_k\right\}\\
&\propto \left(\frac{1}{\tau_k^2}\right)^{\frac{m-1}{2}+1}\exp\left\{-\frac{1}{2\tau_k^2}\u_k^t\V_{11k}^{-1}\u_k\right\}\exp\left\{-a_\tau \tau_k\right\}. 
\end{split}
\]
We draw $\tau_k^2$ using the independent Metropolis-Hastings (MH) algorithm with the proposal distribution $IG((m-1)/2,\u_k^t\V_{11k}^{-1}\u_k/2)$. 
The acceptance probability is given by $\min\left\{1,\exp\left\{-a_\tau\left(\tau_k^{*}-\tau_k\right)\right\}\right\}$ where $\tau_k^*$ is the square root of the proposal from the inverse gamma distribution.

\item
{Sampling $\alpha_k^2$:}
For $k=2,\dots,K$, the full conditional distribution of $\alpha_k^2$ is given by
\[
\begin{split}
    \pi(\alpha_k^2|-)&\propto \left(\frac{1}{\alpha_k^2}\right)^{T/2}\left\{-\frac{1}{2\alpha_k^2}\sum_{t=1}^T(\eta_{tk}-\eta_{t-1,k})^2\right\}
    \left(\frac{1}{\alpha_k^2}\right)^{1/2}\exp\left\{-a_\alpha \alpha_k\right\}\\
    &\propto\left(\frac{1}{\alpha_k^2}\right)^{\frac{T-1}{2}+1}\left\{-\frac{1}{2\alpha_k^2}\sum_{t=1}^T(\eta_{tk}-\eta_{t-1,k})^2\right\}\exp\left\{-a_\alpha \alpha_k\right\}. 
\end{split}
\]
As in the case of $\tau^2_k$, $\alpha_k^2$ is drawn using the independent MH algorithm with the proposal distribution given by $IG((T-1)/2,\sum_{t=1}^T(\eta_{tk}-\eta_{t-1,k})^2/2)$. 
The acceptance probability given by 
 $\min\left\{1, \exp\left\{-a_\alpha\left(\alpha_k^{*}-\alpha_k\right)\right\}\right\}$, where $\alpha_k^{*}$ is the square root of the proposal from the inverse gamma.

\item
{Sampling $\eta_{0k}$:}
For $k=2,\ldots,K$, $\eta_{0k}$ is drawn from $N(V_{\eta_0,k}\eta_{1k}/\alpha_k^2,V_{\eta_0,k})$, where $V_{\eta_0,k}=(1/\alpha_k^2+1/c_\eta)^{-1}$.

\item
{Sampling $\rho_{k}$:}
To sample $\rho_k$, $k=2,\dots,K$, we use a griddy Gibbs sampler to avoid the need for Metropolis--Hastings updates. 
The density of the full conditional distribution of $\rho_k$ is proportional to 
$|\V_{11k}(\rho_k)|^{-1/2}\exp\left(-\frac{1}{2\tau^2_k} \u_k^t\V_{11k}^{-1}(\rho_k)\u_k\right)$, where the dependence of $\V_{11k}$ on $\rho_k$ is explicitly denoted. 
The full conditional density is evaluated at each grid point of a fine grid $(r_1,\dots,r_R)$ on the interval $(0,1)$, from which a grid point is selected with probability proportional to the full conditional density. 
Since $\rho_k$ is the only parameter involved in $\V_{11k}$, $\V_{11k}^{-1}$ and its log-determinant at each grid point can be pre-computed and stored before commencing the Gibbs sampler; they do not need to be recomputed at each iteration. 
In this paper, we use a grid of 99 points $(0.01,0.02,\dots,0.99)$. 

Notably, $\V_{22k}-\V_{21k}\V_{11k}^{-1}\V_{12k}$ and $\V_{21k}\V_{11k}^{-1}$, which are used in spatial interpolation, can also be pre-computed on the grid, $(r_1,\dots,r_R)$ used for sampling $\rho_k$. 

\end{itemize}

\noindent\paragraph*{Sampling steps for $\bbe_k$ and $\sigma^2_k$:}
The sampling steps for the parameters of the component distributions are described below. 
To complete the Gibbs sampler, we simulate latent household incomes given $s_{itgk}$ for $k=1,\dots,K$. 
Recall that $s_{itgk}$ represents the number of households belonging to the $k$-th component of the mixture and the income class, $[Z_{t,g-1},Z_{t,g})$. 
For $i=1,\dots,m$, $t=1,\dots,T$, and $g=1,\dots,G_t$, we obtain $s_{itgk}$ draws from the normal distribution with mean $\x_{it}^t\bbe_k$ and variance $\sigma_k^2$ truncated on $[\log Z_{tg-1},\log Z_{t,g})$. 
Let $\Y_{itk}$ denote the vector obtained by stacking these draws, and let $\X_{itk}$ denote the $s_{itgk}\times p$ matrix with each row equal to $\x_{it}^t$. 
Furthermore, let $\Y_k$ denote the $s_k\times 1$ vector, where $s_k=\sum_{i,t,g}s_{itgk}$, obtained by stacking $\Y_{itk}$ for $i=1,\dots,m$ and $t=1,\dots,T$. 
Similarly, let $\X_k$ denote the $s_k\times p$ matrix obtained by stacking $\X_{itk}$ in the same order. 
Then given $\Y_{k}$, the simple sampling steps for $\bbe_k$ and $\sigma^2_k$ proceed as follows. 
\begin{itemize}
\item{Sampling $\bbe_k$:}
The full conditional distribution of $\bbe_k$ is $N(\V_{\beta_k}\X_k^t\Y_k/\sigma^2_k, \V_{\beta_k})$ where  $\V_{\beta_k}=(\X_k^t\X_k/\sigma^2_k+\I_p/c_\beta )^{-1}$.

\item{Sampling $\sigma^2_k$:}
The full conditional distribution of $\sigma_k^2$ is given by
\[
\begin{split}    
\pi(\sigma_k^2|-)&\propto \left(\frac{1}{\sigma^2_k}\right)^{s_k/2}\left\{-\frac{1}{2\sigma_k^2} (\Y_k-\X_k\bbe_k)^t(\Y_k-\X_k\bbe_k)\right\}
\left(\frac{1}{\sigma_k^2}\right)^{1/2}\exp\left\{-a_\sigma \sigma_k^2\right\}\\
&\propto \left(\frac{1}{\sigma^2_k}\right)^{\frac{s_k-1}{2}+1}\left\{-\frac{1}{2\sigma_k^2} (\Y_k-\X_k\bbe_k)^t(\Y_k-\X_k\bbe_k)\right\}
\exp\left\{-a_\sigma \sigma_k^2\right\}
\end{split}
\]
where $\pi(\sigma_k^2)$ is the prior density. 
As in the cases of $\tau^2_k$ and $\alpha^2_k$, $\sigma_k^2$ is sampled using the independent MH algorithm with the proposal distribution given by $IG(\frac{s_k-1}{2},(\Y_k-\X_k\bbe_k)^t(\Y_k-\X_k\bbe_k))$ and the acceptance probability given by $\min\left\{1,\exp\left\{-a_\sigma\left(\sigma_k^*-\sigma_k\right)\right\}\right\}$, where $\sigma_k^*$ is the square root of the proposal from the inverse gamma.

\end{itemize}

\section{Additional figures for the HLS data application}\label{sup:additional}

\subsection{Tax data}
\begin{figure}[H]
    \centering
    \begin{tabular}{ccc}
    \includegraphics[scale=0.18]{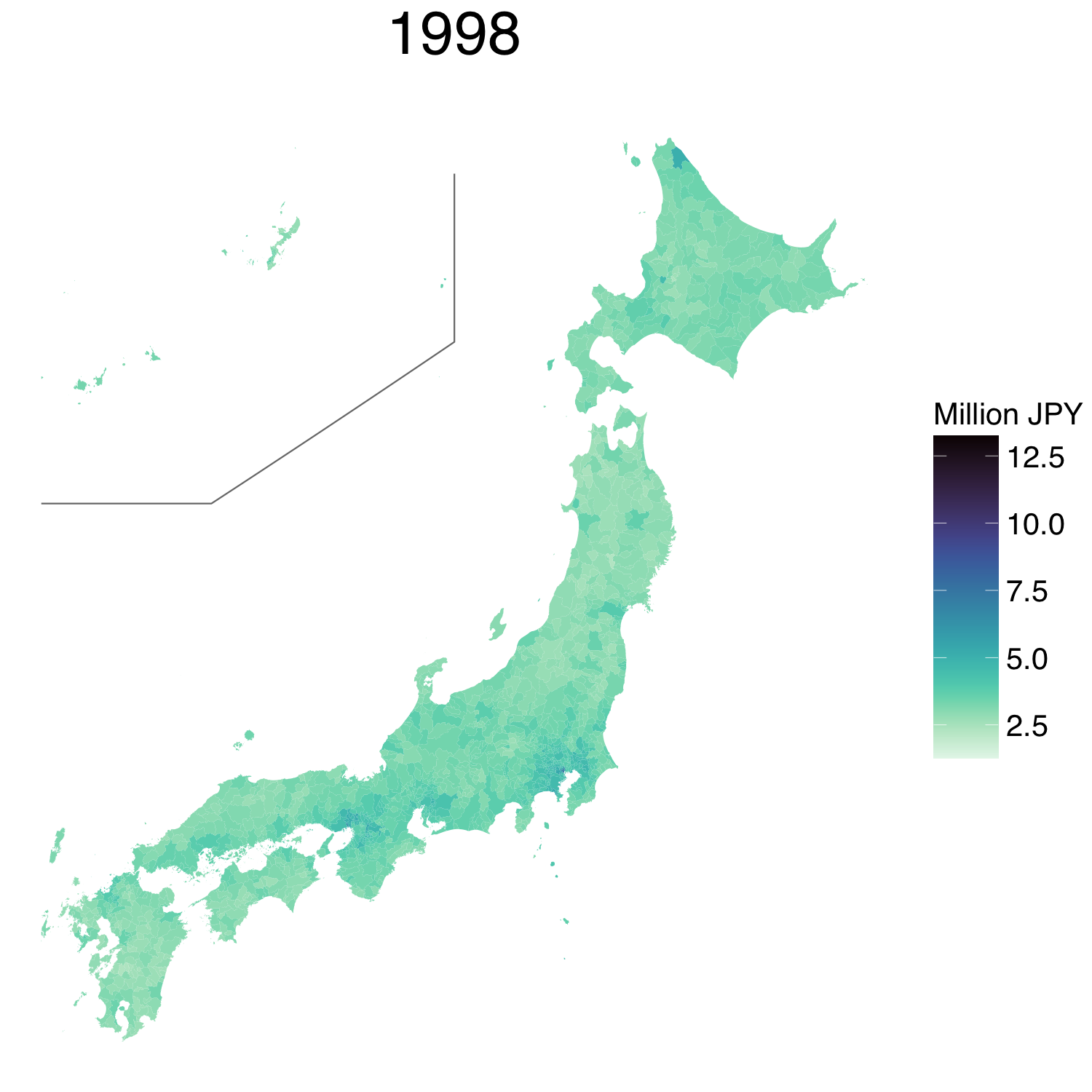}&
    \includegraphics[scale=0.18]{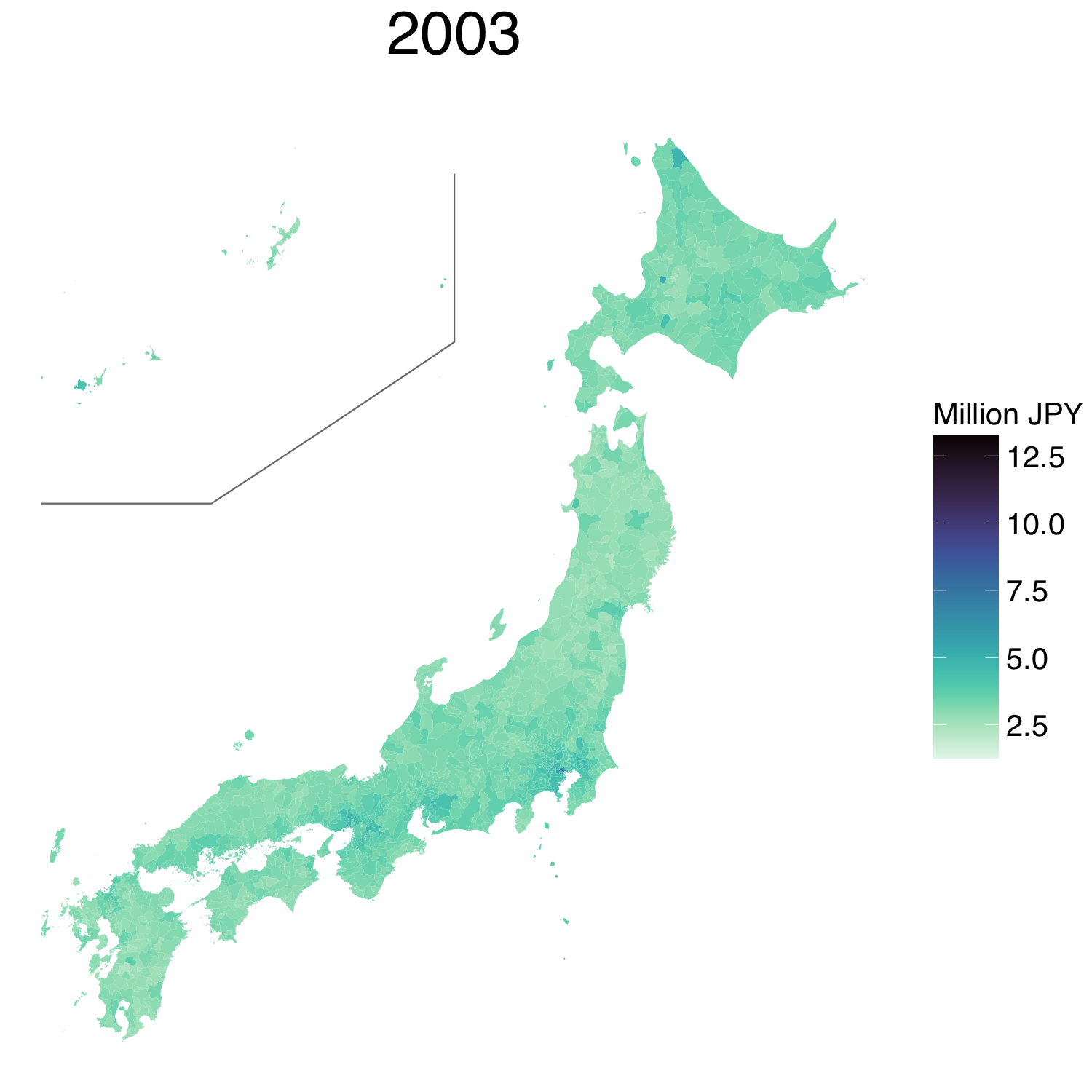}&
    \includegraphics[scale=0.18]{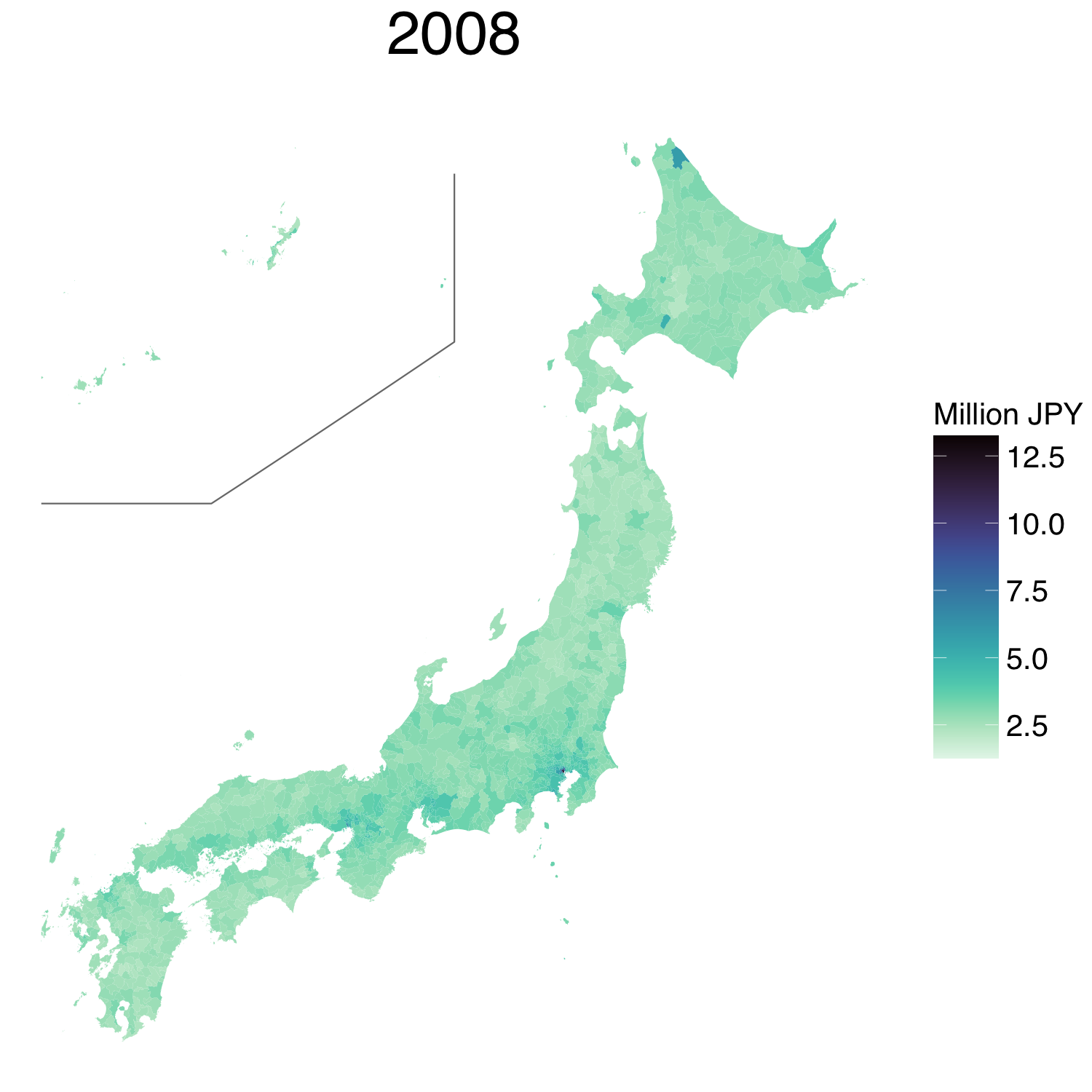}\\
    \includegraphics[scale=0.18]{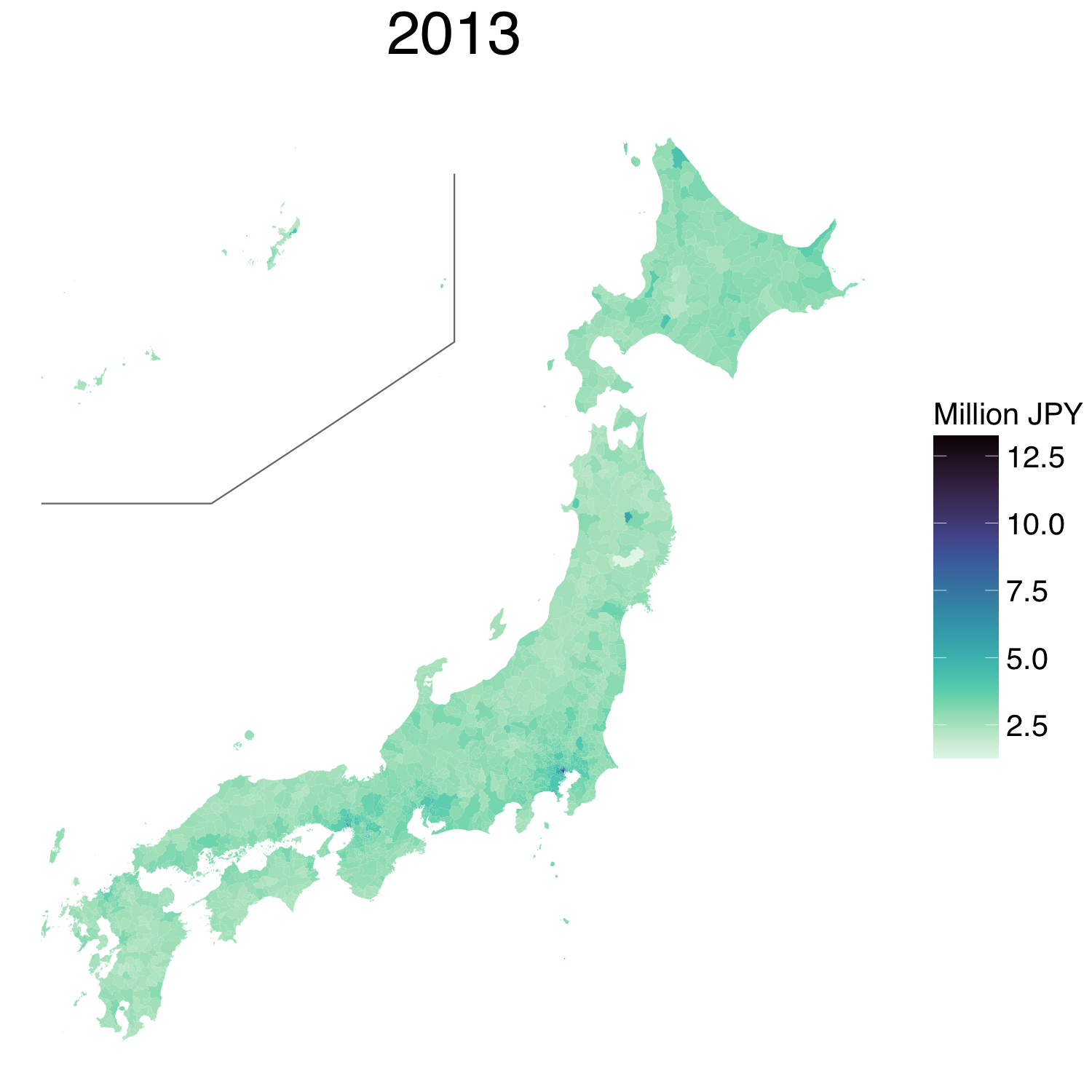}&
    \includegraphics[scale=0.18]{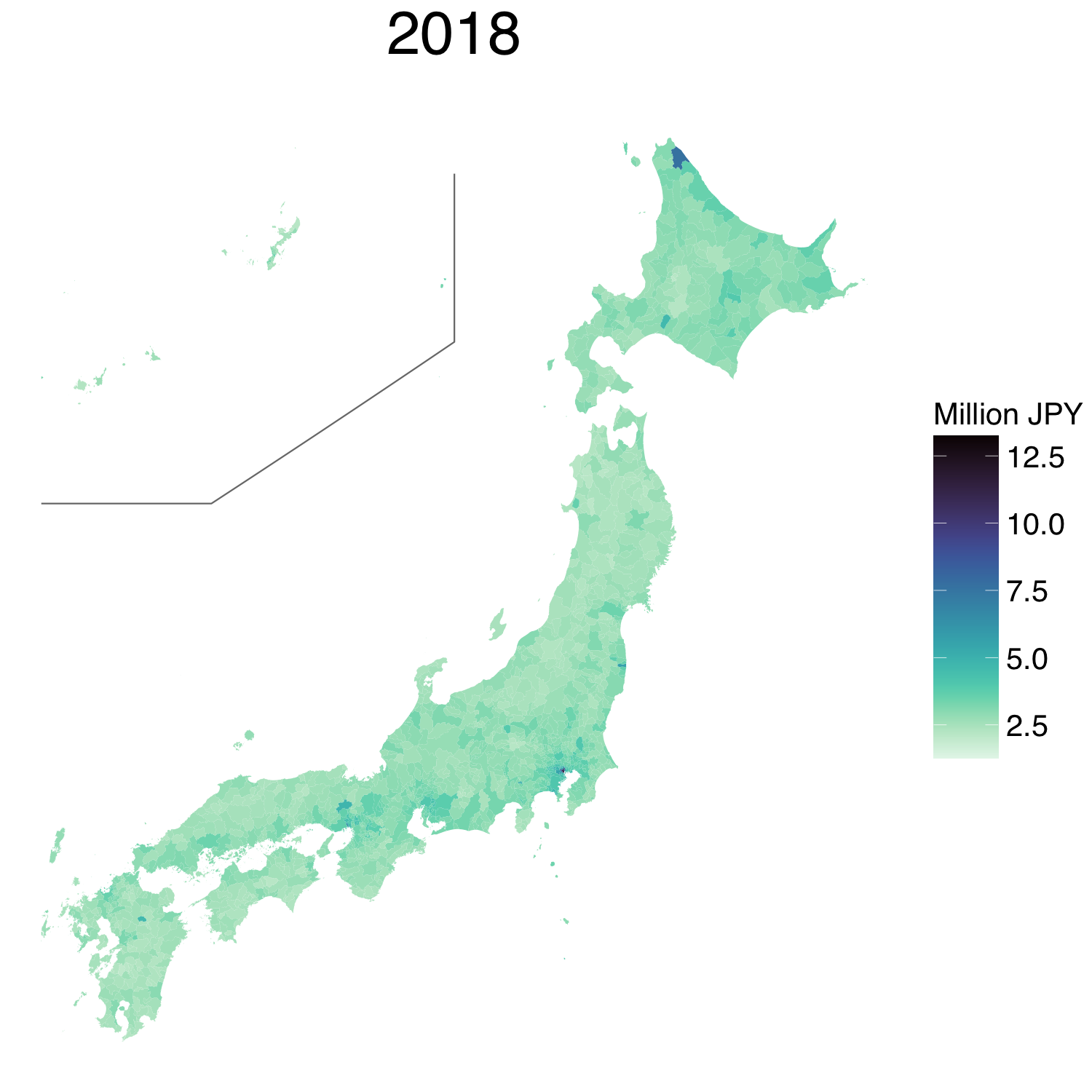}&
    \includegraphics[scale=0.18]{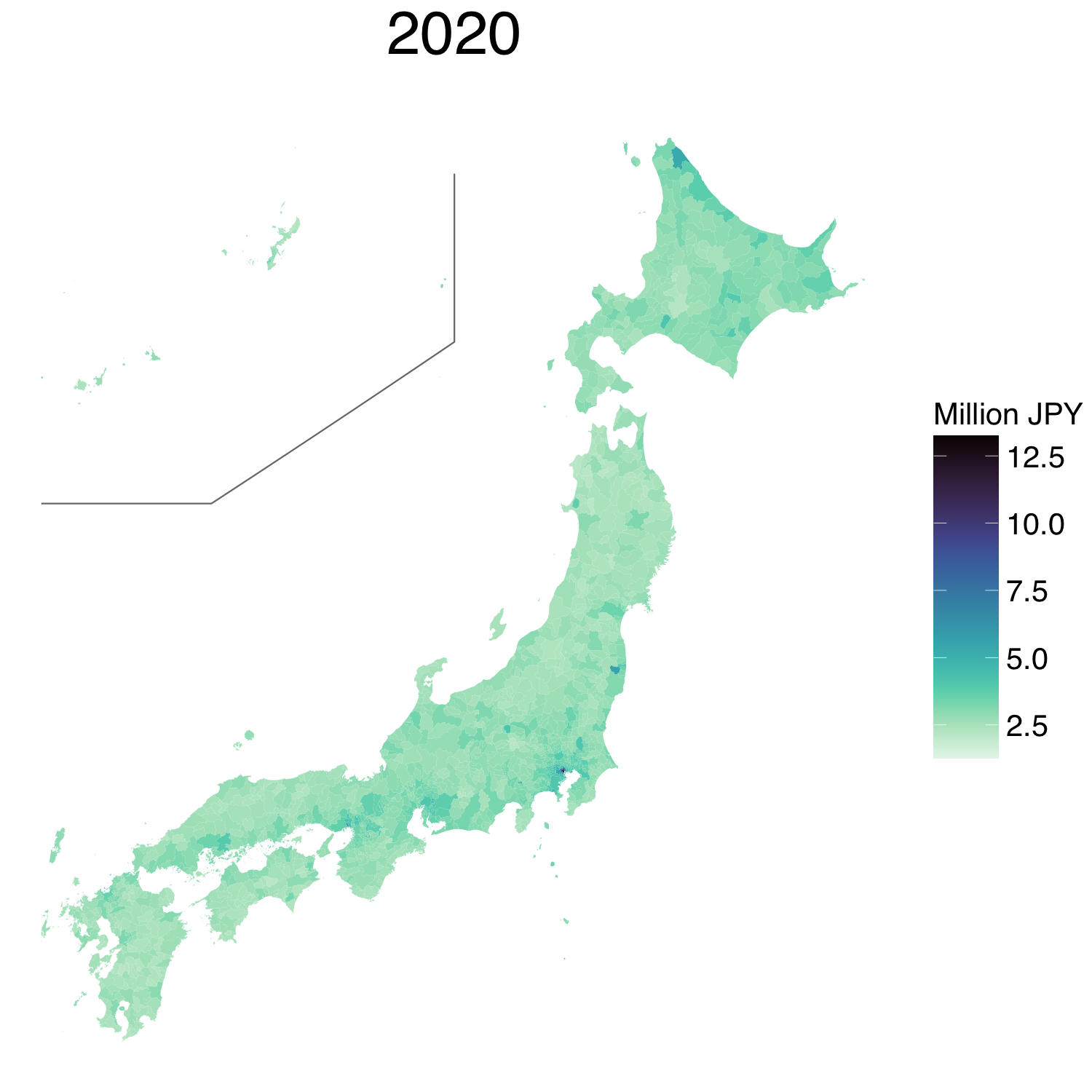}
    \end{tabular}
    \caption{Taxable income per taxpayer}
    \label{fig:tax}
\end{figure}

Figure~\ref{fig:tax} illustrates the spatial distributions of taxable income per taxpayer from 1998 to 2020. 
We observe that municipalities in metropolitan areas tend to exhibit higher taxable income levels, as indicated by the darker shading. 
Additionally, certain municipalities in rural areas report remarkably higher taxable income levels compared to their neighbouring regions.

\subsection{Parameter estimates}\label{sup:para}
\begin{figure}[H]
    \centering
    \includegraphics[width=\textwidth]{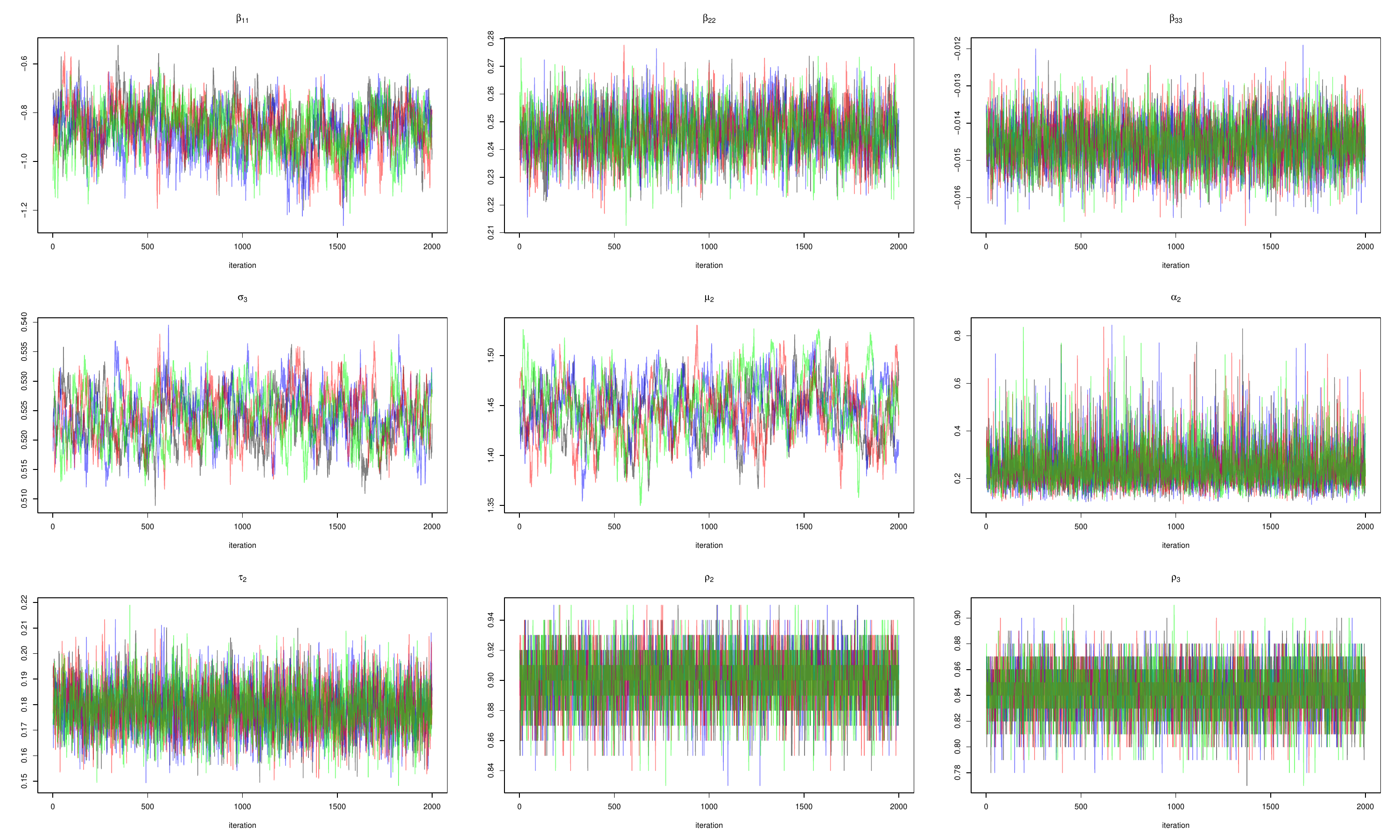}
    \caption{Trace plots of the four Markov chains initiated from different starting values.}
    \label{fig:trace}
\end{figure}

\begin{figure}[H]
    \centering
    $k=1$
    \includegraphics[width=\textwidth]{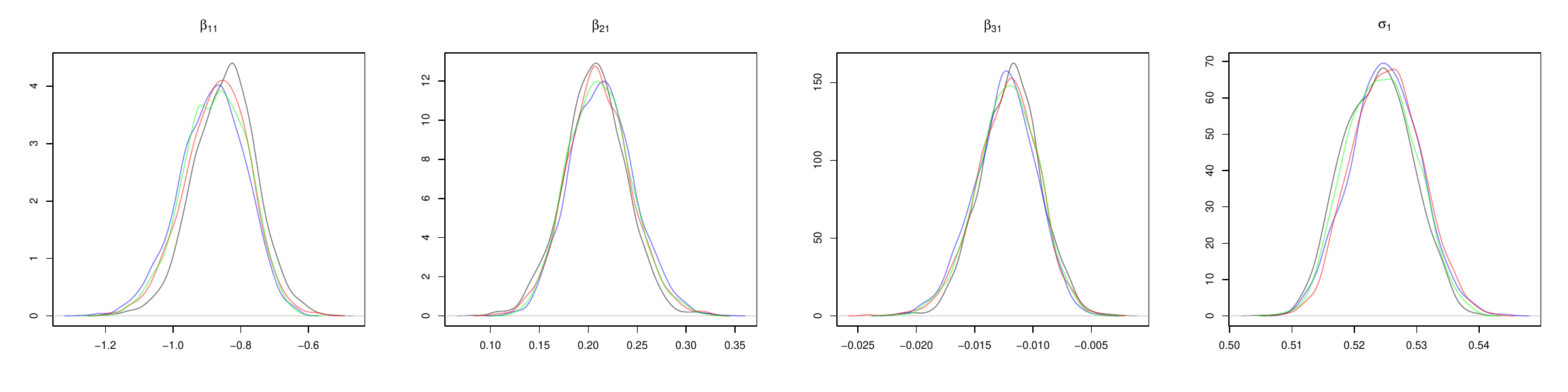}\\
    $k=2$
    \includegraphics[width=\textwidth]{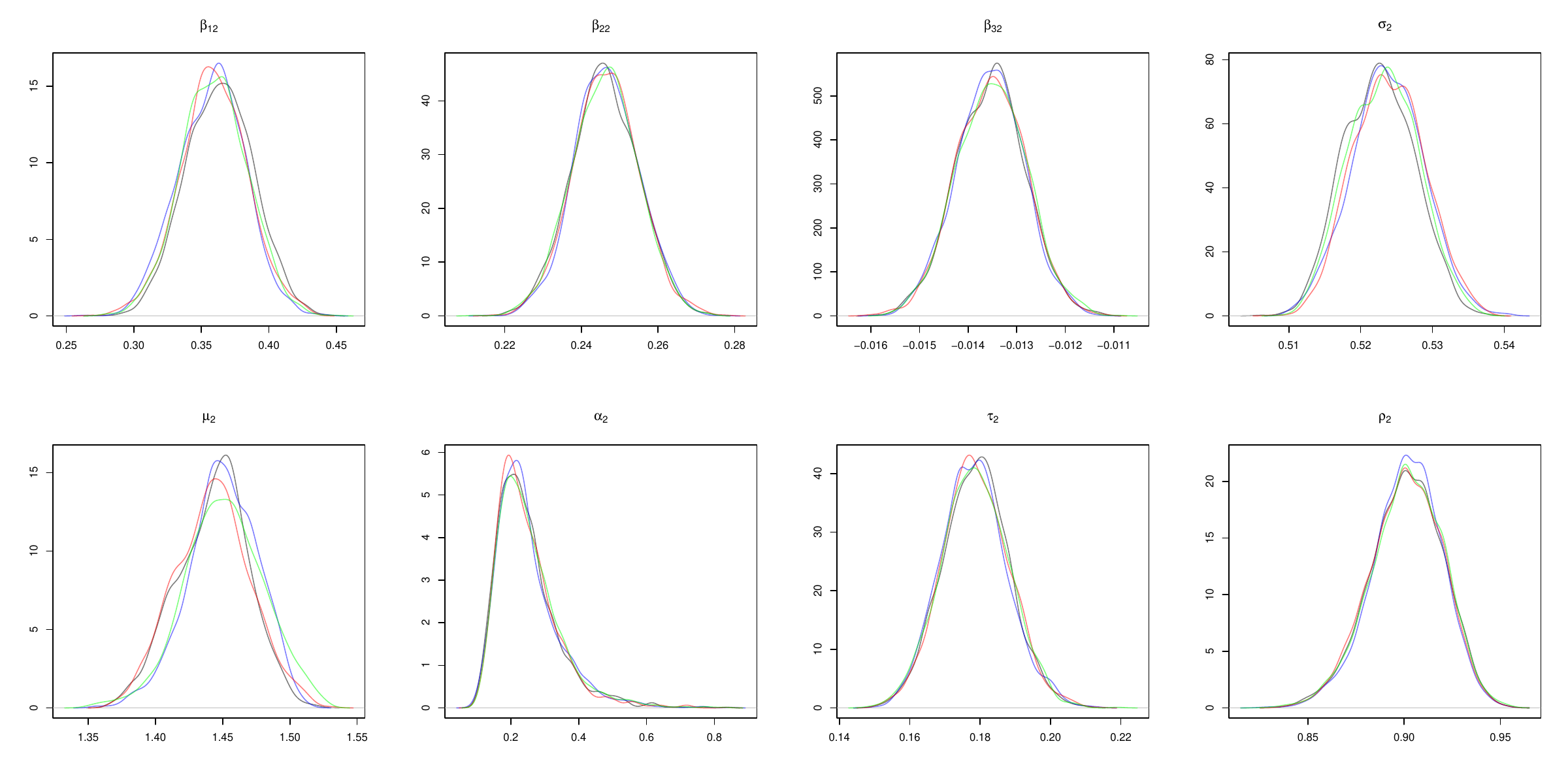}\\
    $k=3$
    \includegraphics[width=\textwidth]{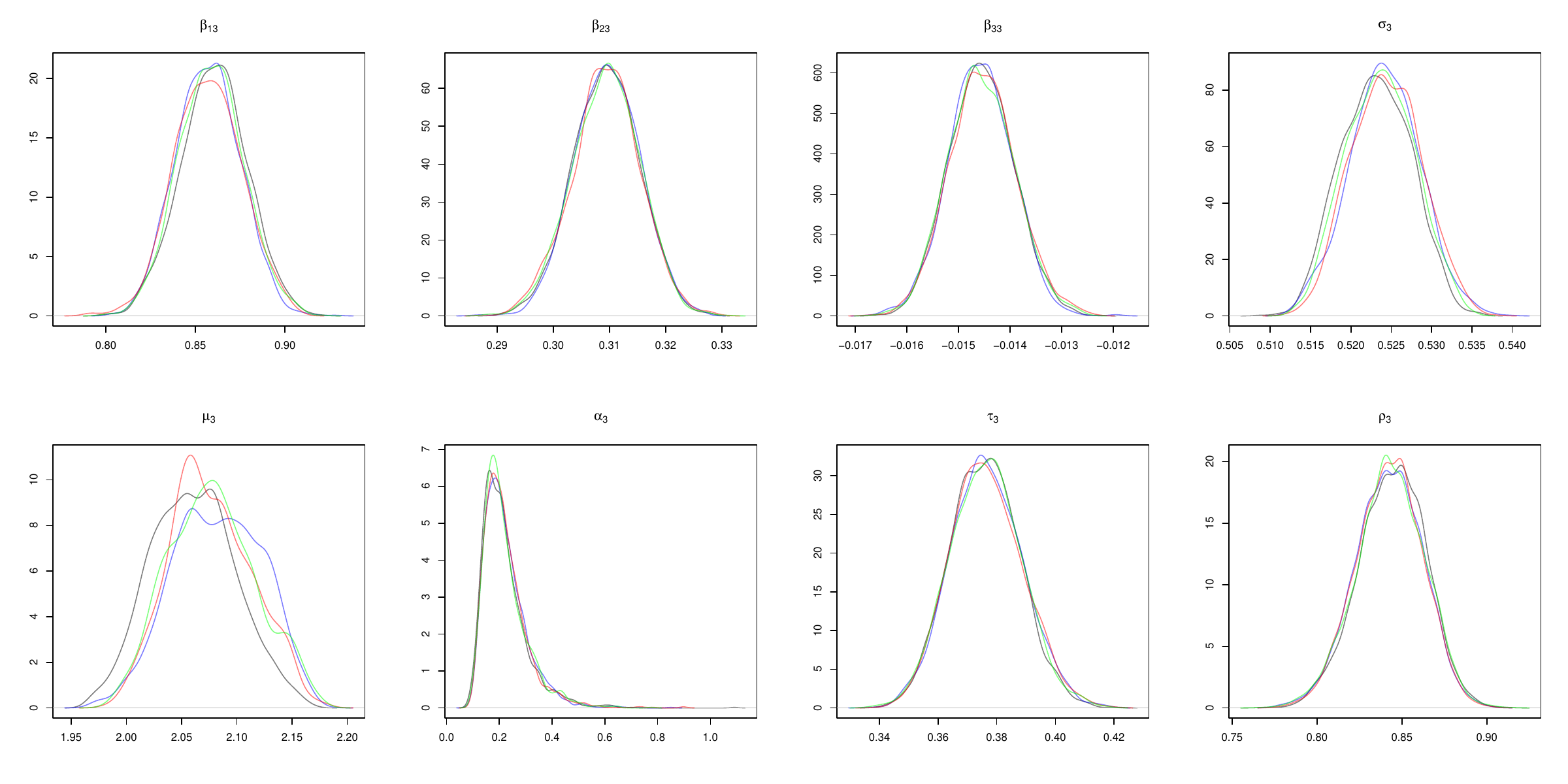}
    \caption{Posterior distributions of the parameters from the four Markov chains initiated from different starting values.}
    \label{fig:para}
\end{figure}

Figure~\ref{fig:trace} presents the trace plots of the four parallel MCMC chains for selected parameters of the mixture model with $K=3$. 
Figure~\ref{fig:para} displays the posterior distributions of the parameters for each component. 
Each chain was initiated from randomly chosen starting values and  run for 40,000 iterations with the initial burn-in period of 10,000 iterations. 
Every 15th draw was retained, yielding a total of 8,000 pooled smaples from the four chains for posterior inference. 
To facilitate the comparison and combination of the MCMC draws from the independent chains, we imposed an order constraint on the $\beta_{1k}$ values. 
This is justified by the fact that their posterior distributions are well-separated, as illustrated in Figure~\ref{fig:para}.  
The figures demonstrate that the chains starting from different initial values converged to the same regions, with the posterior distributions showing substantial overlap.

The posterior distributions of $\sigma_k$ are almost identical, with each concentrated around a posterior mean of $0.525$. 
Since the Gini index of the log-normal distribution is given by $2\Phi(\sigma_k/\sqrt{2})-1$, the Gini index of each component distribution is approximately $0.290$. 
It should be noted that the Gini index of a mixture model is not a convex combination of the Gini indices of its individual components. 
Nevertheless, we consider these values to be reasonable. 
In general, sampling scale parameters tend to be less efficient than sampling location parameters, and this is particularly evident in the case of $\sigma_k$. 

The Gibbs sampler for the proposed model includes many latent variables, which may lead to certain computational inefficiencies. 
We also attempted to implement the Metropolis--Hastings (MH) algorithm for $\bbe_k$ and $\sigma_k$ with a fixed or adaptive proposal distribution without the use of data augmentation.   
However, in the context of our mixture model for the grouped data, the MH algorithm failed to explore the parameter space effectively and remained stuck near the starting values.  
Therefore, the proposed Gibbs sampler is considered a robust and reasonable MCMC approach, particularly as it does not require algorithmic tuning often associated with MH schemes.

Figure~\ref{fig:para} also demonstrates that the posterior distributions of $\rho_2$ and $\rho_3$ are primarily concentrated above $0.8$ and $0.85$, respectively. 
These high values indicate that the spatial effects are strongly correlated across the municipalities.

\begin{figure}[H]
    \centering
    \includegraphics[width=\textwidth]{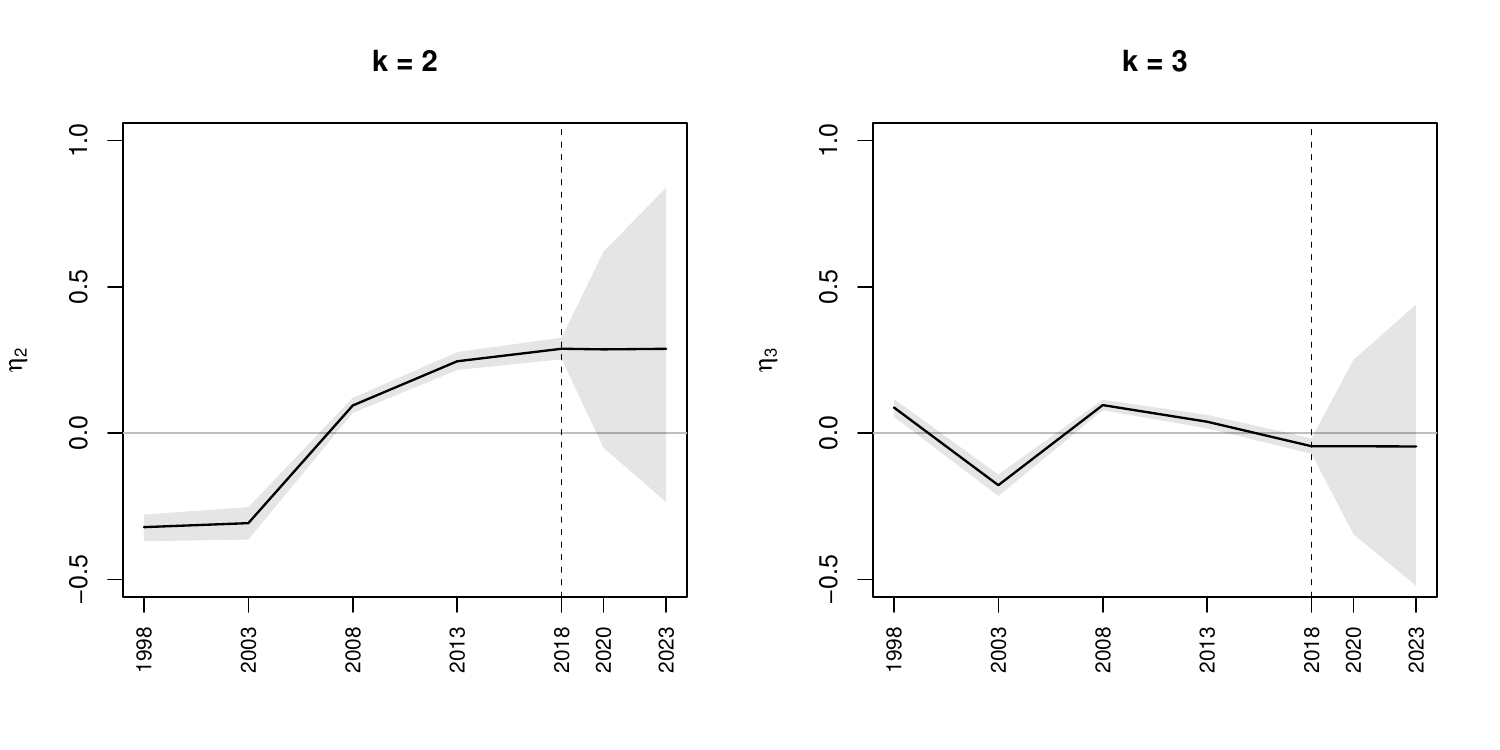}
    \caption{Posterior means, and 95\% intervals for $\eta_{kt}$}
    \label{fig:eta}
\end{figure}

Figure~\ref{fig:eta} presents the posterior means and 95\% intervals for $\eta_{kt}$ for $k=2$ and $3$. 
It is evident that $\eta_{2t}$ exhibits a consistent upward trend over time.

\begin{figure}[H]
    \centering
    \includegraphics[width=\linewidth]{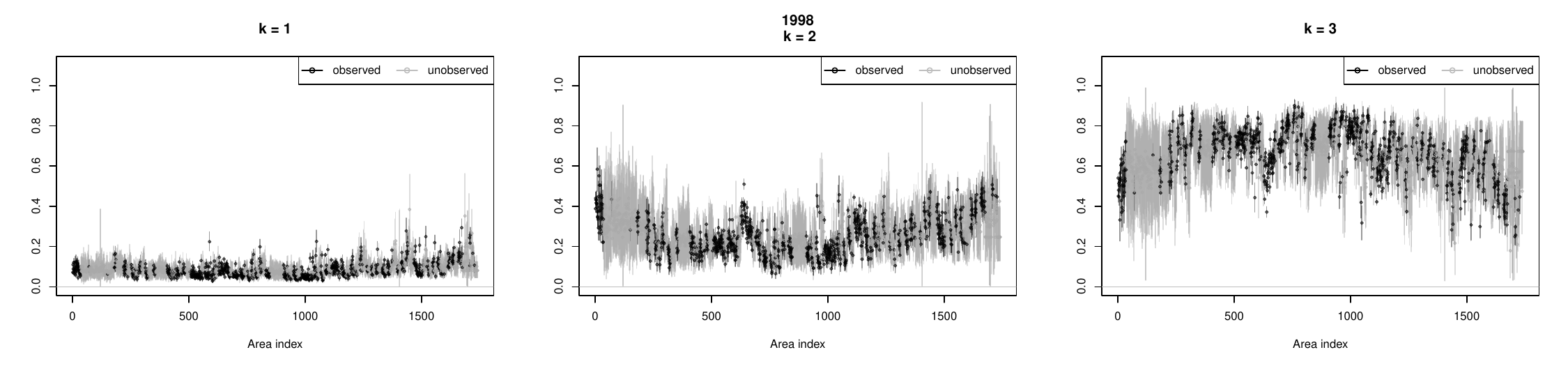}
    \includegraphics[width=\linewidth]{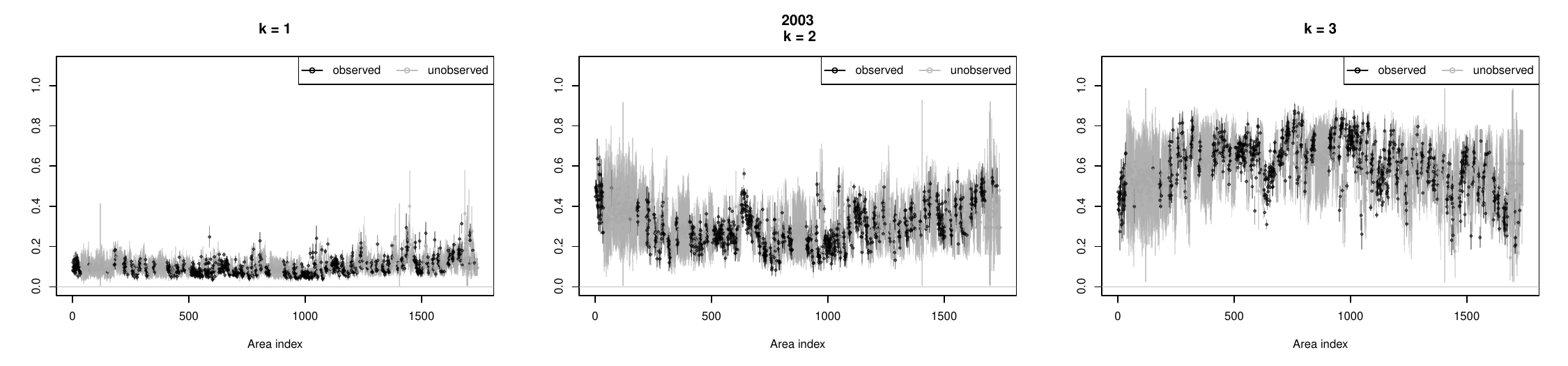}
    \includegraphics[width=\linewidth]{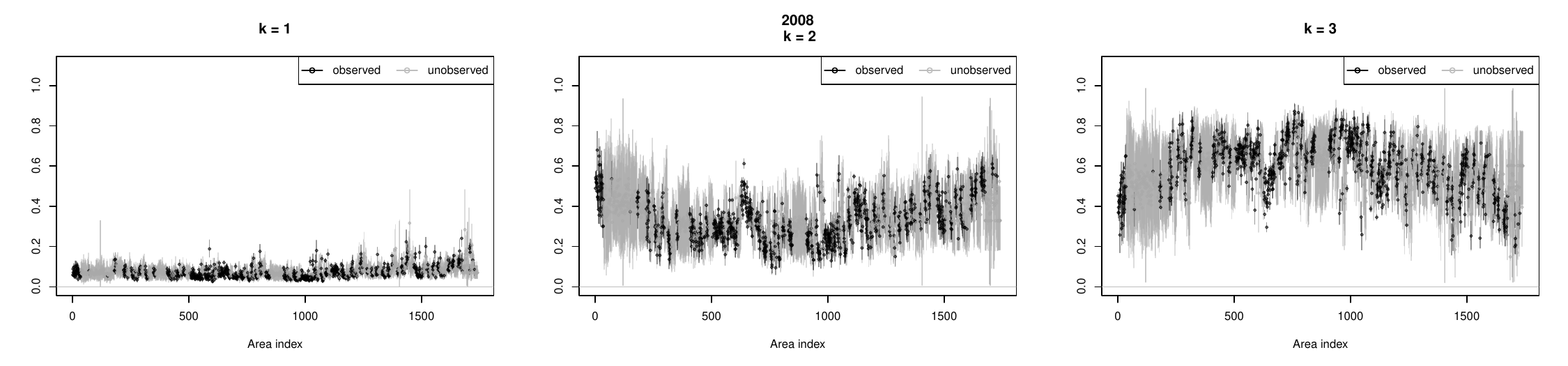}
    \includegraphics[width=\linewidth]{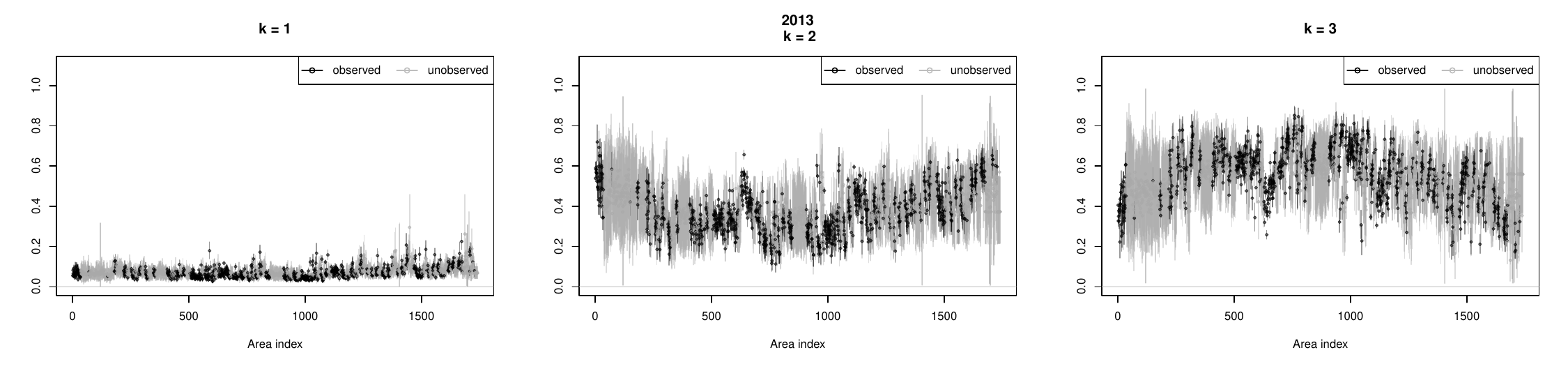}
    \includegraphics[width=\linewidth]{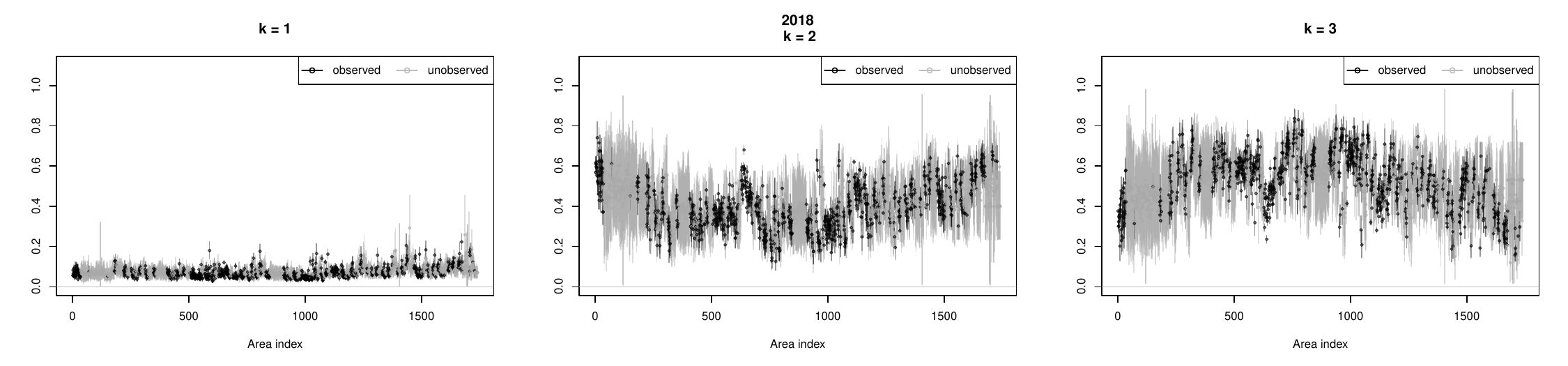}
    \includegraphics[width=\linewidth]{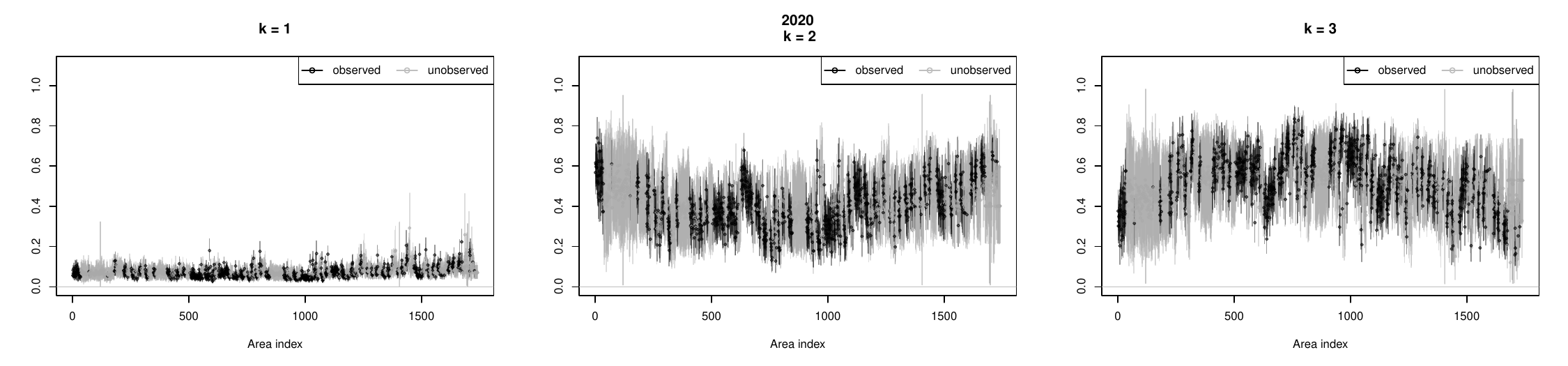}
    \caption{{Posterior and posterior predictive means (points), along with 95\% credible and prediction intervals (line segments), for the mixing proportions for the observed (black) and the unobserved (grey) municipalities.}}
\label{fig:pi_ci}
\end{figure}

\subsection{Maps of the median incomes}
\begin{figure}[H]
    \centering
    \begin{tabular}{ccc}
    \includegraphics[scale=0.18]{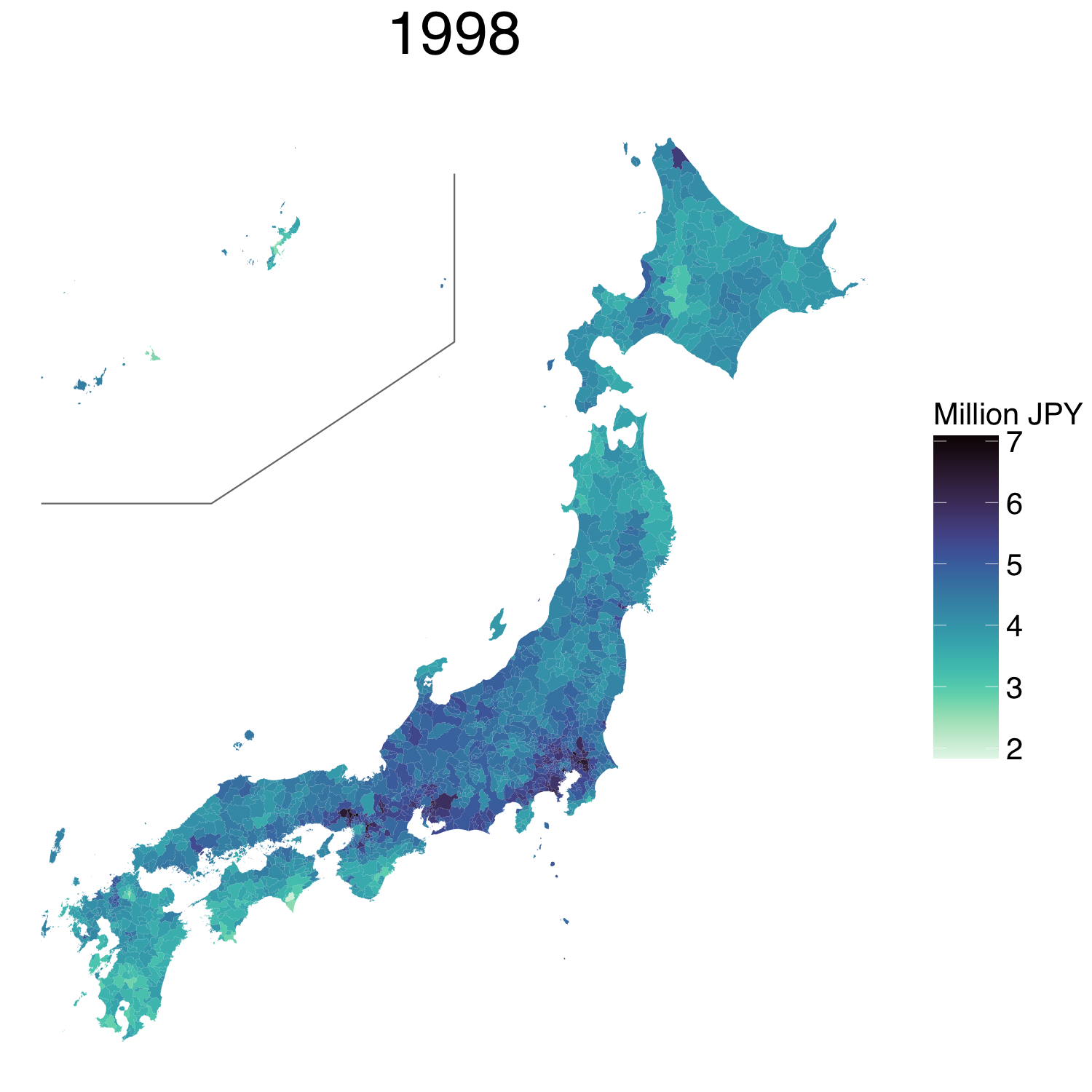}&
    \includegraphics[scale=0.18]{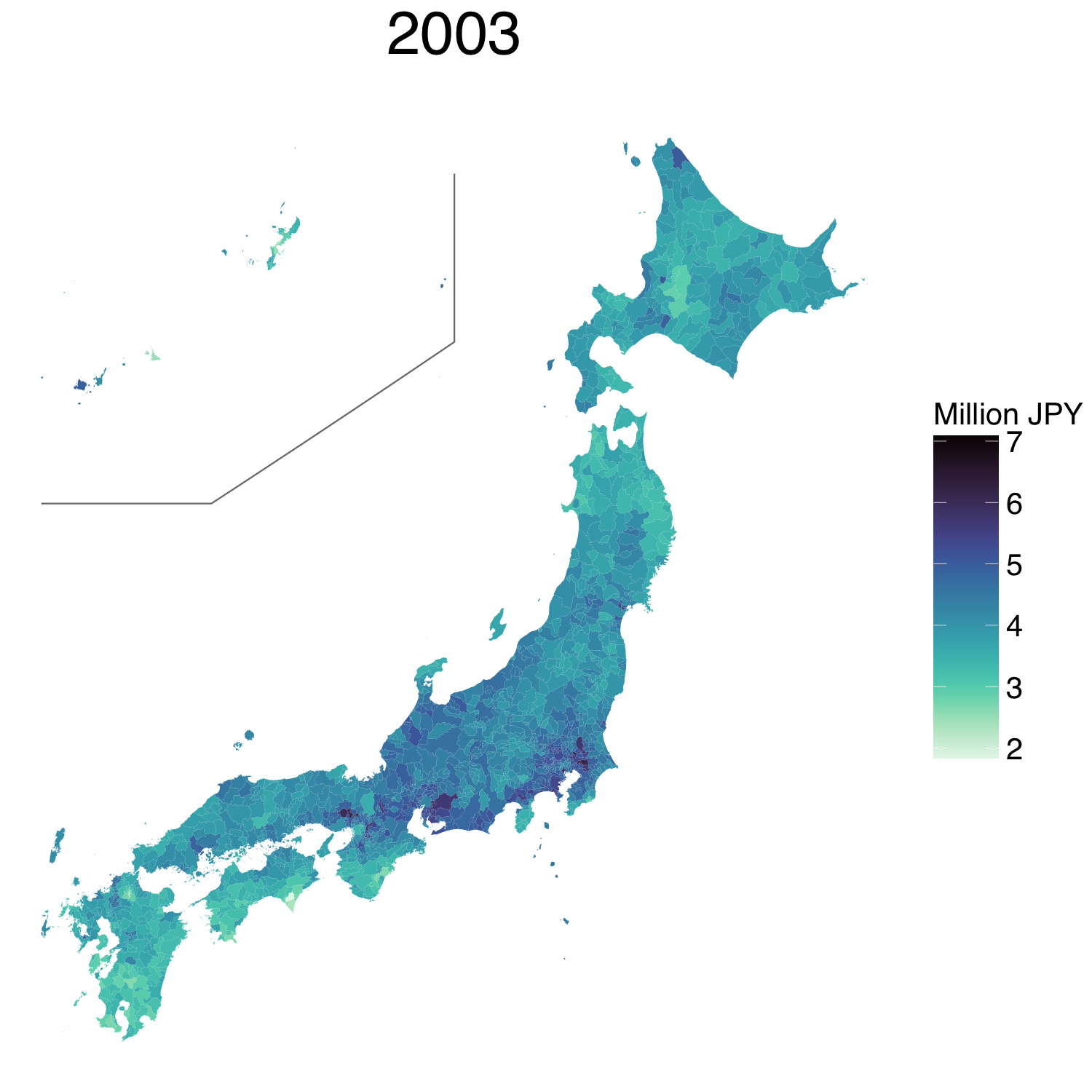}&
    \includegraphics[scale=0.18]{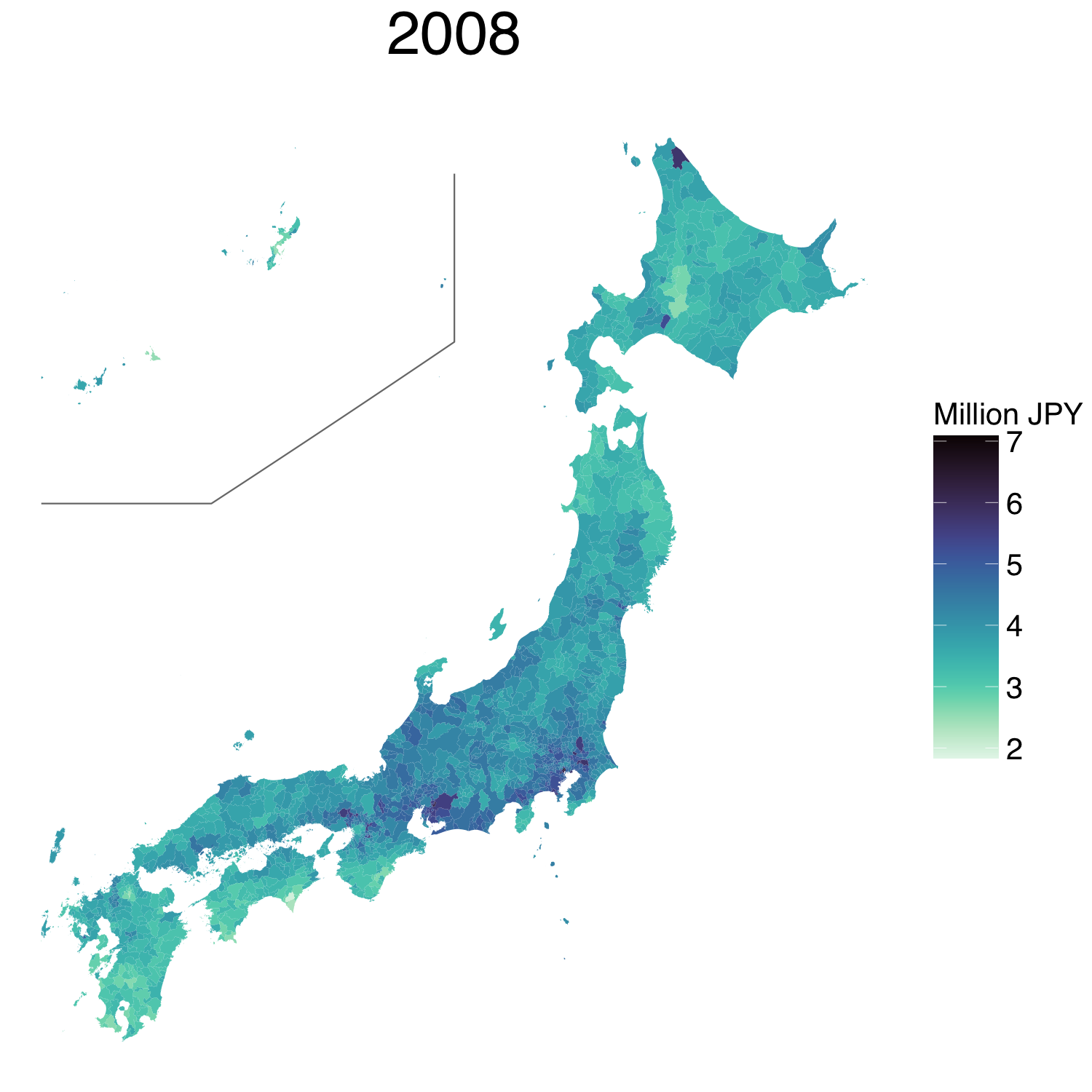}\\
    \includegraphics[scale=0.18]{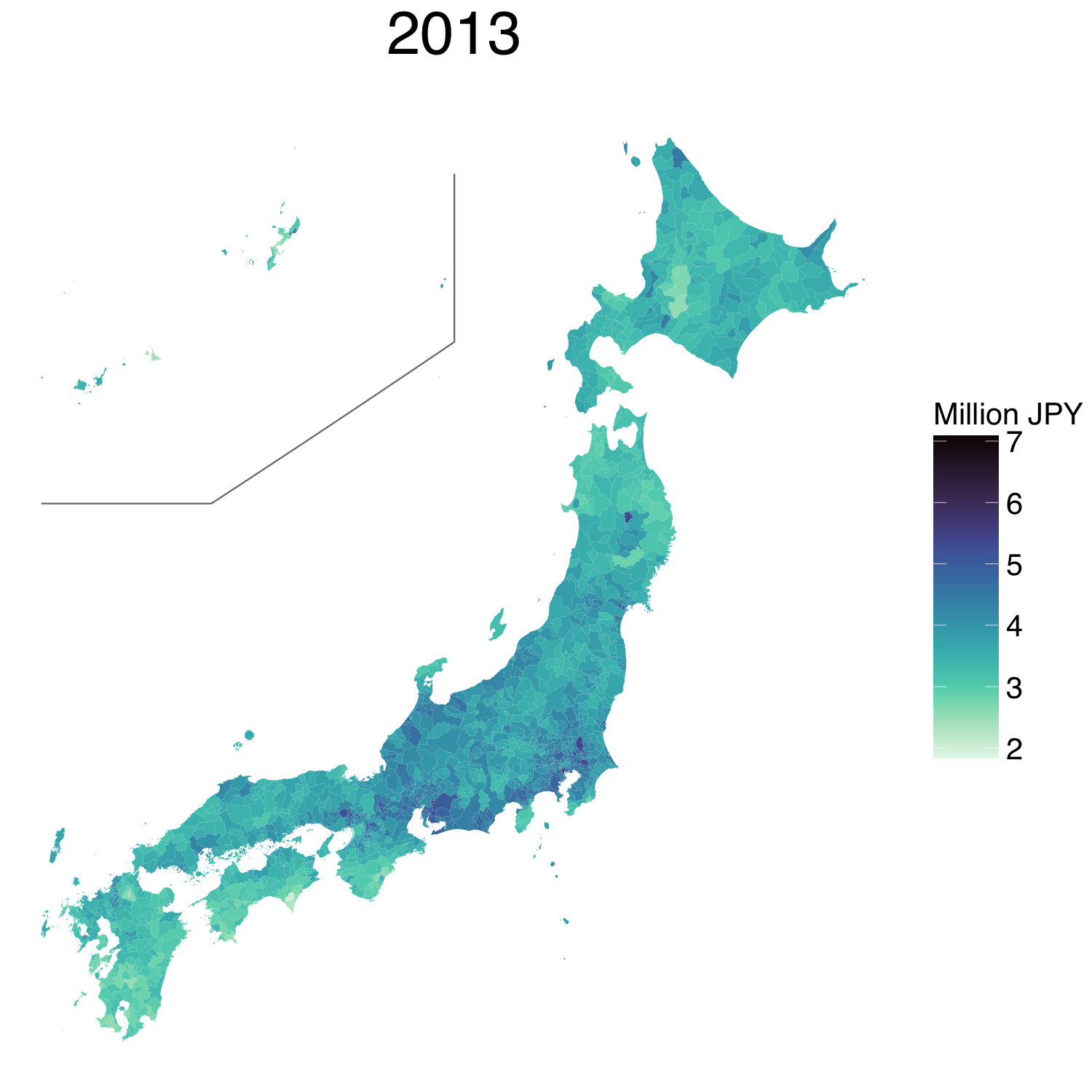}&
    \includegraphics[scale=0.18]{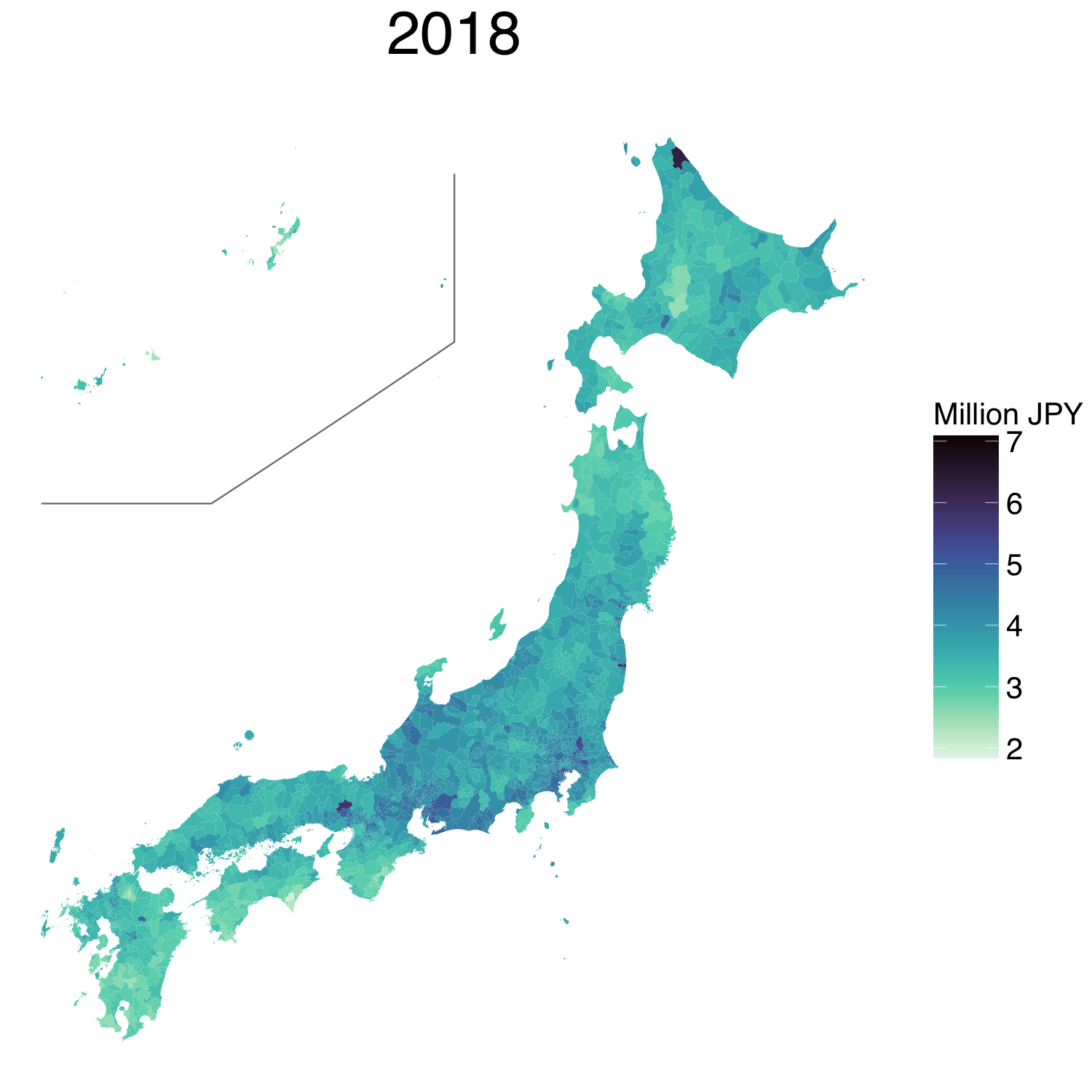}&
    \includegraphics[scale=0.18]{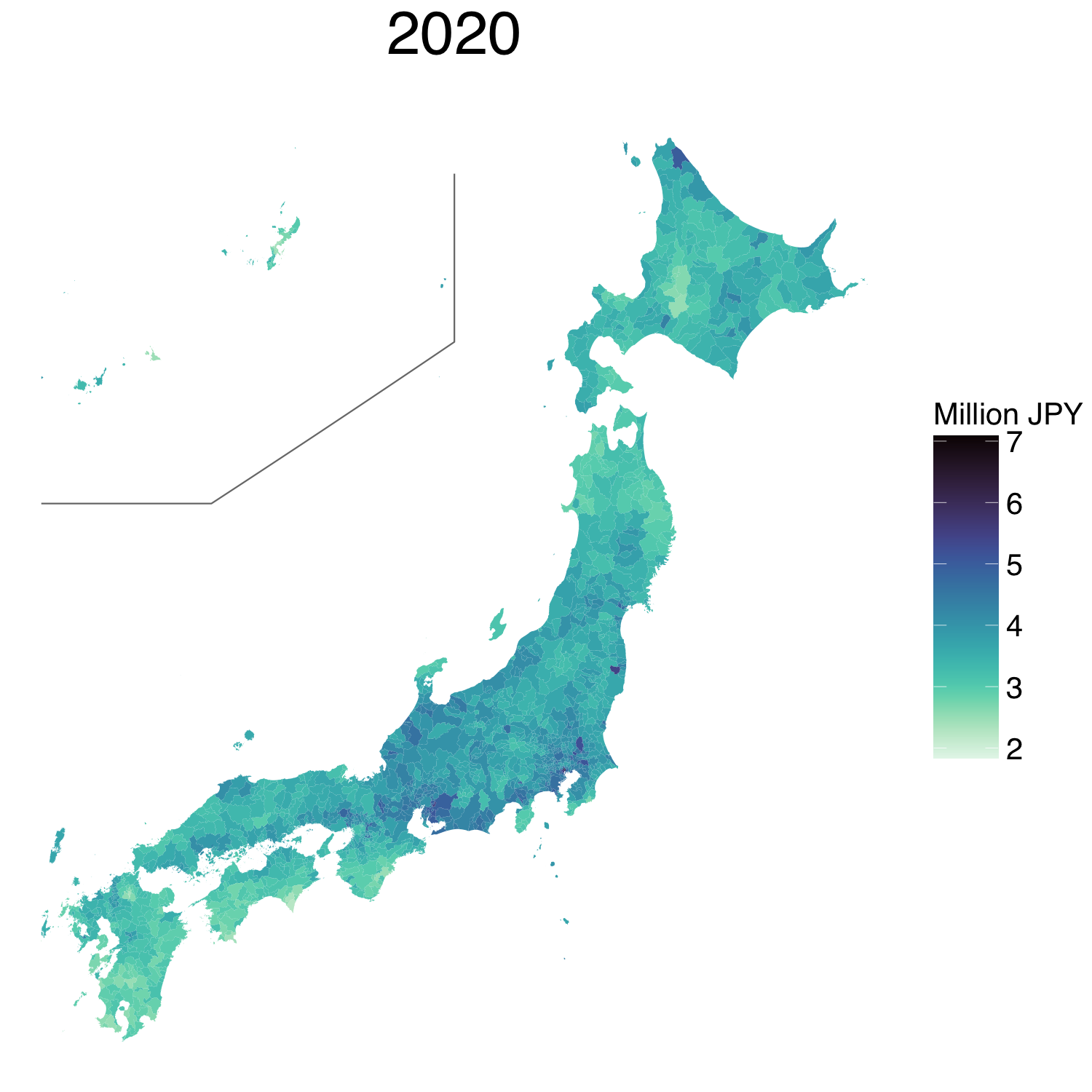}
    \end{tabular}
    \caption{Complete maps of median income}
    \label{fig:medi_map}
\end{figure}

\begin{figure}[H]
    \centering
    \includegraphics[width=\linewidth]{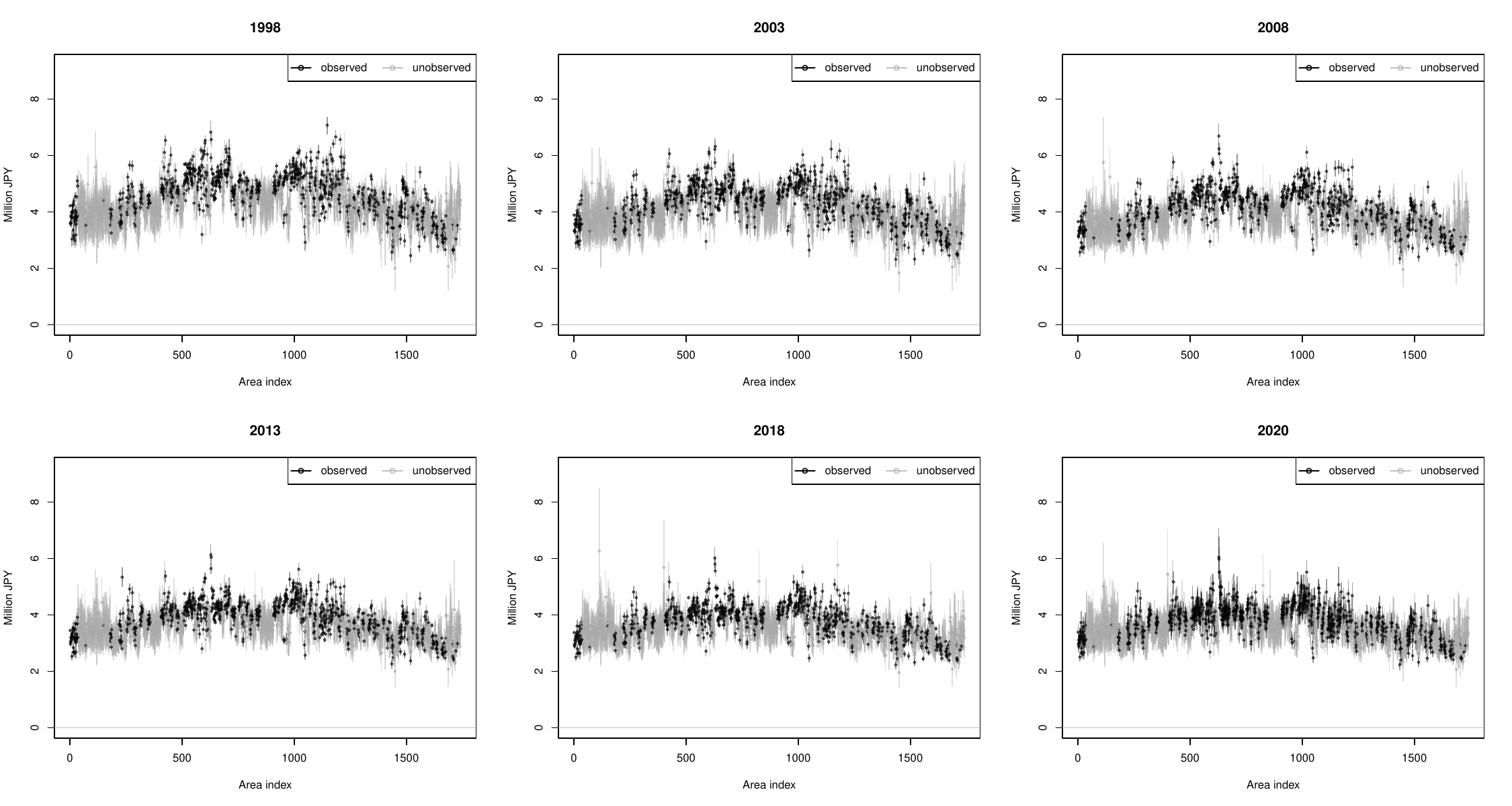}
    \caption{{Posterior and posterior predictive means (points) and 95\% credible and prediction intervals (line segments) for median income for the observed (black) and the unobserved (grey) municipalities.}}
    \label{fig:ci_mi}
\end{figure}

Figure~\ref{fig:medi_map} presents the complete maps of median income. 
{{Figure~\ref{fig:ci_mi}} displays the associated 95\% credible and prediction intervals. }
The figures reveal that the spatial and temporal patterns for  median income levels and their 95\% intervals follow closely mirror those for average income described in the main text.

\subsection{Locations of the eight selected municipalities}
\begin{figure}[H]
    \centering
    \includegraphics[scale=0.3]{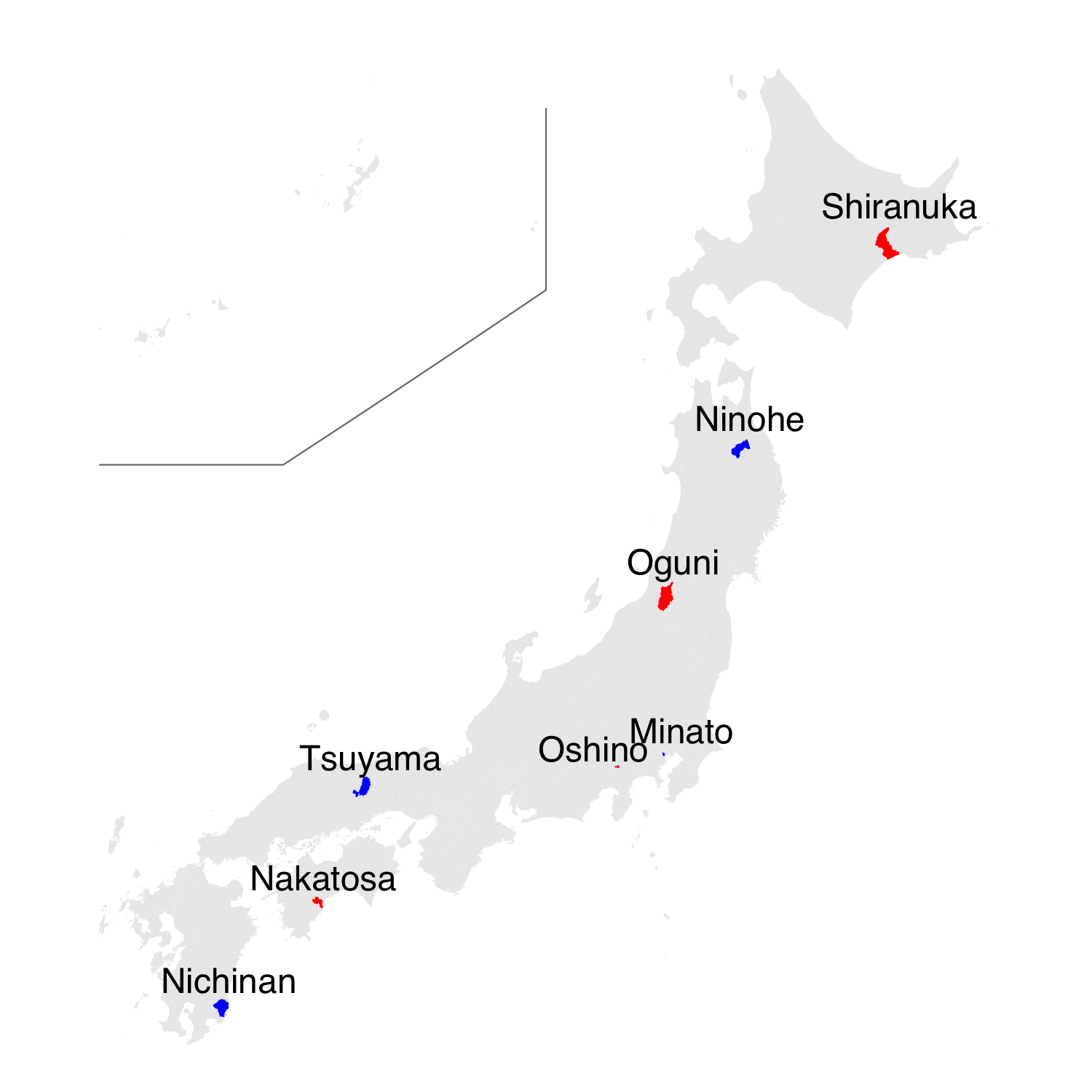}
    \caption{Locations of the eight selected municipalities (blue: observed; red: unobserved).}
    \label{fig:map_id}
\end{figure}

Figure~\ref{fig:map_id} presents the locations of the eight selected municipalities. 
The observed and unobserved municipalities are indicated in blue and red, respectively. 

\subsection{Additional figures for prior sensitivity check}\label{sup:alt}
In Section~\ref{sec:alt} of the main text, we compared the posterior distributions for average and median income levels and the Gini index under the default and alternative prior specifications. 
The default priors are $\sigma_k\sim Exp(1)$, $\alpha_k\sim Exp(5)$, and $\tau_k\sim Exp(5)$.
The alternative priors considered here are $\sigma_k\sim Exp(0.5)$,  $\alpha_k \sim Exp(2.5)$, and $\tau_k\sim Exp(2.5)$. 
The prior means are all doubled, thereby allowing for a broader range of standard deviation values. 

\begin{figure}[H]
    \centering
    \includegraphics[width=\textwidth]{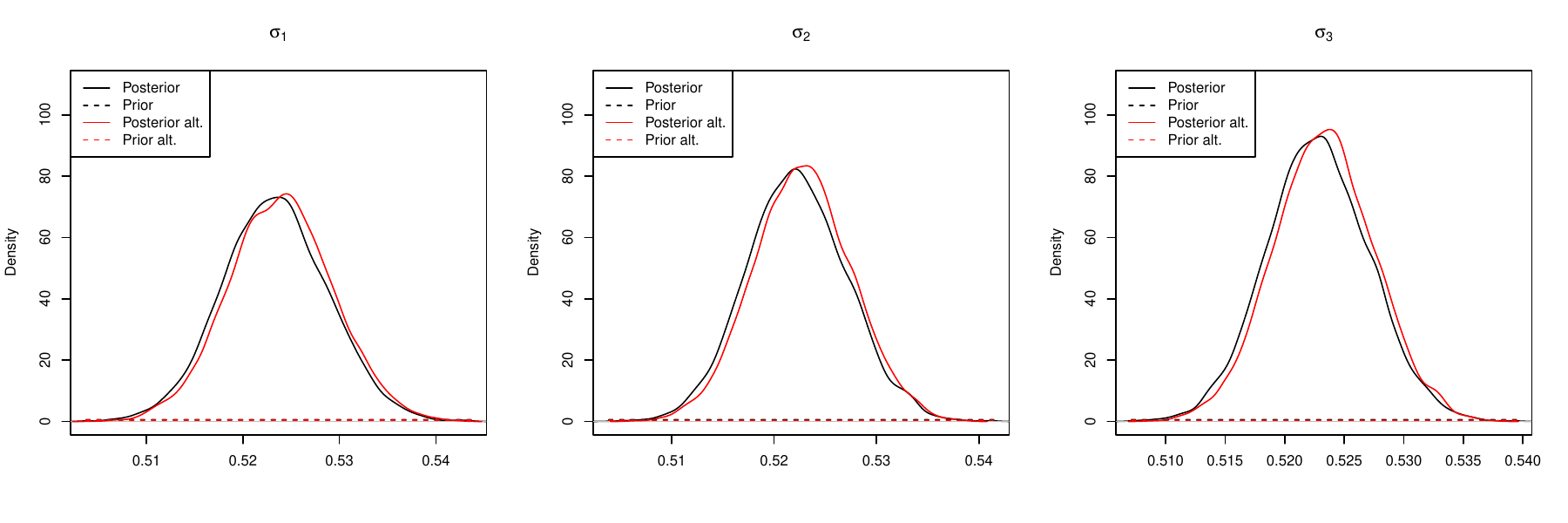}\\
    \includegraphics[scale=0.45]{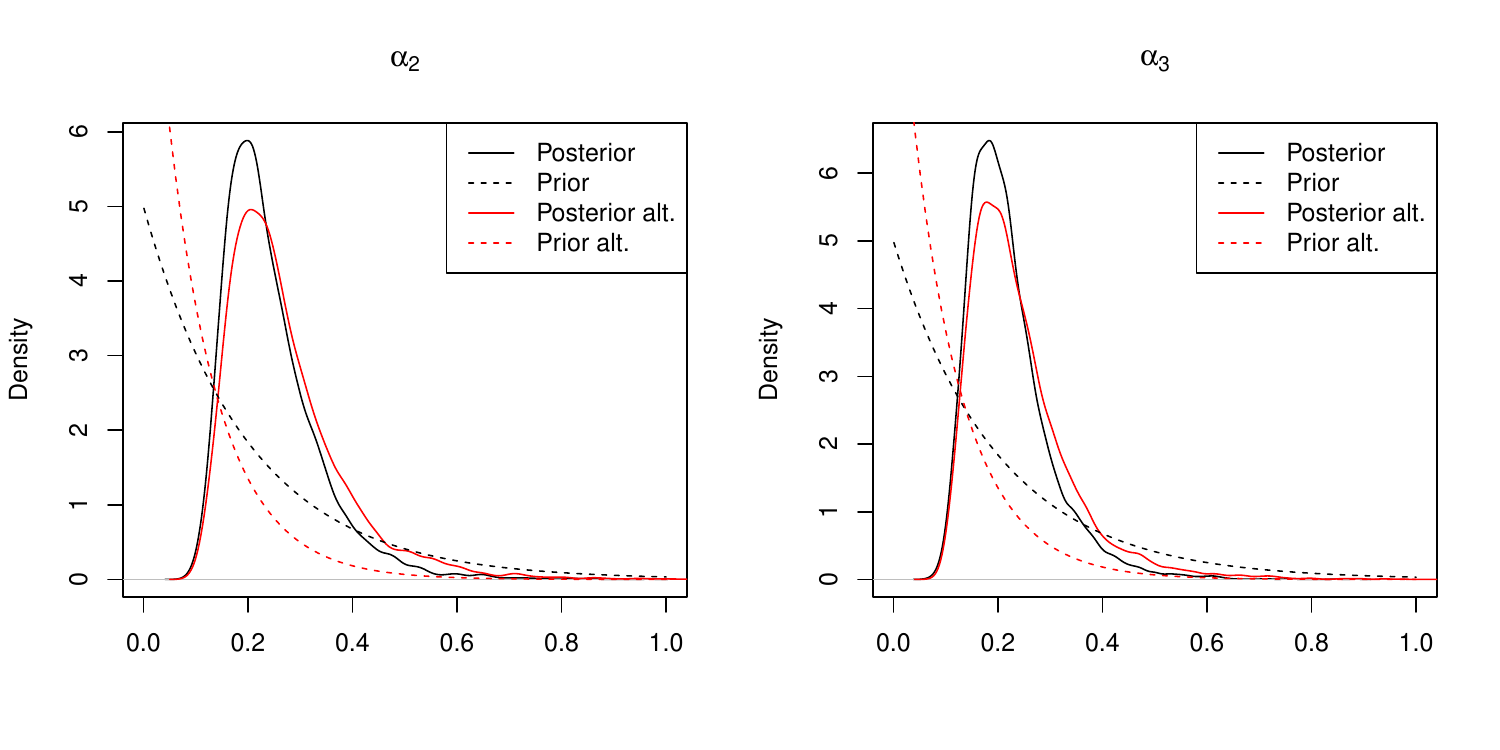}\\
    \includegraphics[scale=0.45]{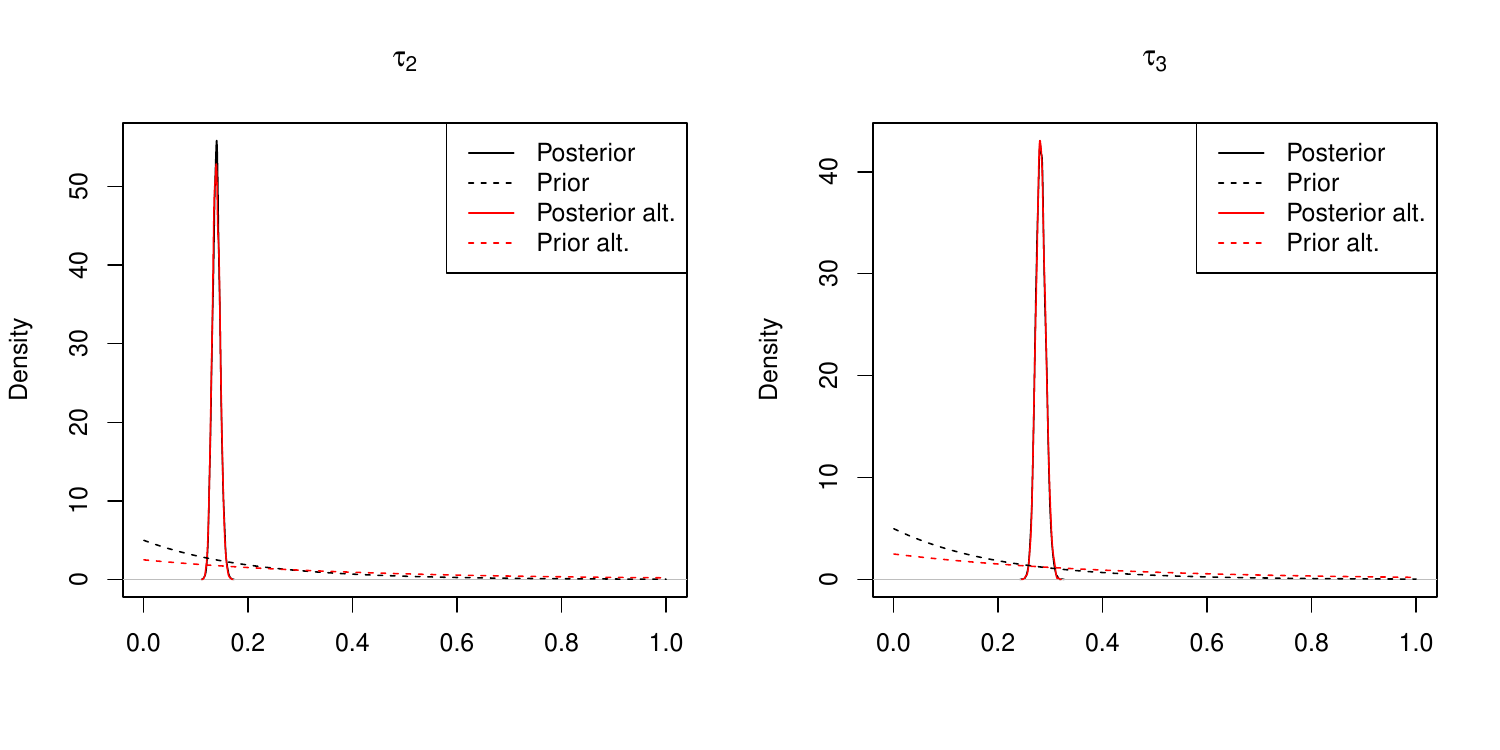}
    \caption{Posterior and prior distributions of $\sigma_k$, $\alpha_k$, and $\tau_k$ under the default and alternative prior specifications.}
    \label{fig:check_alt}
\end{figure}

Figure~\ref{fig:check_alt} presents the default and alternative priors, along with the resulting posterior distributions for $\sigma_k$, $\alpha_k$, and $\tau_k$. 
The posterior distributions of $\sigma_k$ and $\tau_k$ under the two prior settings are nearly identical. 
For instance, the posterior mean and 95\% credible interval (CI) for $\tau_2$ are $0.179$ and $(0.161, 0.198)$ under the default setting, compared to $0.179$ and $(0.160, 0.198)$ under the alternative setting. 
Similarly, for $\sigma_1$, the posterior mean and 95\% CI are $0.525$ and $(0.514, 0.535)$ under the default setting, and $0.525$ and $(0.515, 0.536)$ under the alternative setting.

While the posterior distributions of $\alpha_k$ under the alternative priors cover slightly larger values than those under the default priors, the overall impact is minimal. 
For example, the posterior mean and 95\% CI of $\alpha_3$ are $0.222$ and $(0.118, 0.433)$ under the default setting, and $0.242$ and $(0.121, 0.491)$ under the alternative setting. 
As demonstrated in the main text, the influence of these prior specifications on the primary quantities of interest is minuscule.

\section{Details of the small area models} \label{sup:sae}
\subsection{Space-time model}
In Setting~1, the model for the crude average income is given by 
\[
\hat{y}_{it} = \x^t_{it}\bbe+u_i+\eta_t+e_{it}, \quad i=1,\dots,M, \quad t=1,\dots,T,
\]
where $e_{it}$ is the sampling error following $N(0,\sigma^2_{it})$ with $\sigma^2_{it}$ assumed known. 
The term $u_i$ represents the spatial effect following the SAR model, as in the proposed model, while $\eta_t$ is the temporal effect following the random walk process, $\eta_t|\eta_{t-1}\sim N(\eta_{t-1},\alpha^2)$. 

The average income $\hat{y}_{it}$ is computed as $N_{it}^{-1}\sum_{g=1}^GN_{itg}{\rm mid}_{tg}$, where ${\rm mid}_{tg}$ is the mid-point of the $g$-th income class. 
The known sampling variance is calculated as $\sigma_{it}^2=N_{it}^{-2}\sum_{g=1}^GN_{itg}({\rm mid}_{tg}-\hat{y}_{it})^2$. 
For the top-income class ($g=G$) with an open-ended interval, ${\rm mid}_{tG}$ is set to $Z_G+\frac{1}{2}(Z_{G}-Z_{G-1})=17.5$. 

As the present simulation study focuses on evaluating the performance in spatial interpolation and temporal prediction, we assume a common temporal effect across all areas; this contrasts with the approach of \cite{marhuenda2013small}, who considered area-specific temporal effects. 
Regarding the prior distributions, we assume $\bbe\sim N(\zero, \S_\beta)$, $\tau^2\sim IG(n_\tau,s_\tau)$, $\alpha^2\sim IG(n_\alpha,s_\alpha)$, $\eta_0\sim N(0,c_\eta)$, and $\rho\sim U(0,1)$. 
We set $\S_\beta=10^8\I$, $n_\tau=s_\tau=n_\alpha=s_\alpha=1$, $c_\eta=10$. 

To simulate from the posterior distribution, the following Gibbs sampler is implemented:

\begin{itemize}
    \item 
    Sampling $\bbe$: $\bbe$ is drawn from $N(\m_\beta,\V_\beta)$ where
    \[
    \V_\beta = \left[\S_\beta^{-1}+\sum_{i=1}^m\sum_{t=1}^T\frac{\x_{it}\x_{it}^t}{\sigma_{it}^2}\right]^{-1},\quad
    \m_\beta=\V_\beta\left[ \sum_{i=1}^m\sum_{t=1}^T\frac{\x_{it}}{\sigma_{it}^2}(\hat{y}_{it}-u_i-\eta_t)\right]. 
    \]

    \item 
    Sampling $\u$: the $\u=(u_1,\dots,u_m)^t$ is drawn from $N(\m_u, \V_u)$ where
    \[
    \V_u = \left[\sum_{t=1}^T\bSigma_t^{-1} + \frac{1}{\tau^2}\V_{11}^{-1}\right]^{-1},\quad 
    \m_u=\V_u\left[\sum_{t=1}^T\bSigma_t^{-1}(\hat{\y}_t-\X_t\bbe-\eta_t\biota_m)\right], 
    \]
    where $\bSigma_t=\diag(\sigma_{1t}^2,\dots,\sigma_{mt}^2)$, $\hat{\y}_{t}=(\hat{y}_{1t},\dots,\hat{y}_{mt})^t$, $\X_t=[\x_{1t},\dots,\x_{mt}]^t$, $\V_{11}$ is defined in the same manner as in the proposed model. 

    \item 
    Sampling $\eta_t$: For $t=1,\dots,T$, $\eta_{t}$ is sampled from $N(m_{\eta,t},V_{\eta,t})$, where 
    \begin{eqnarray*}
    V_{\eta,t}&=&\left\{
    \begin{array}{ll}
    \left(\sum_{i=1}^m\frac{1}{\sigma_{it}^2} + \frac{2}{\alpha^2}\right)^{-1}    &  \text{if}\quad t\neq T,\\
\left(\frac{1}{\sigma_{it}^2} + \frac{1}{\alpha^2}\right)^{-1}    &  \text{if}\quad t= T
\end{array}
\right.\\
m_{\eta,t}&=&V_{\eta,t}\left[\sum_{i=1}^m \frac{\hat{y}_{it}-\x_{it}^t\bbe-u_i}{\sigma_{it}^2}+\frac{e_t}{\alpha^2}\right],\\
e_t&=&\left\{
\begin{array}{ll}
\eta_{t-1}    &  \text{if}\quad t=T,\\
\eta_{t-1}+\eta_{t+1}    &  \text{otherwise}.
\end{array}
\right.\\
\end{eqnarray*}

\item 
Sampling $\eta_0$ and $\rho$ is done in the same manner as in the case of the proposed model. 

\item 
Sampling $\tau^2$: $\tau^2$ is sampled from $IG(n_\tau+m/2, s_\tau+\u^t\V_{11}^{-1}\u/2)$. 

\item 
Sampling $\alpha^2$: $\alpha^2$ is sampled from $IG(n_\alpha+T/2, s_\alpha+\sum_{t=1}^T(\eta_t-\eta_{t-1})^2/2)$. 

\end{itemize}

\subsection{Two-fold model}
In Setting~2 of the simulation study, where a sub-area structure is considered, we fit the following two-fold model, considered by \cite{Torabi}, independently for each period. 
In this model, the crude average income $\hat{y}_{bj}$ is expressed as  
\[
\hat{y}_{bj} = \x^t_{bj}\bbe+v_b+u_{bj}+e_{bj},\quad b=1,\dots,B,  \quad j=1,\dots,M_b,
\]
where $v_b\sim N(0,\sigma^2_{v})$ represents the block effect, $u_{bj}\sim N(0,\sigma^2_{u})$ is the sub-area effect, and $e_{bj}\sim N(0,\sigma^2_{ebj})$ is the sampling error.

For the $j$-th sub-area within the $b$-th block, which corresponds to the $i$-th area such that $b=h_i$, the average income $\hat{y}_{bj}$ is calculated as $N_{it}^{-1}\sum_{g=1}^GN_{itg}{\rm mid}_{tg}$. 
The fixed sampling variance $\sigma^2_{ebj}$ is calculated as $N_{it}^{-2}\sum_{g=1}^GN_{itg}({\rm mid}_{tg}-\hat{y}_{bj})^2$.
Here, $M_b$ denotes the number of sub-areas in the $b$-th block and $B$ is the number of blocks.
In the simulation study in the main text, $B=49$ and $\sum_{b=1}^B M_b=1700$.

Similar to the space-time model, for prior distributions of $\bbe$, $\sigma_v^2$ and $\sigma_u^2$, we assign conditionally conjugate priors, $\bbe\sim N(0, \S_{\beta})$, $\sigma_v^2\sim {\rm IG}(n_v, s_v)$, and $\sigma_u^2\sim {\rm IG}(n_u, s_u)$. 
In our simulation study,  we set $\S_{\beta}=10^8 \I$ and $n_v=n_u=s_v=s_u=1$.

To generate posterior samples of the random effects and model parameters, we employ the Gibbs sampler described as follows: 

\begin{itemize}
\item 
Sampling $\bbe$: $\bbe$ is drawn from $N(\m_\beta, \V_\beta)$, where 
$$
\V_\beta=\left[\S_\beta^{-1} + \sum_{b=1}^B\sum_{j=1}^{M_b}\frac{\x_{bj}\x_{bj}^t}{\sigma^2_{ebj}}\right]^{-1}, \ \ \ \  
\m_\beta = \V_\beta \left[\sum_{b=1}^B\sum_{j=1}^{M_b} \frac{\x_{bj}}{\sigma^2_{ebj}}(\hat{y}_{bj}-v_b-u_{bj})\right].
$$

\item 
Sampling $v_b$: For $b=1,\ldots,B$, $v_b$ is drawn from $N(m_{v,b}, V_{v,b})$, where 
$$
V_{v,b}=\left(\frac{1}{\sigma_v^2} + \sum_{j=1}^{M_b}\frac{1}{\sigma^2_{ebj}} \right)^{-1}, \ \ \ \ 
m_{v,b}=V_{v,b} \sum_{j=1}^{M_b} \frac{1}{\sigma^2_{ebj}} (\hat{y}_{bj}-\x_{bj}^t \bbe - u_{bj}).
$$

\item 
Sampling $\sigma_v^2$: $\sigma_v^2$ is drawn from $IG(n_v+B/2, n_v+\sum_{b=1}^B v_b^2/2)$.

\item 
Sampling $u_{bj}$: For $b=1,\ldots,B$ and $j=1,\ldots,M_b$, $u_{bj}$ is drawn from $N(m_{u,b,j}, V_{u,b,j})$, where 
$$
V_{u,b,j}=\left(\frac{1}{\sigma_u^2} + \frac{1}{\sigma^2_{ebj}} \right)^{-1}, \ \ \ \ 
m_{u,b,j}= \frac{V_{v,b}}{\sigma^2_{ebj}} (\hat{y}_{bj}-\x_{bj}^t \bbe - v_b).
$$

\item 
Sampling $\sigma_u^2$: $\sigma_u^2$ is drawn from $IG(s_u+\sum_{b=1}^B M_b/2, s_u+\sum_{b=1}^B \sum_{j=1}^{M_b}u_{bj}^2/2)$.

\end{itemize}

\bibliographystyle{chicago}
\bibliography{Ref}

\end{document}